\documentclass[fleqn,usenatbib]{mnras}

\usepackage[T1]{fontenc}
\usepackage[normalem]{ulem}
\DeclareRobustCommand{\VAN}[3]{#2}
\let\VANthebibliography\thebibliography
\def\thebibliography{\DeclareRobustCommand{\VAN}[3]{##3}\VANthebibliography}

\usepackage{graphicx, subfigure}	
\usepackage{amsmath}	
\usepackage{amssymb}	
\usepackage{makecell}
\usepackage{xcolor}
\usepackage{comment}
\usepackage{multirow}
\usepackage{lineno}
\usepackage{newtxtext,newtxmath}
\def \paul{\color{black}}
\def \p{\color{black}}

\newcommand{\teff}{T_{\rm eff}}
\newcommand{\logg}{\log g}
\newcommand{\mh}{\rm{[M/H]}}
\newcommand{\feh}{\rm{[Fe/H]}}
\newcommand{\vsini}{v\sin{i}}
\newcommand{\vmac}{v_{\rm mac}}
\newcommand{\vbroad}{v_{\rm b}}
\newcommand{\kms}{\rm km.s^{-1}}
\newcommand{\mann}{M15}
\newcommand{\pass}{P19}

\title[Estimating fundamental parameters]{Estimating fundamental parameters of nearby M dwarfs from SPIRou spectra}

\author[Cristofari et al.]{
P. I. Cristofari$^{1}$\thanks{E-mail: paul.cristofari@irap.omp.eu (IRAP)},
J.-F. Donati$^{1}$,
T. Masseron$^{2,3}$,
P. Fouqué$^{1,4}$,
C. Moutou$^{1}$,
X. Delfosse$^{5}$,
E. Artigau$^{6}$,
\newauthor
C. P. Folsom$^{7}$,
A. Carmona$^{5}$,
E. Gaidos$^{8}$,
J.-D.~do~Nascimento,~Jr.$^{9, 10}$,
F. Jahandar$^{6}$,
G. Hébrard$^{11, 12}$
\\
$^{1}$Univ. de Toulouse, CNRS, IRAP, 14 av. Belin, 31400 Toulouse, France\\
$^{2}$Instituto de Astrofísica de Canarias, E-38205 La Laguna, Tenerife, Spain\\
$^{3}$Departamento de Astrofísica, Universidad de La Laguna, E-38206 La Laguna, Tenerife, Spain\\
$^{4}$Canada-France-Hawaii Telescope, CNRS, Kamuela, HI 96743, USA\\
$^{5}$Univ. Grenoble Alpes, CNRS, IPAG, 38000 Grenoble, France\\
$^{6}$Universit\'e de Montr\'eal, D\'epartement de Physique, IREX, Montr\'eal, QC H3C 3J7, Canada\\
$^{7}$Tartu Observatory, University of Tartu, Observatooriumi 1, Tõravere, 61602 Tartumaa, Estonia\\	
$^{8}$Department of earth sciences, University of Hawai'i at M\=anoa, Honolulu, HI 96822, USA\\
$^{9}${Dept. of  Physics, Univ. Federal of Rio Grande do Norte (UFRN), 59078-970, Natal, RN, Brazil}\\
$^{10}${Center for Astrophysics | Harvard \& Smithsonian, 60 Garden Street, Cambridge MA 02138, USA}\\
$^{11}$Institut d’astrophysique de Paris, UMR7095 CNRS, Universit\'ee Pierre \& Marie Curie, 98bis boulevard Arago, 75014 Paris, Frances \\
$^{12}$Observatoire de Haute-Provence, CNRS, Universit\'ee d’Aix-Marseille, 04870 Saint-Michel-l’Observatoire, France \\
}

\date{Accepted 2021 December 10. Received 2021 December 10; in original form 2021 November 08}
\pubyear{2015}

\begin{document}
\label{firstpage}
\pagerange{\pageref{firstpage}--\pageref{lastpage}}
\maketitle

\begin{abstract}
	We present the results of a study aiming at retrieving the fundamental parameters of M dwarfs from spectra secured with SPIRou, the near-infrared high-resolution spectropolarimeter installed at the Canada-France-Hawaii Telescope (CFHT), in the framework of the SPIRou Legacy Survey (SLS). Our study relies on comparing observed spectra with two grids of synthetic spectra, respectively computed from PHOENIX and MARCS model atmospheres, with the ultimate goal of optimizing the precision at which fundamental parameters can be determined.
	In this first step, we applied our technique to 12 inactive M dwarfs with effective temperatures ($\teff$) ranging from 3000 to 4000~K. We implemented a benchmark to carry out a comparison of the two models used in this study. We report that the choice of model has a significant impact on the results and may lead to discrepancies in the derived parameters of 30~K in $\teff$ and 0.05~dex to 0.10~dex in surface gravity ($\logg$) and metallicity ($\mh$), as well as systematic shifts of up to 50~K in $\teff$ and 0.4~dex $\logg$ and $\mh$. The analysis is performed on high signal-to-noise ratio template SPIRou spectra, averaged over multiple observations corrected from telluric absorption features and sky lines, using both a synthetic telluric transmission model and principal component analysis.
	With both models, we retrieve $\teff$, $\logg$ and $\mh$ estimates in good agreement with reference literature studies, with internal error bars of about 30~K, 0.05~dex and 0.1~dex, respectively.
\end{abstract}

\begin{keywords}
stars: fundamental parameters -- stars: low-mass -- infrared: stars -- techniques: spectroscopic
\end{keywords}



\section{Introduction}
\label{sec:introduction}

\begin{table*}
\centering
\caption{Stellar properties of the studied targets. For each target, column 2 presents the spectral type and columns 3-6 respectively list the distance, the absolute K magnitude, the mass derived from the mass-luminosity relation of~\citet{mann_2019}, and the corresponding radius using~\citet{baraffe_2015} models.
Column 7 lists the surface gravity derived from columns 5 and 6. 
Column 8 reports $\logg$ values from angular diameters $\theta_{\rm LD }$ computed from interferometric data~\citep{boyajian_2012}, assuming the distances and masses reported in columns 3 and 5.
Columns~9-11 list the stellar properties from literature; (1):~\mann{}, (2):~\pass{} (nIR), (3):~\pass{} (nIR + optical).
Spectral types, magnitudes and parallaxes where obtained through SIMBAD (\url{http://simbad.u-strasbg.fr/simbad/}) and used to compute absolute magnitudes.
}
\resizebox{\textwidth}{!}{
\begin{tabular}[h]{cccccccccccc}
\\
\hline
Star & \thead{Spectral \\ type} & Distance (pc) & $M_{\rm K}$ & $M_{\star}/M_{\sun}$ & $R_{\star}/R_{\sun}$ & \thead{$\logg$ (dex) \\ from $M_{\star}$ and $R_{\star}$} & \thead{$\logg$ (dex) \\ from interferometry} & $\teff$ (K) & $\logg$ (dex) & $\mh$ (dex)  & Ref. \\
\hline
Gl~846 & M0.5V & 10.555 $\pm$ 0.016 & 5.205 $\pm$ 0.023 & 0.444 $\pm$ -0.004 & 0.416 $\pm$ 0.007 & 4.846 $\pm$ 0.004 &   & 3848 $\pm$ 60 & 4.73 $\pm$ 0.12 & 0.02 $\pm$ 0.08 & (1) \\
&  &   &   &   &   &   &   & 3826 $\pm$ 56 & 4.65 $\pm$ 0.04 & 0.40 $\pm$ 0.16 & (2) \\
&  &   &   &   &   &   &   & 3911 $\pm$ 54 & 4.64 $\pm$ 0.06 & 0.29 $\pm$ 0.19 & (3) \\
Gl~880 & M1.5V & 6.868 $\pm$ 0.002 & 5.339 $\pm$ 0.016 & 0.422 $\pm$ -0.002 & 0.397 $\pm$ 0.004 & 4.866 $\pm$ 0.003 & 4.584 $\pm$ 0.005 & 3720 $\pm$ 60 & 4.72 $\pm$ 0.12 & 0.21 $\pm$ 0.08 & (1) \\
&  &   &   &   &   &   &   & 3784 $\pm$ 56 & 4.65 $\pm$ 0.04 & 0.53 $\pm$ 0.16 & (2) \\
&  &   &   &   &   &   &   & 3810 $\pm$ 60 & 4.65 $\pm$ 0.06 & 0.38 $\pm$ 0.19 & (3) \\
Gl~15A & M2V & 3.563 $\pm$ 0.001 & 6.261 $\pm$ 0.020 & 0.301 $\pm$ -0.002 & 0.300 $\pm$ 0.004 & 4.963 $\pm$ 0.003 & 4.745 $\pm$ 0.005 & 3603 $\pm$ 60 & 4.86 $\pm$ 0.12 & -0.30 $\pm$ 0.08 & (1) \\
&  &   &   &   &   &   &   & 3628 $\pm$ 56 & 4.77 $\pm$ 0.04 & -0.18 $\pm$ 0.16 & (2) \\
&  &   &   &   &   &   &   & 3606 $\pm$ 54 & 4.77 $\pm$ 0.06 & -0.14 $\pm$ 0.19 & (3) \\
Gl~411 & M2V & 2.547 $\pm$ 0.004 & 6.310 $\pm$ 0.050 & 0.295 $\pm$ -0.005 & 0.295 $\pm$ 0.009 & 4.968 $\pm$ 0.008 & 4.722 $\pm$ 0.011 & 3563 $\pm$ 60 & 4.84 $\pm$ 0.12 & -0.38 $\pm$ 0.08 & (1) \\
&  &   &   &   &   &   &   & 3603 $\pm$ 56 & 4.79 $\pm$ 0.04 & -0.21 $\pm$ 0.16 & (2) \\
&  &   &   &   &   &   &   & 3569 $\pm$ 54 & 4.75 $\pm$ 0.06 & -0.01 $\pm$ 0.19 & (3) \\
Gl~752A & M3V & 5.912 $\pm$ 0.002 & 5.814 $\pm$ 0.020 & 0.355 $\pm$ -0.003 & 0.342 $\pm$ 0.004 & 4.921 $\pm$ 0.003 &   & 3558 $\pm$ 60 & 4.76 $\pm$ 0.12 & 0.10 $\pm$ 0.08 & (1) \\
&  &   &   &   &   &   &   & 3633 $\pm$ 56 & 4.66 $\pm$ 0.04 & 0.44 $\pm$ 0.16 & (2) \\
&  &   &   &   &   &   &   & 3583 $\pm$ 54 & 4.69 $\pm$ 0.06 & 0.25 $\pm$ 0.19 & (3) \\
Gl~849 & M3.5V & 8.803 $\pm$ 0.004 & 5.871 $\pm$ 0.017 & 0.347 $\pm$ -0.002 & 0.336 $\pm$ 0.003 & 4.927 $\pm$ 0.003 &   & 3530 $\pm$ 60 & 4.78 $\pm$ 0.13 & 0.37 $\pm$ 0.08 & (1) \\
&  &   &   &   &   &   &   & 3633 $\pm$ 56 & 4.68 $\pm$ 0.04 & 0.54 $\pm$ 0.16 & (2) \\
&  &   &   &   &   &   &   & 3427 $\pm$ 54 & 4.80 $\pm$ 0.06 & 0.09 $\pm$ 0.19 & (3) \\
Gl~436 & M3.5V & 9.756 $\pm$ 0.009 & 6.127 $\pm$ 0.016 & 0.316 $\pm$ -0.002 & 0.312 $\pm$ 0.003 & 4.951 $\pm$ 0.003 &   & 3479 $\pm$ 60 & 4.79 $\pm$ 0.13 & 0.01 $\pm$ 0.08 & (1) \\
&  &   &   &   &   &   &   & 3571 $\pm$ 56 & 4.69 $\pm$ 0.04 & 0.30 $\pm$ 0.16 & (2) \\
&  &   &   &   &   &   &   & 3472 $\pm$ 54 & 4.77 $\pm$ 0.06 & 0.03 $\pm$ 0.19 & (3) \\
Gl~725A & M3V & 3.522 $\pm$ 0.001 & 6.698 $\pm$ 0.020 & 0.256 $\pm$ -0.002 & 0.263 $\pm$ 0.003 & 5.005 $\pm$ 0.003 & 4.746 $\pm$ 0.008 & 3441 $\pm$ 60 & 4.87 $\pm$ 0.12 & -0.23 $\pm$ 0.08 & (1) \\
Gl~725B & M3.5V & 3.523 $\pm$ 0.001 & 7.266 $\pm$ 0.023 & 0.208 $\pm$ -0.002 & 0.224 $\pm$ 0.003 & 5.054 $\pm$ 0.004 & 4.739 $\pm$ 0.016 & 3345 $\pm$ 60 & 4.96 $\pm$ 0.13 & -0.30 $\pm$ 0.08 & (1) \\
Gl~699 & M4V & 1.827 $\pm$ 0.001 & 8.216 $\pm$ 0.020 & 0.150 $\pm$ -0.001 & 0.175 $\pm$ 0.001 & 5.128 $\pm$ 0.002 & 5.071 $\pm$ 0.005 & 3228 $\pm$ 60 & 5.09 $\pm$ 0.12 & -0.40 $\pm$ 0.08 & (1) \\
&  &   &   &   &   &   &   & 3278 $\pm$ 56 & 4.93 $\pm$ 0.04 & -0.13 $\pm$ 0.16 & (2) \\
&  &   &   &   &   &   &   & 3243 $\pm$ 54 & 4.96 $\pm$ 0.06 & -0.09 $\pm$ 0.19 & (3) \\
Gl~15B & M3.5V & 3.561 $\pm$ 0.001 & 8.190 $\pm$ 0.024 & 0.151 $\pm$ -0.001 & 0.176 $\pm$ 0.002 & 5.127 $\pm$ 0.003 &   & 3218 $\pm$ 60 & 5.07 $\pm$ 0.13 & -0.30 $\pm$ 0.08 & (1) \\
&  &   &   &   &   &   &   & 3264 $\pm$ 56 & 4.94 $\pm$ 0.04 & -0.05 $\pm$ 0.16 & (2) \\
&  &   &   &   &   &   &   & 3261 $\pm$ 54 & 4.96 $\pm$ 0.06 & -0.12 $\pm$ 0.19 & (3) \\
Gl~905 & M5.0V & 3.155 $\pm$ 0.001 & 8.434 $\pm$ 0.020 & 0.142 $\pm$ -0.001 & 0.167 $\pm$ 0.001 & 5.143 $\pm$ 0.002 &   & 2930 $\pm$ 60 & 5.04 $\pm$ 0.13 & 0.23 $\pm$ 0.08 & (1) \\
&  &   &   &   &   &   &   & 3143 $\pm$ 56 & 4.97 $\pm$ 0.04 & 0.00 $\pm$ 0.16 & (2) \\
&  &   &   &   &   &   &   & 3069 $\pm$ 54 & 4.97 $\pm$ 0.06 & 0.11 $\pm$ 0.19 & (3) \\
\hline
\end{tabular}
}
\label{tab:literature_values}
\end{table*}

M dwarfs are the most numerous stars of the solar vicinity~\citep{reyle_2021}, and have recently attracted increasing attention in the search for exoplanets located in the habitable zone of their host stars~\citep{gaidos_2016, bonfils_2013, dressing_2013}. 
Determining the fundamental parameters of host stars is a mandatory step for characterizing planets orbiting M dwarfs~\citep{mann_2015, passegger_2019}.

In particular, the goal is to estimate as accurately as possible the effective temperature ($\teff$), surface gravity ($\logg$) and metallicity ($\mh$) of the host stars. 
These parameters are essential to derive accurate masses and radii of the orbiting companions, as these depend on the masses and radii of the stars when relying on indirect detection methods. 

\begin{figure*}
    \centering
    \includegraphics[width=0.8\linewidth, trim={2cm 1.5cm 2cm 0}, clip]{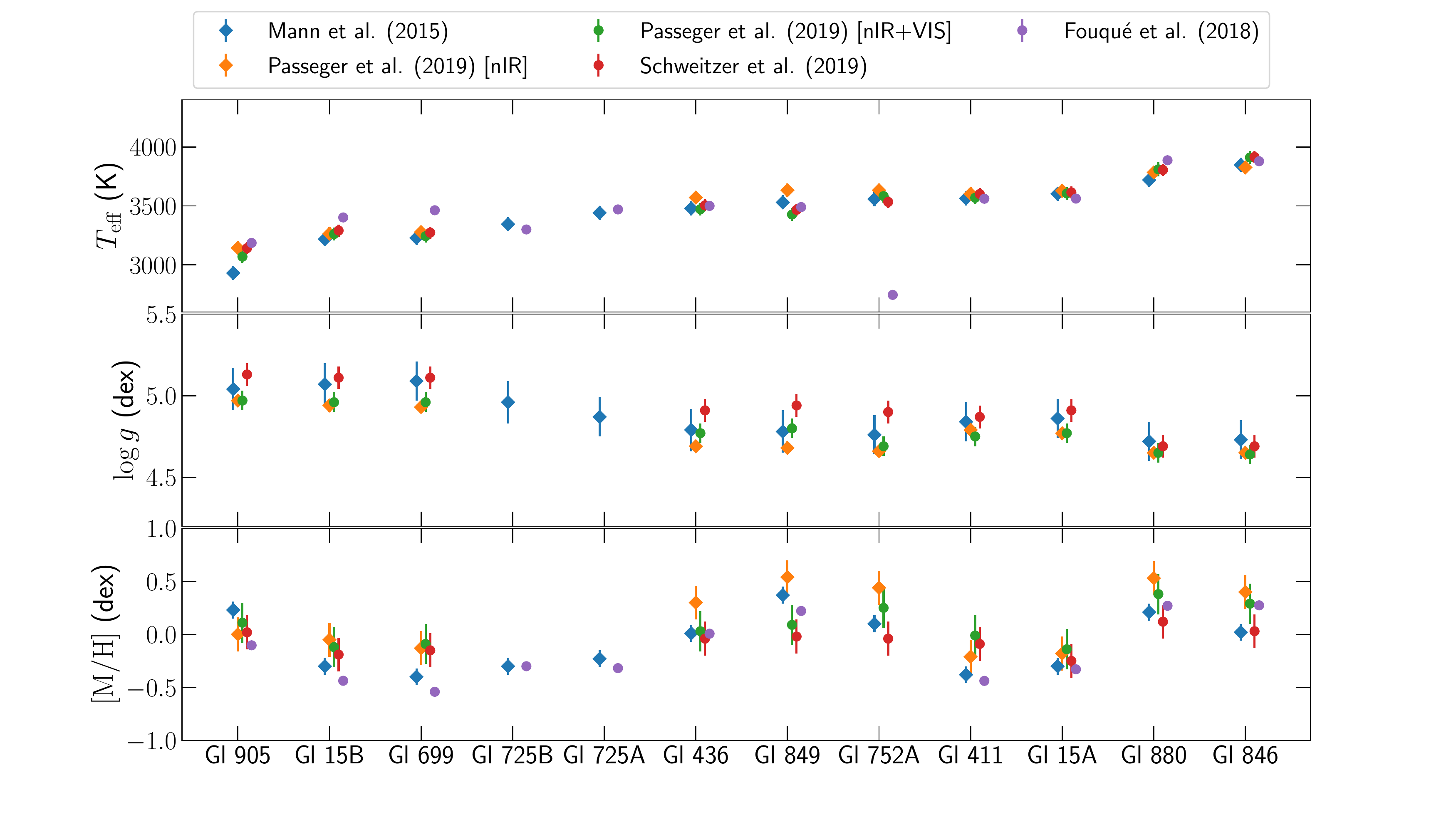}
    \caption{$\teff$, $\logg$ and $\mh$ values extracted from the reference studies of~\citet{mann_2015},~\citet{passegger_2019},~\citet{schweitzer_2019} and~\citet{fouque_2018}. The typical RMS with respect to the mean is of 45~K in $\teff$, 0.07~dex in $\logg$ and 0.15~dex in $\mh$. }
    \label{fig:all_literature}
\end{figure*}

\begin{figure*}
\centering
    \subfigure{
        \centering
    \includegraphics[scale=0.4]{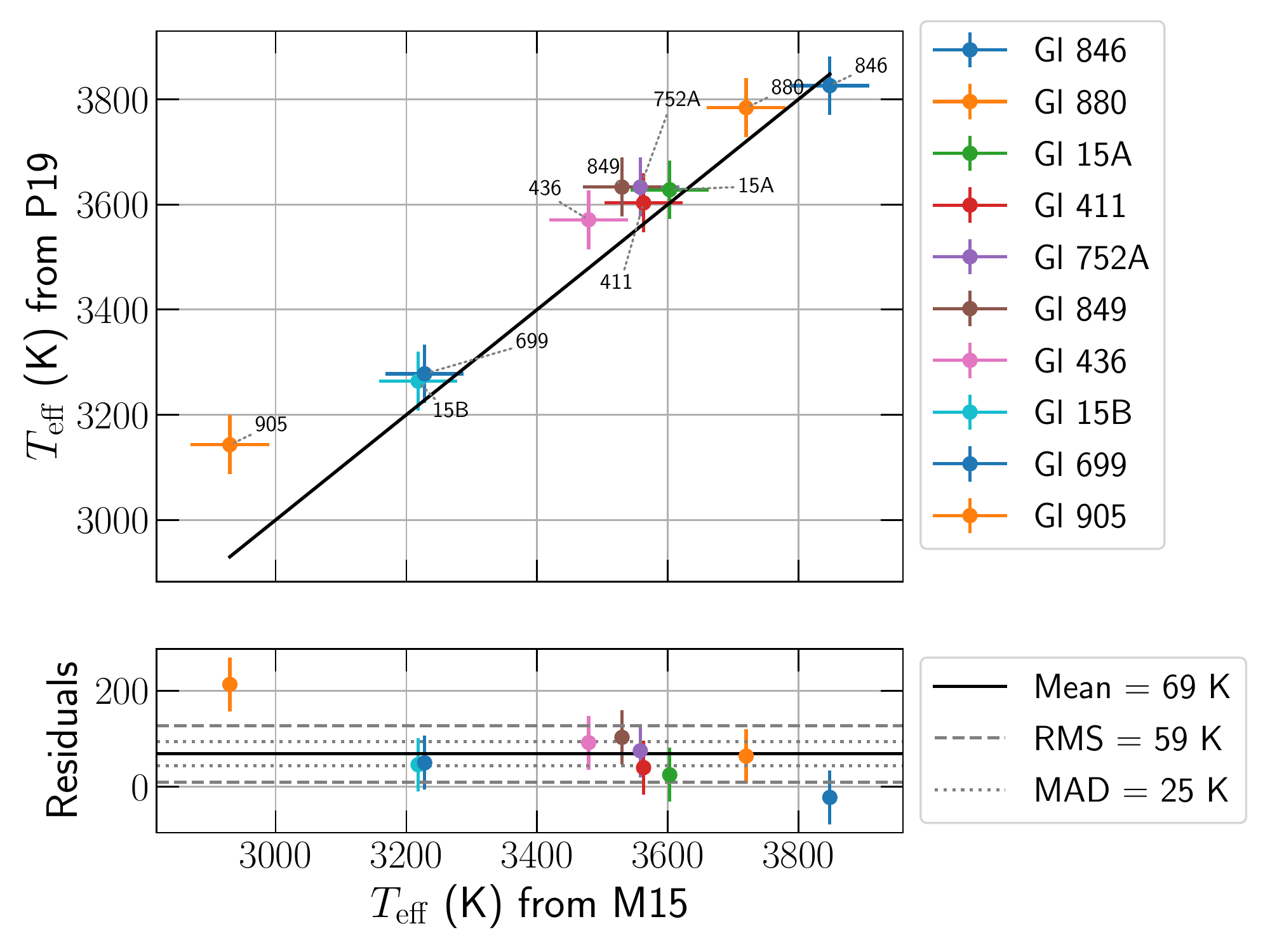}
    }
    \subfigure{
        \centering
    \includegraphics[scale=0.4]{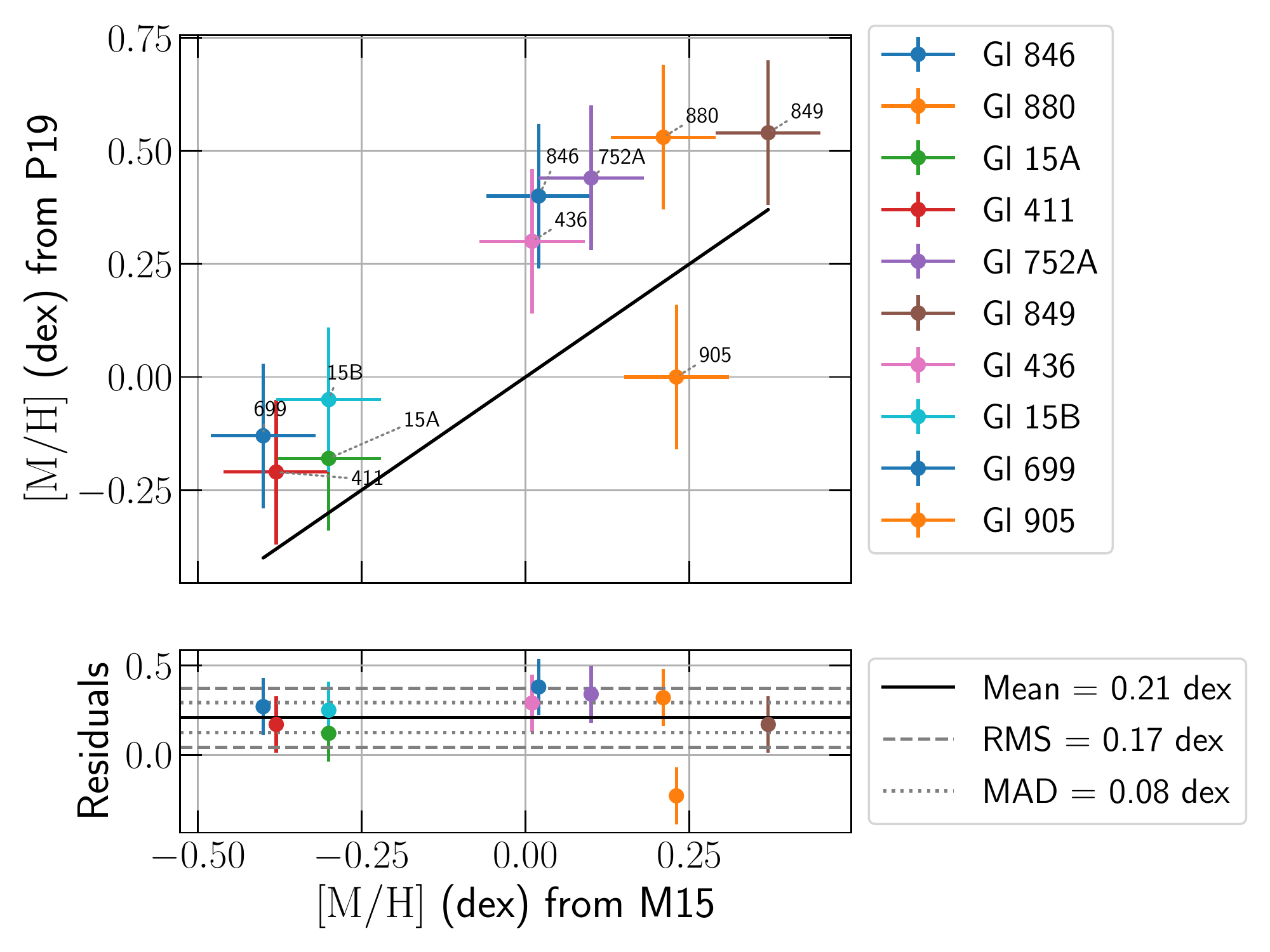}
    }
    \caption{Comparison between parameters published by~\pass{} (using near-infrared regions only) and~\mann{}. Top plots present the values retrieved by~\pass{} plotted against the values retrieved by~\mann{} for effective temperature (left) and metallicity (right). Bottom plots present the residuals, i.e. values retrieved by~\pass{} minus values retrieved by~\mann{}. We additionally display the mean value, standard deviation and median absolute deviation of the residuals.}
    \label{fig:pass_mann}
\end{figure*}

Multiple techniques have been developed to study these parameters  by, e.g., adjusting equivalent widths of spectral lines~\citep{rojas-ayala_2010, neves_2014,fouque_2018}, fitting spectral energy distributions (SEDs) to low to mid-resolution spectra~\citep{mann_2013},
or fitting synthetic models to high resolution spectra~\citep{passegger_2018,passegger_2019,schweitzer_2019}. For instance, \citet[hereafter \mann{}]{mann_2015} derived $\teff$, $\feh$, masses and radii of M dwarfs using empirical mass--magnitude relations, equivalent widths and BT-settl PHOENIX models with low resolution spectra (R $\simeq$ 1000). In contrast, \citet[hereafter \pass{}]{passegger_2019} performed fits of synthetic models on high resolution CARMENES spectra, computing $\logg$ from empirical $\teff$--$\logg$ relations. These different approaches typically result in different parameter values, as illustrated in Fig.~\ref{fig:all_literature} for the 12 inactive nearby M dwarfs which this paper will focus on. In particular, we compare the values published by~\mann{} and~\pass{} in Fig.~\ref{fig:pass_mann}, and recall the estimates derived by these two references in Table~\ref{tab:literature_values}\footnote{In this paper, we assume that the overall metallicity $\mh$ = $\feh$, considering no alpha enhancement as a simplifying assumption, and we therefore use the label $\mh$ in all circumstances.}.

The ultimate goal of the study we embark on, of which the present paper is a first step, is to optimize the determination of these fundamental parameters taking advantage of the large homogeneous collection of SPIRou spectra recorded in the framework of the SPIRou Legacy Survey (SLS).
Comparing high-resolution spectra of observed M dwarfs to dense grids of synthetic spectra derived from theoretical model atmospheres is presumably the most promising approach to this problem. However, the high complexity of the spectra, featuring large amounts of molecular and atomic lines, renders this approach challenging.
For such studies to be attempted, high-resolution spectroscopy is mandatory, in order to resolve individual spectral features and their profile shapes, and thereby guide us to a more reliable spectral modeling of M dwarfs.

In practice, this requires accurate synthetic spectra that can be compared with observations.  
Throughout the last decade, multiple codes have been developed to produce synthetic spectra based on observational and experimental data (e.g., the properties of atomic and molecular lines). 
Codes such as MOOG~\citep{sneden_2012}, SME~\citep{valenti_2012}, SYNTHE~\citep{kurucz_2005} or Turbospectrum~\citep{plez_1998, plez_2012} can compute synthetic spectra for different types of stars. These tools typically rely on pre-computed atmosphere models, such as MARCS~\citep{gustafsson_2008}, or ATLAS~\citep{kurucz_1970}, and use radiative transfer codes to compute the emergent high-resolution spectra. In contrast, PHOENIX {\paul performs the computation of both the
} model atmosphere and the emergent spectrum.
These models are usually based on a number of assumptions, such as Local Thermodynamic Equilibrium (LTE) or Non-Local Thermodynamic equilibrium (NLTE), plane-parallel atmospheres or spherical geometry, and the way the micro-turbulence is taken into account.

PHOENIX is widely considered as one of the most advanced tool for computing stellar atmospheres of M dwarfs and the corresponding emergent spectra. The most recent grid of atmosphere models and synthetic spectra, baptized PHOENIX-ACES models, was published in 2013~\citep{husser_2013}, updated in 2015, and covers a temperature range from 2300 to 12000~K, suitable for the studies of various objects, such as M dwarfs and giants.
MARCS models have been used in several studies focusing on FGK stars~\citep{blanco_2014, tabernero_2019}, and more recently on M dwarfs~\citep{sarmento_2021}.
In particular, recent publications~\citep{passegger_2018, passegger_2019, rajpurohit_2018, flores_2019, sarmento_2021} have reported the use of PHOENIX and MARCS models to derive stellar properties of M dwarfs and young low-mass stars from high-resolution spectra secured with various instruments such as CARMENES~\citep{nowak_2020}, iSHELL~\citep{rayner_2016} or APOGEE~\citep{wilson_2019}, in the near-infrared (nIR) and/or optical domains. The study of the nIR domain, and the development of high-resolution spectrographs working in this spectral range, is mainly motivated by the hunt for planets orbiting very--low--mass stars that are often too faint to be observed in the optical domain.
The most up-to-date models are however quite far from precisely reproducing every single line across the entire wavelength range. This is particularly true for the nIR domain, 
for which data is still limited.


In the present study, we analyze nIR high-resolution spectra acquired with the SpectroPolarimètre Infra-Rouge ~\citep[SPIRou,][]{donati_2020} installed at the Canada-France-Hawaii Telescope (CFHT) to determine the fundamental parameters of twelve M dwarfs with effective temperatures ranging from about 3000 to 4000~K, using both PHOENIX-ACES and MARCS synthetic spectra.
With this work, we push forward the efforts of previous studies
and try to improve the accuracy on parameters measurements from nIR spectroscopy.
In particular, we take advantage of the high resolving power (R$\sim$70~000) of SPIRou, which covers a spectral range in a single exposure spanning 980--2350~nm, allowing to observe spectral lines in nIR bands for which few high-resolution observations are currently available.
By collecting spectra of M dwarfs at different epochs, we are able to accurately correct for telluric absorption features and sky lines throughout the nIR domain, and to obtain high quality stellar spectra even in regions dominated by telluric absorption lines.
Furthermore, {\paul SPIRou monitored about 70 M dwarfs, which will allow 
} us to construct a self-consistent database of stellar parameters for these targets.
In the rest of the paper, we typically choose to confront our results to those published by~\mann{}, because this reference study based its results on techniques that are largely different from ours, reducing the risk of potential biases.




In Sec~\ref{sec:observations} we outline the SPIRou observations used in this paper, and detail in Sec.~\ref{sec:telluric_correction} the way
reference stellar spectra (called `template spectra' in this paper) are derived from {\paul 40 to 80 
} individual spectra recorded at different epochs and corrected for telluric absorption and sky lines.
In Sec.~\ref{sec:spectral_anaysis}, we present the method we developed to retrieve the fundamental parameters of the host stars from their template SPIRou spectra. We discuss our results in Sec.~\ref{sec:results}, and conclude on the performances of our method and future steps to further extend its application (see Sec.~\ref{sec:conclusions}).

\section{SPIRou Observations}
\label{sec:observations}
\subsection{Targets selection}
\label{sec:targets_description}
We focus our study on the 12 inactive targets outlined in Sec~\ref{sec:introduction}, selected on the basis of three main criteria. More specifically, we chose stars that were observed {\paul at least 40}
times with SPIRou,
for which the parameters were determined by previous studies in order to have reference values for comparison,
and whose effective temperatures range from 3000 to 4000~K. 
The sample also include 2 binary stars for which $\mh$ values are expected to be similar.

For each M dwarf of our sample, we select 
40 to 80 spectra among the best quality ones collected with SPIRou at different Barycentric Earth Radial Velocities (BERV). This data set allows us to construct high signal-to-noise ratio (SNR) telluric-corrected template spectra of the selected targets from sets of SPIRou observations (see Sec~\ref{sec:telluric_correction}).
The number of SPIRou spectra used to build the templates of each star, and the typical SNR levels of these spectra, are listed in Table~\ref{tab:snrs}.

\subsection{Observations}

Observations were collected using SPIRou,
mostly in the framework of the large program called the SPIRou Legacy Survey~(SLS) that was allocated 300 nights at CFHT over 3.5 years. The two main science goals of the SLS are the search for exoplanets orbiting nearby M dwarfs, and the study of the impact of magnetic fields on star / planet formation.
Data is processed through the SPIRou reduction pipeline, APERO~(version 0.6.131, Cook et al., in prep). APERO also provides a blaze function estimated from flat-field exposures acquired prior to observations, which is used to flatten observation spectra. {\p Circularly polarized spectra were also recorded for the 12 stars in our sample but were not used in this analysis.}

The spectra are then normalized using a low order polynomial fitted through the points of the continuum. Because SPIRou spectra are not flux calibrated, the normalization steps are mandatory to properly compare the acquired spectra to the synthetic ones. Both telluric correction steps (described in Sec~\ref{sec:telluric_correction}) and normalization steps are performed independently from APERO.

\begin{table}
    \centering
    \caption{Number of spectra and typical SNR per pixel in the H band used to build template spectra.}
    \begin{tabular}{ccc}
        \hline
        Star & Number of spectra & Median SNR [SNR range]\\
        \hline
        Gl~846 & 54 & 160 [150 -- 220] \\ 
        Gl~880 & 47 & 220 [150 -- 245] \\ 
        Gl~15A & 38 & 285 [185 -- 505] \\ 
        Gl~411 & 36 & 385 [310 -- 435] \\ 
        Gl~752A & 38 & 200 [145 -- 230] \\ 
        Gl~849 & 51 & 125 [105 -- 140] \\ 
        Gl~436 & 37 & 150 [100 -- 225] \\ 
        Gl~725A & 64 & 230 [190 -- 255] \\ 
        Gl~725B & 56 & 180 [160 -- 190] \\ 
        Gl~699 & 46 & 210 [165 -- 240] \\ 
        Gl~15B & 77 & 105 [80 -- 180] \\ 
        Gl~905 & 79 & 125 [90 -- 130] \\
        \hline
    \end{tabular}
    \label{tab:snrs}
\end{table}

\subsection{Alternative \texorpdfstring{$\logg$}{TEXT} estimation.}
\label{sec:empirical_logg}
As estimating $\logg$ from stellar spectra is notoriously tricky (e.g. \pass{}), we also summarized alternative estimates obtained with 2 independent techniques.

The first method consists in computing $\logg$ from the radius and mass of the stars derived from empirical relations and models. This particular approach presents the advantage of not relying on the retrieved $\teff$ or $\mh$. For the twelve stars in our sample, we obtained photometric measurements from the SIMBAD service\footnote{\url{http://simbad.u-strasbg.fr/simbad/}}.
We compute $\logg$ from the mass--luminosity relation of~\citet{mann_2019} in the Ks band and theoretical mass-radius relations from~\citet{baraffe_2015}
assuming an age of 5~Gyr for all stars in our sample. The mass--radius relations show little deviation with respect to metallicity for low mass stars, and solar metallicity is therefore assumed. The $\logg$ values thus computed show little deviation from those estimated by \mann{} (RMS of 0.02~dex).

A second option is to compute $\logg$ from interferometry~\citep{boyajian_2012}.
This technique allows to accurately determine the radius of a given star, and to  therefore derive $\logg$ for a given mass.
However, interferometric data of M dwarfs remain rare, and~\citet{boyajian_2012} published angular diameters for only 6 stars in our sample.
A comparison between the values obtained using interferometry and those derived from evolutionary models leads to a RMS of the residuals of 0.06~dex and a median absolute deviation (MAD) of 0.04~dex, smaller than the typical computed uncertainties on $\logg$.

All $\logg$ values mentioned above are reported in Table~\ref{tab:literature_values}.

\section{Constructing templates from SPIRou spectra}
\label{sec:telluric_correction}

Template spectra of our target stars are constructed through an iterative two-step process. We first correct tellurics from observed spectra, then derive the template spectra by computing the median of individual corrected spectra in the barycentric reference frame. This step is repeated until proper convergence is achieved (see Sec.~\ref{sec:optimization_process}). In a second step, we apply Principal Component Analysis (PCA) to the residuals of all individual spectra with respect to the median, to refine the telluric correction and {\paul remove} emission lines from atmospheric airglow. 

\subsection{TAPAS correction of telluric lines}
\label{sec:tapas_model_description}
To correct telluric lines, we use TAPAS ~\citep[Transmissions of the \mbox{AtmosPhere} for AStronomical data,][]{berteaux_2014}, a tool capable of computing the atmosphere transmission in the line-of-sight of a given target. The computation of the transmission relies on the LBLRTM software~\citep{clough_1995}, using line lists provided by the HITRAN database~\citep{rothman_2009, rothman_2013}.

The TAPAS web-server {\paul 
provides} the transmission spectrum for a given epoch, site and air mass, and for individual atmospheric molecules. 
For our purpose, we
retrieved a typical theoretical spectrum for the 6 molecules primarily responsible for telluric absorption, i.e., $\rm O_2$, $\rm H_2O$, $\rm O_3$, $\rm CO_2$, $\rm CH_4$ and $\rm NO_2$. Each contribution is adjusted by a power law, and the resulting atmospheric transmission T is expressed as follows:

\begin{equation}
\label{eq:tell_transmission}
\begin{aligned}
    T = {} & \Big(T_{\rm 1}^{p_{\rm 1}}\ T_{\rm 2}^{p_{2}}\ T_{\rm 3}^{p_{\rm 3}}\ T_{\rm 4}^{p_{\rm 4}}\ T_{\rm 5}^{p_{\rm 5}} \ T_{\rm 6}^{p_{\rm 6}} \Big) * G_\sigma
\end{aligned}
\end{equation}

\noindent where $T_{\rm X}$ is the absorption spectrum, $p_{\rm X}$ is the adjusting exponent for molecule of index $\rm X$ (1: $\rm H_2O$, 2: $\rm CH_4$, 3: $\rm CO_2$, 4: $\rm NO_2$, 5: $\rm O_2$, 6: $\rm O_3$). $\rm{G}_\sigma$ is  a Gaussian broadening function of standard deviation $\sigma=1.83\ \kms$ (corresponding to a full-width at half maximum of 4.3~$\kms$) appropriate for the instrumental broadening of SPIRou~\citep{donati_2020}.

We also allow for radial velocity shifts of the entire telluric spectrum, as well as for a specific velocity shift of water lines with respect to the rest of the spectrum because of the less homogeneous spatial distribution of this molecule within the atmosphere and thereby its higher sensitivity to weather conditions~\citep{moll_2019}. The synthetic telluric transmission model therefore depends on 8 parameters.

To minimize the number of free parameters, we use the simplifying assumption that the powers $p_{\rm O_{2}}$, $p_{\rm CO_{2}}$ and $p_{\rm CH_{4}}$ are proportional to the air mass so that $p_{\rm X} = a_{\rm X} \ A$, with A denoting the air mass. We derived the values and error bars of the $a_{\rm X}$ slopes for the three molecules by fitting the model on telluric standards spectra acquired at various epochs, yielding:
\begin{equation}
    \text{}
        \begin{cases}
            a_{\rm CH_4} = 1.027\pm 0.004 \\
            a_{\rm CO_2} = 1.059 \pm 0.003\\
            a_{\rm O_2} = 0.998 \pm 0.006
        \end{cases}
\end{equation}

$\rm NO_2$ and $\rm O_3$ having negligible impact on the resulting telluric absorption spectrum in the SPIRou domain, we chose to set these coefficients to a standard value (of 1).
The resulting model thus requires to fit three parameters: $p_{\rm H_{2}O}$ and the two radial velocities.

\begin{figure}
    \centering
    \includegraphics[width=\columnwidth]{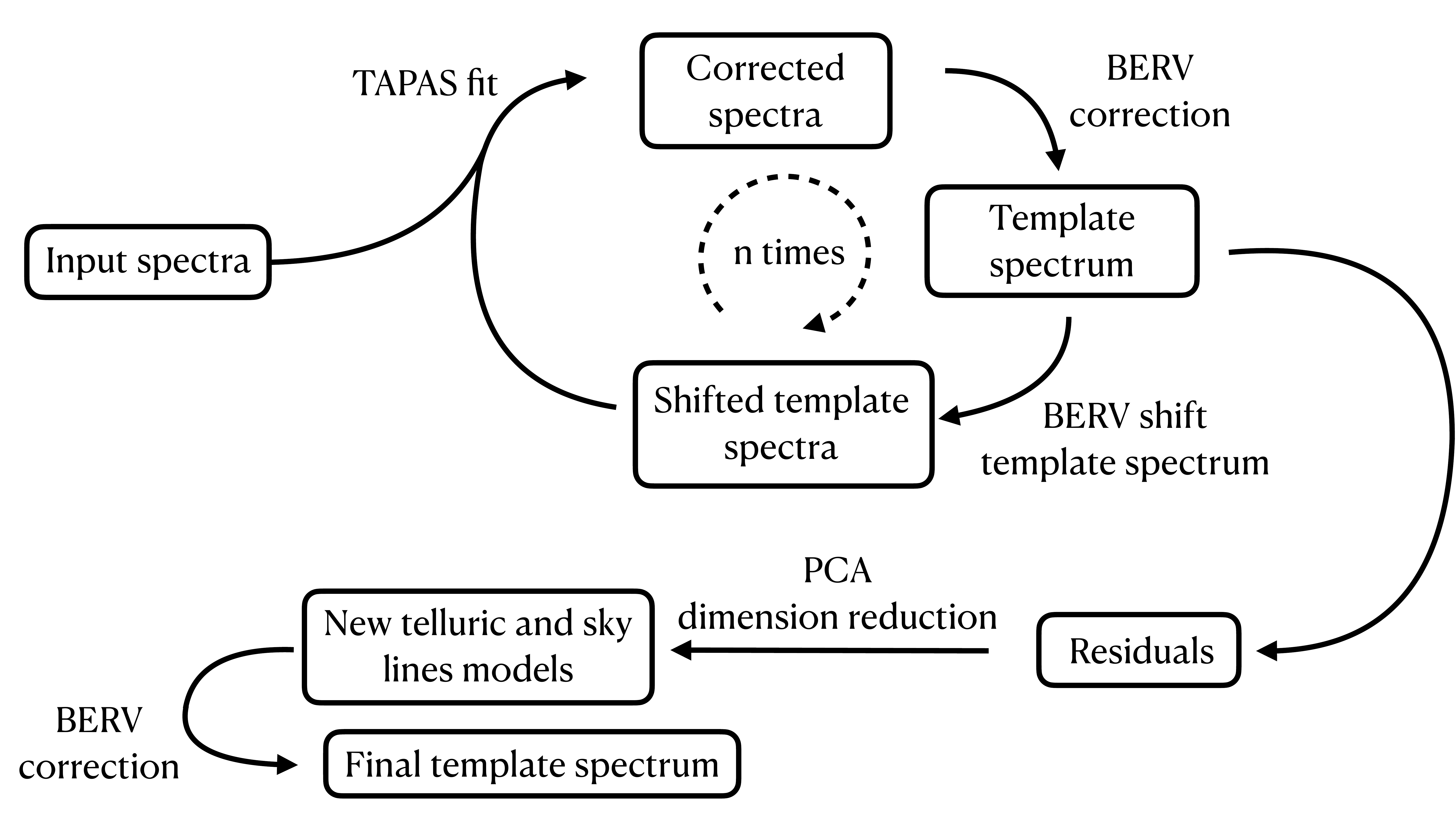}
    \caption{Dual-step iterative scheme used to derive template spectra from individual SPIRou spectra. TAPAS models are fitted on the input spectra using an iterative procedure. A PCA analysis is then applied on the residuals to improve upon the initial TAPAS correction. The stellar template is obtained by taking the median on the full set of TAPAS and PCA corrected spectra.}
    \label{fig:sketch_iterative_procedure}
\end{figure}

\begin{figure*}
\subfigure{
    \includegraphics[width=0.66\columnwidth]{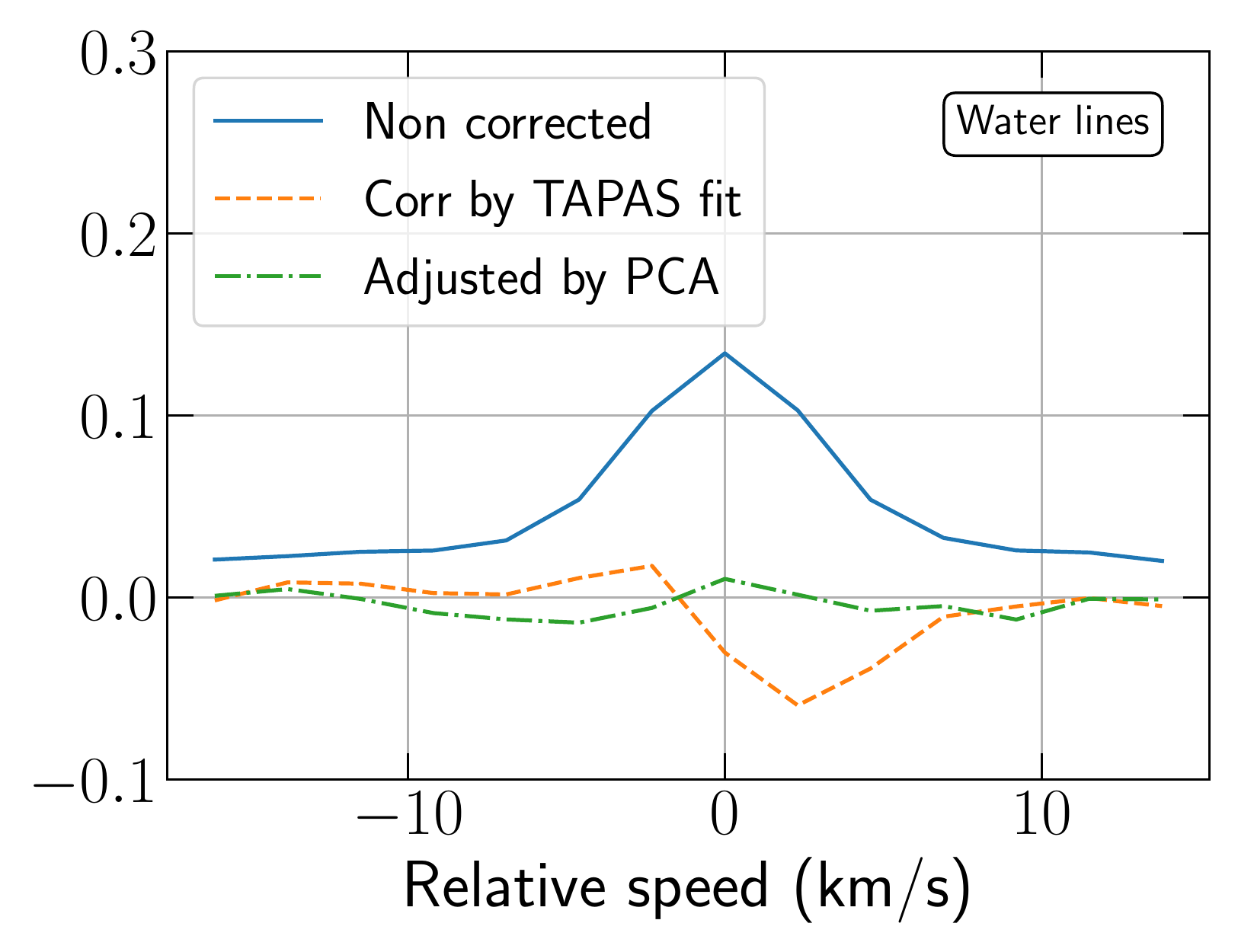}
}
\subfigure{
    \includegraphics[width=0.66\columnwidth]{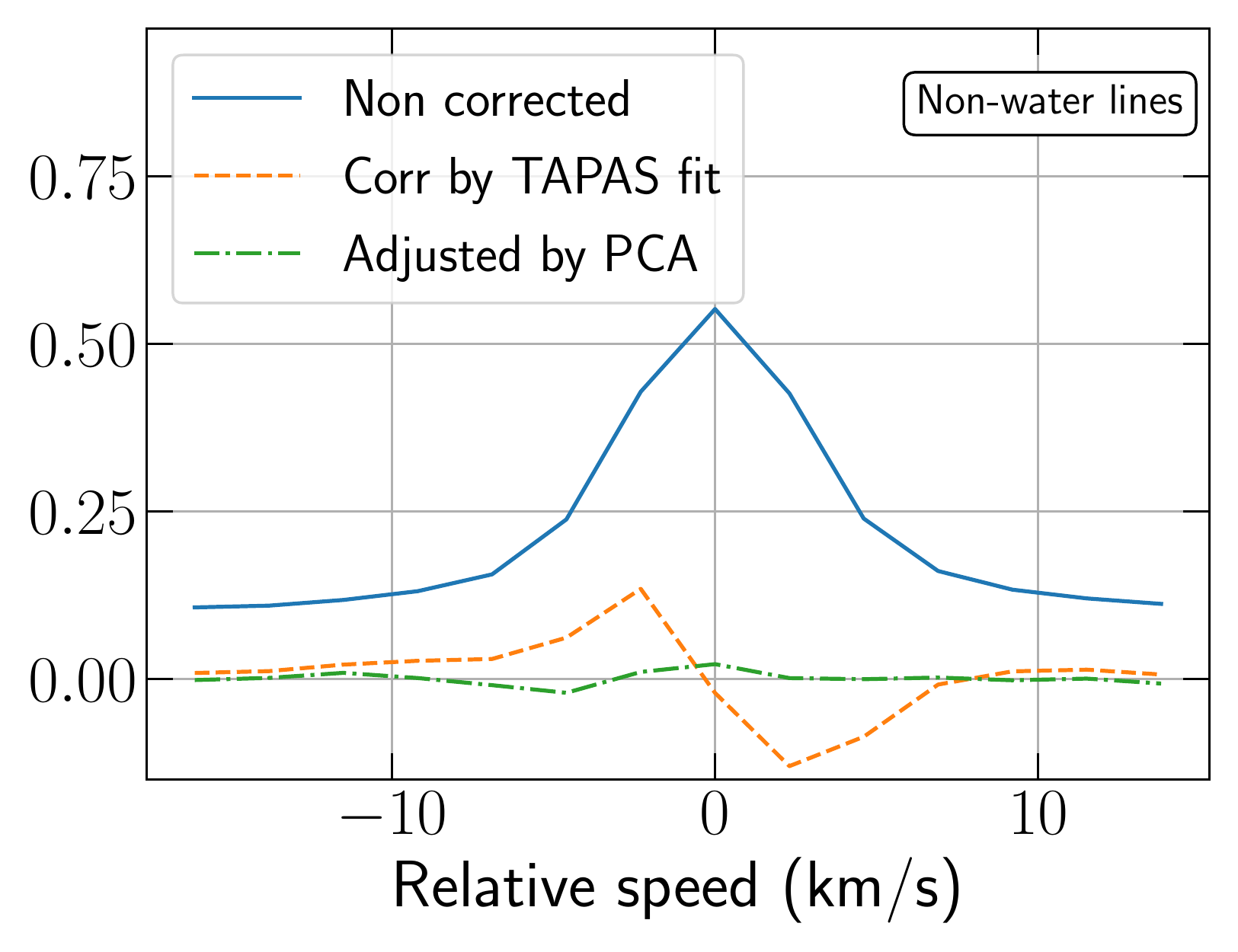}
}
\subfigure{
    \includegraphics[width=0.66\columnwidth]{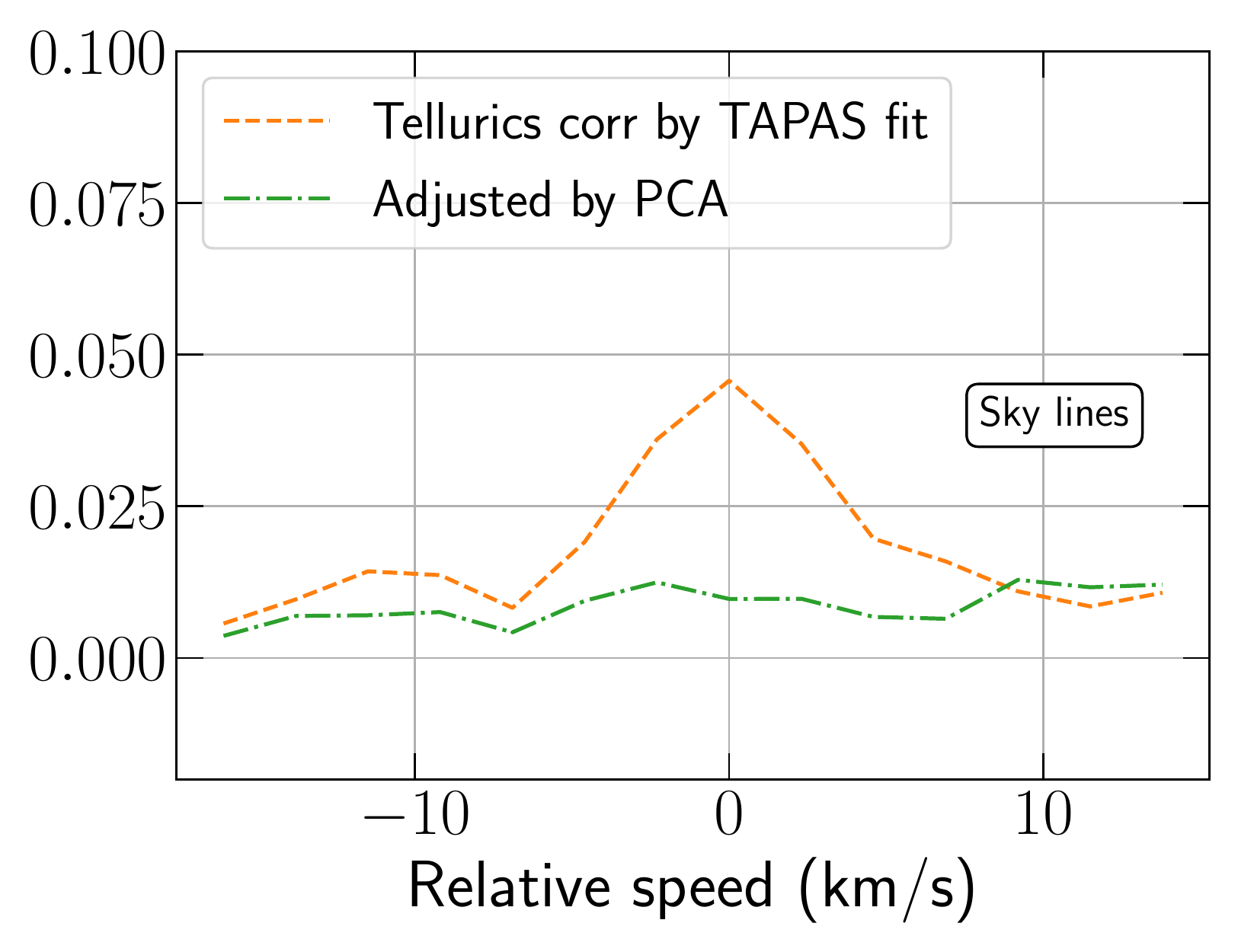}
}
\caption{Left: cross--correlation profile computed between the residuals for one of our Gl~699 spectra and a mask of water absorption lines. Middle: same as left panel but with a mask containing lines for all the telluric absorbers but water. Right: same as left panel but with a mask containing OH emission lines.}
    \label{fig:cross_corr}
\end{figure*}

\subsection{Template construction procedure}
\label{sec:optimization_process}

The template spectra are built through the iterative procedure illustrated in Fig.~\ref{fig:sketch_iterative_procedure}.
We fit TAPAS models on the input spectra with a Levenberg Mardquardt algorithm, and correct the template spectra with the resulting transmissions.
The corrected spectra are shifted to account for the BERV, interpolated on the the SPIRou wavelength grid, and a first template spectrum is computed by taking the median of the corrected spectra in the barycentric frame.
For each value of the BERV, the template is shifted back in the observer frame and used to correct the original spectra from the stellar spectrum itself. The resulting spectra contain less stellar features and mostly telluric lines, allowing to perform a better fit of the TAPAS model.
The process can be repeated multiple times, and we find that 5 iterations are sufficient to reach satisfactory convergence for the stars in our sample i.e. for the coefficients to remain stable from iteration to iteration.

At the end of the iterative process, residuals are computed by correcting the original spectra by the TAPAS models and the \mbox{template} spectrum shifted to the geocentric frame.
PCA is then applied to the residuals to extract the components accounting for most of the spectrum-to-spectrum variations. We found that the 3 components associated with the highest eigenvalues typically contain most of the variance and spectral line features.
We therefore filter the residuals using these 3 components only and obtain improved model spectra of non-stellar features to correct stellar spectra with. In particular, this last PCA step allows one to correct for emission lines from the sky (atmospheric airglow) that are not included in the TAPAS models, but show up in the SPIRou spectra.
All corrected spectra are then shifted to the barycentric reference frame, and the final stellar template is obtained by taking the median of all corrected spectra.
The stellar templates computed with the described procedure have a typical SNR per pixel in the H band in the range  500--2000.

\begin{figure*}
    \centering
    \includegraphics[scale=0.4]{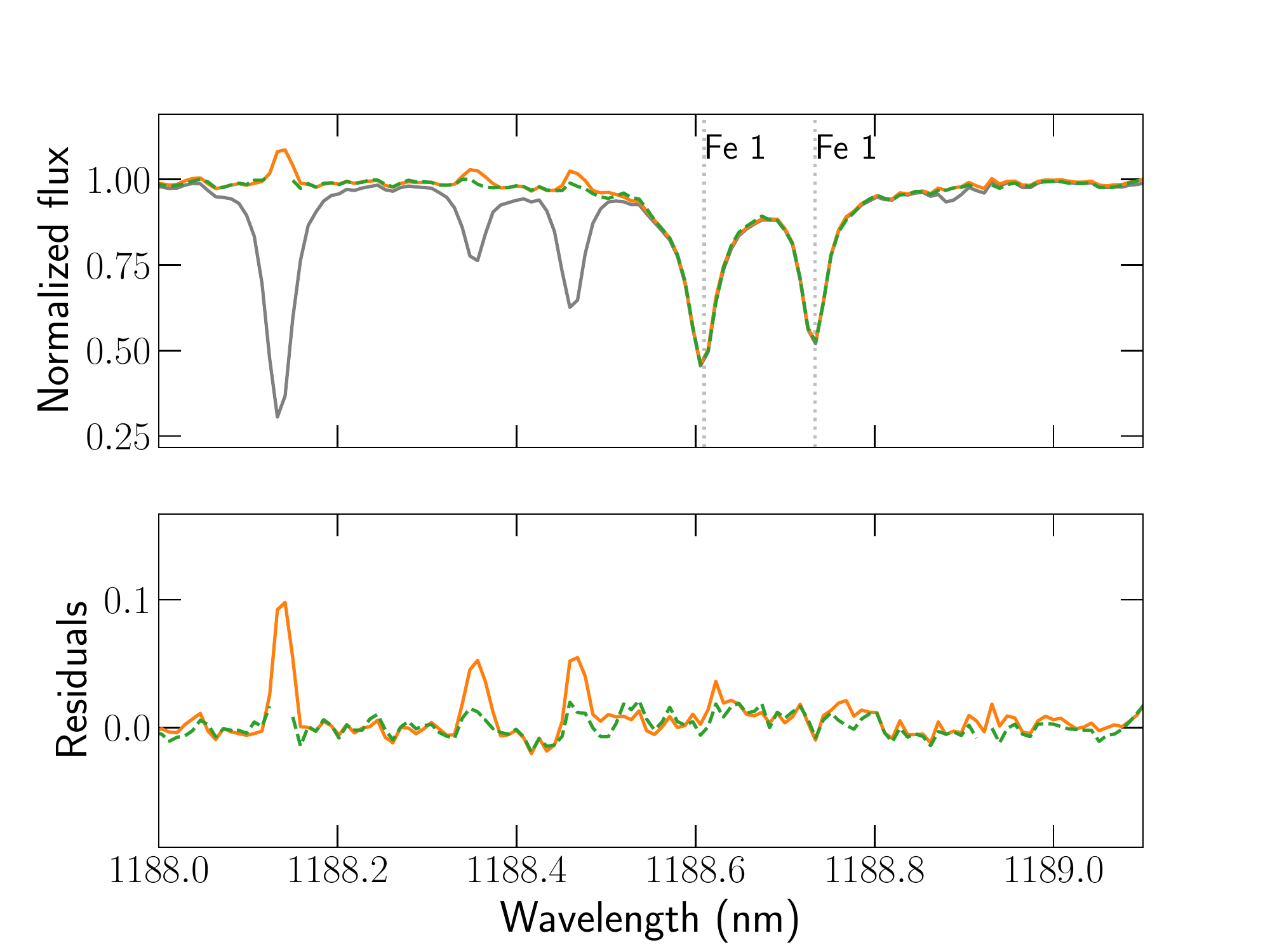}
    \includegraphics[scale=0.4]{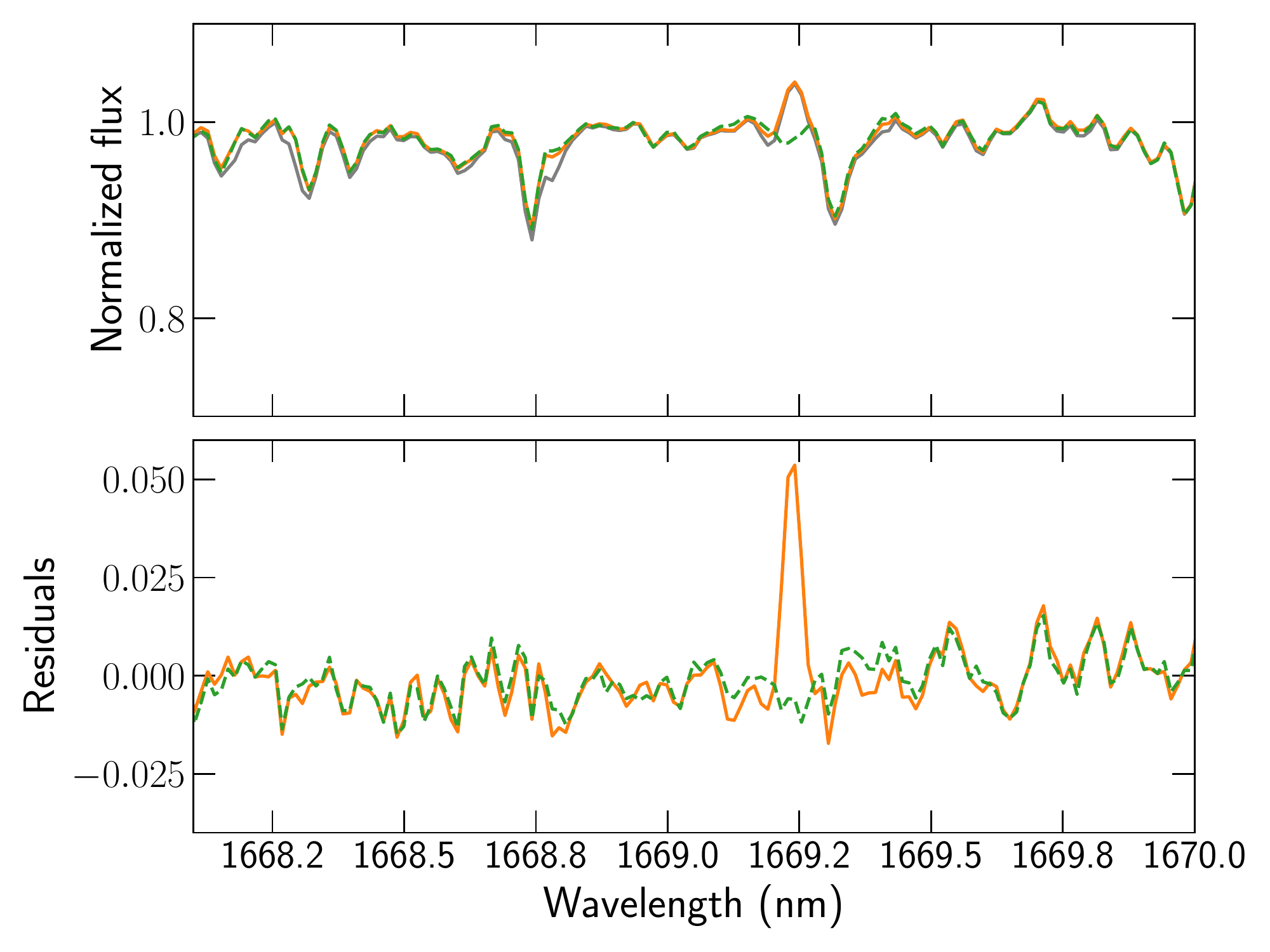}
    \caption{Examples of one of our Gl~15A spectra before and after correction of telluric and sky lines, the left and right panels showing examples of telluric and sky line correction respectively. Top panel: The uncorrected spectrum (gray) features telluric lines approximately removed following the TAPAS correction (orange). Applying PCA on the residuals yields an improved correction (green). Spectral points with telluric absorption larger than 60\% of the local continuum (like those around 1188.14~nm) are excluded prior to applying PCA, to optimize correction on the weak to medium-strength telluric features. Bottom panel: Corresponding residuals before (orange) and after (green) applying PCA.}
    \label{fig:corrections_comparison}
\end{figure*}

We assess the quality of the telluric correction by performing cross--correlations between telluric absorption line masks and \mbox{residuals}.
The cross--correlation profile shows a peak in the case of non-corrected spectra, which mostly vanishes with a proper correction of telluric and sky lines (see Fig.~\ref{fig:cross_corr} for example). Fig.~\ref{fig:corrections_comparison} illustrates the successive correction steps for one of our Gl~15A spectra.

We checked that the telluric- and sky- line-corrected template spectra generated with our direct approach, only applicable to stars for which tens of spectra are available for a wide range of BERV values,  agree well with the nominal ones produced by the (more general) correction procedure implemented within APERO.


\section{Spectral analysis of SPIRou template spectra}
\label{sec:spectral_anaysis}

Our analysis then consists in comparing template spectra (derived as outlined in Sec.~\ref{sec:telluric_correction}) to grids of synthetic spectra computed from model atmospheres and radiative transfer codes. In this section we describe how this comparison is achieved (Sec.~\ref{sec:model_description}), how spectral regions to be compared are selected (Sec~\ref{sec:line_selection}), and how the parameters of interest (i.e. $\teff$, $\logg$ and $\mh$) are obtained along with their associated error bars (Sec.~\ref{sec:param_determination}).

\subsection{Comparing observed template spectra with synthetic spectra}
\label{sec:model_description}

For this study, we gathered synthetic spectra computed with two different model atmospheres, namely PHOENIX~\citep{allard_1995} and MARCS~\citep{gustafsson_2008}.
We rely on the most recent grid of PHOENIX spectra available in the published literature~\citep{husser_2013}, computed with a sampling rate of about $0.6$~$\kms$ for various $\teff$, $\logg$ and $\mh$. MARCS synthetic spectra were computed from the latest available MARCS model atmospheres and the Turbospectrum radiative transfer code~\citep{plez_1998, plez_2012}, for a spectral sampling of $0.0025$~nm (corresponding to about 0.5~$\rm km s^{-1}$ at 1400~nm). The range of parameters covered by the computed grid of PHOENIX and MARCS synthetic spectra is summarized in Table~\ref{tab:marcs_range}.
The latest version of PHOENIX was specifically developed to improve the modeling of M dwarfs spectra at temperatures 3000~K and below, and is therefore expected to be more reliable than MARCS models on the low side of our temperature range. 

To compare the models to template spectra, the synthetic spectra are integrated on the wavelength grid associated with the template spectra.
We then adjust the continuum of the observed spectrum locally by matching the continuum points (defined as the highest 5\% points of each spectral window) of the observed spectrum to those of the synthetic spectrum.

We consider 4 main spectral-line broadeners: one of them to account for the instrument itself, and 3 associated with the star (micro-turbulence $v_{\rm mic}$, macro-turbulence $\vmac$ and rotation). We account for the instrumental broadening by applying a convolution with a Gaussian profile of full width at half maximum (FWHM) of 4.3~$\rm km.s^{-1}$~\citep{donati_2020}. The value of $v_{\rm mic}$ is set to 1~$\kms$ for MARCS models. The PHOENIX models were computed with values of $v_{\rm mic}$ varying from 0.04~$\rm km.s^{-1}$ to 0.6~$\rm km.s^{-1}$ for the range of parameters covered in this study (with the lowest values of $v_{\rm mic}$ corresponding to the coolest stars). 
Subsequent tests involving the computation of MARCS models with a $v_{\rm mic}$ set to 0.3~$\rm km.s^{-1}$ showed that the influence of micro-turbulence is small compared to the differences observed between the two models.
The effect of rotation is expected to be small compared to the other line broadeners for the inactive M dwarfs in our sample~\citep{reiners_2018}, and difficult to disentangle from macro-turbulence~\citep{brewer_2016}. We chose to account for the joint contribution of rotation and macro-turbulence by convolving all the synthetic spectra of the grid with a Gaussian profile of FWHM $\vbroad$. 
In the rest of the paper, we will be assigning to $\vbroad$ the value of the FWHM of the Gaussian profile, which may differ from conventional values reported for macroturbulence, often given as $\xi = {\rm FWHM} / (2\sqrt{\ln{2}})$ $\simeq$ 0.6 FWHM.



The radial velocity (RV) of each star is first estimated by performing a cross--correlation of each template spectrum with a line mask generated from the VALD database~\citep{pakhomov_2019}. The RV is then finely adjusted by minimizing a $\chi^2$ with the help of a Levenberg--Marquardt algorithm for each individual synthetic spectrum. 


\begin{table}
    \centering
    \caption{Parameter range covered by the PHOENIX and MARCS synthetic spectral grids. The interpolation factor are chosen based on the typical uncertainties retrieved with our analysis, and indicate the level to which the models are interpolated for the analysis.}
    \resizebox{\columnwidth}{!}{
        \begin{tabular}{cccccc}
            \hline
            Variable & Range (and step size) & Range (and step size) & Interpolation factor & final \\
             & \multicolumn{1}{c}{PHOENIX} & \multicolumn{1}{c}{MARCS} & PHOENIX/MARCS & step size \\
            \hline
            $\teff$~(K) & 2300 -- 7000 (100) & 3000 -- 4000 (100) & 20/20 & 5\\
            $\logg$~(dex) & 0.0 -- +6.0 (0.5) & 3.5 -- 5.5 (0.5) & 50/50 & 0.01 \\
            $\mh$~(dex) & -2.0 -- +1.0 (0.5) & -1.5 -- +1.0 (0.25) & 50/25 & 0.01 \\
            \hline
        \end{tabular}
        }
    \label{tab:marcs_range}
\end{table}

\begin{figure*}
    \centering
    \includegraphics[width=\linewidth, trim={3cm 0cm 3cm 0cm}, clip]{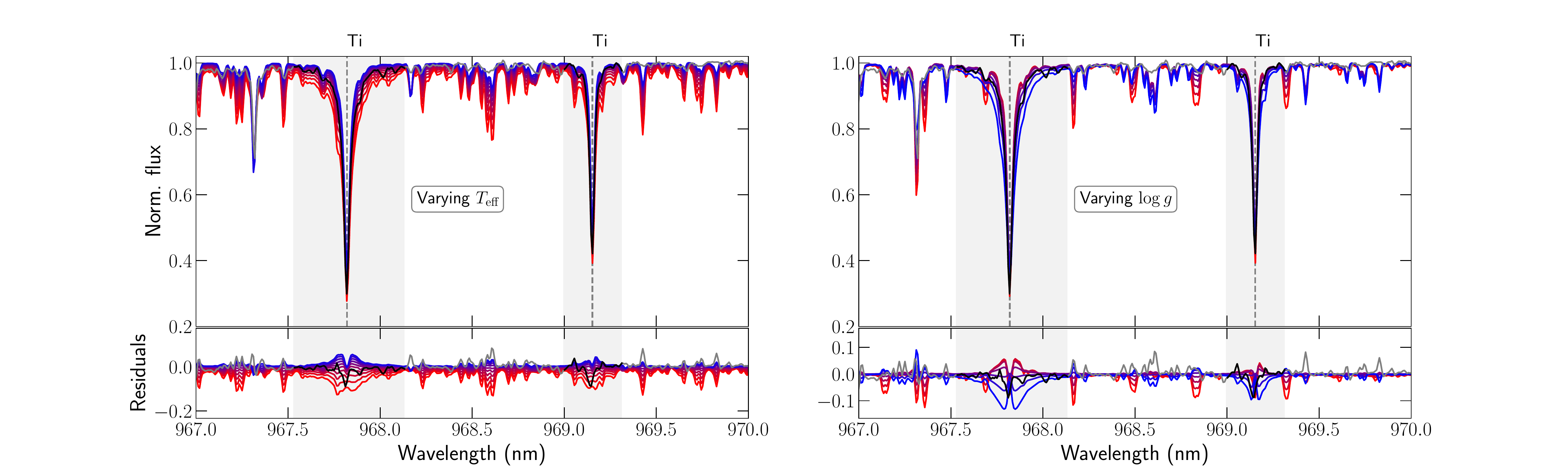}
    \caption{Comparison between PHOENIX synthetic spectra and the template spectrum of Gl~15A (gray). In the regions selected for the analysis, outlined with a grey background, the observed spectrum is displayed as a black line.
    The colored spectra correspond to synthetic spectra for different
    values of the parameters, with red being the lowest and blue the highest. 
    The associated parameters vary from 3000~K to 4000~K in steps of 100~K in $\teff$ (left panel), and from 3. to 6.0~dex in steps of 0.5~dex in $\logg$ (right panel). 
    The bottom plots show the residuals, i.e., the synthetic spectra and template spectrum minus the synthetic spectrum corresponding to the parameters of~\mann{}.}
    \label{fig:parameters_exploration}
\end{figure*}

\subsection{Selecting spectral windows}
\label{sec:line_selection}

\begin{figure*}
    \includegraphics[scale=0.4]{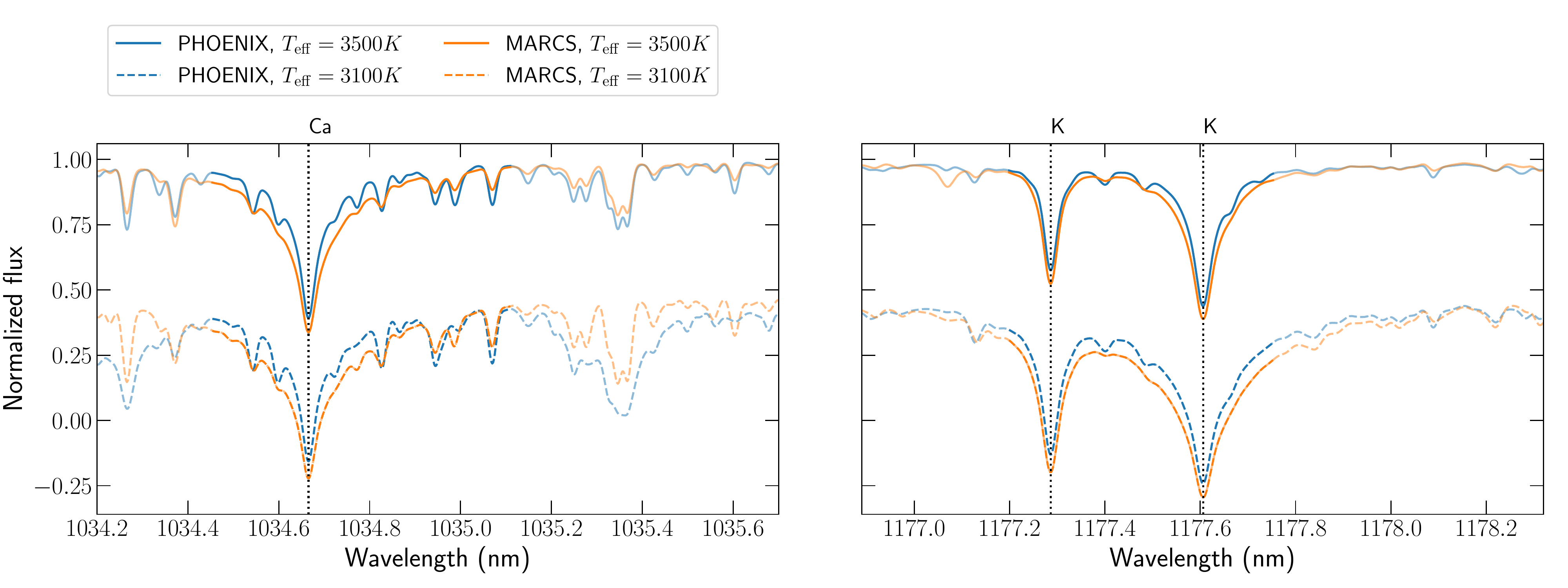}
    \caption{Comparison between synthetic spectra computed from PHOENIX and MARCS model atmospheres. The spectra are compared for two temperatures, and an offset is applied to the spectra for better readability. For all the models, $\logg$ = 5.0~dex and $\mh$ = 0.0~dex}
    \label{fig:comparison_models}
\end{figure*}

Prior to the analysis, we need to identify the lines that are best reproduced by the models,
that are sensitive to a least one of the fundamental parameters we aim at characterizing (i.e. $\teff$, $\logg$ and $\mh$), and for which the correction of telluric and sky lines is reliable.
A number of such lines were identified in previous studies~\citep{passegger_2019, rajpurohit_2018, flores_2019, valdivia_2019}, and we used them as a starting point for the line selection.
This was achieved by comparing SPIRou spectra to synthetic spectra, assuming the parameters published by~\mann{}. We began by selecting the lines that deviate from the observed spectrum by less than an arbitrary RMS threshold of 0.02, and for which the depth with respect to the continuum is expected to be greater than 20\%. A visual inspection was then carried out on each line to reject those heavily blended with nearby features. We also looked at the effect of varying $\teff$, $\logg$ and $\mh$ on the lines to investigate how strong a role they can play for pinpointing these parameters (see Fig.~\ref{fig:parameters_exploration} for example). 
{\paul The final list of selected lines is given in Table~\ref{tab:lines}.}
This list contains about 30 atomic lines, and about 40 molecular lines, the latter being primarily CO lines redward of 2293~nm.
Table~\ref{tab:line_list} summarizes the fundamental properties of the lines used with the PHOENIX and MARCS models, when available. Significant differences can be found in the line lists, which may partially explain the observed differences illustrated in Fig.~\ref{fig:comparison_models}
\footnote{We double checked that adjusting the van der Waals coefficients of the lines used in our study to the values proposed by~\citet{petit_2021} have little to no impact on the results detailed below.}.
{\paul  Fig.~\ref{fig:lines_1} shows a comparison of the SPIRou template spectra for the 12 M dwarfs in our sample along with the best fitted MARCS and PHOENIX models, for 8 selected lines. 
{\p Fig. A2 (available as supplementary material) presents a similar comparison for all the lines used for the analysis.}
}

\begin{table}
\center
\caption{Selected lines for the analysis. Vacuum wavelengths were extracted from the VALD database.}
\begin{tabular}[h]{cc}
\hline
Species & Wavelength (nm)\\
\hline
Ti I & 967.8198, 969.15274, 970.83269, 972.16252 \\ & 1058.7534, 1066.4544, 1189.6132, 1197.7124 \\ & 1281.4983, 1571.9867, 2296.9597 \\
Ca I & 1034.6654 \\
Fe I & 1169.3173, 1197.6323 \\
K I & 1169.342, 1177.2861, 1177.6061, 1243.5675, 1516.7211 \\
Mn I & 1297.9459 \\
Al I & 1315.435, 1672.3524, 1675.514 \\
Mg I & 1504.4357 \\
Na I & 2206.242, 2208.969 \\
OH & 1672.3418, 1675.3831, 1675.6299 \\
CO & 2293.5233, 2293.5291, 2293.5585, 2293.5754 \\ & 2293.6343, 2293.6627, 2293.7511, 2293.7900 \\ & 2293.9094, 2293.9584, 2294.1089, 2294.1668 \\ & 2294.3494, 2294.4163, 2294.6311, 2294.7059 \\ & 2294.9544, 2295.3195, 2295.4059, 2295.7263 \\ & 2295.8159, 2296.1743, 2296.2671, 2296.6648 \\ & 2296.7576, 2297.1971, 2297.2884, 2297.7719 \\ & 2297.8596, 2298.3888, 2298.4707, 2299.0488 \\ & 2299.1222, 2311.2404, 2312.4542, 2315.0029, 2316.3381 \\
\hline
\end{tabular}
\label{tab:lines}
\end{table}

\begin{figure}
    \centering
    \includegraphics[width=\columnwidth, trim={0 1cm 0 0}, clip]{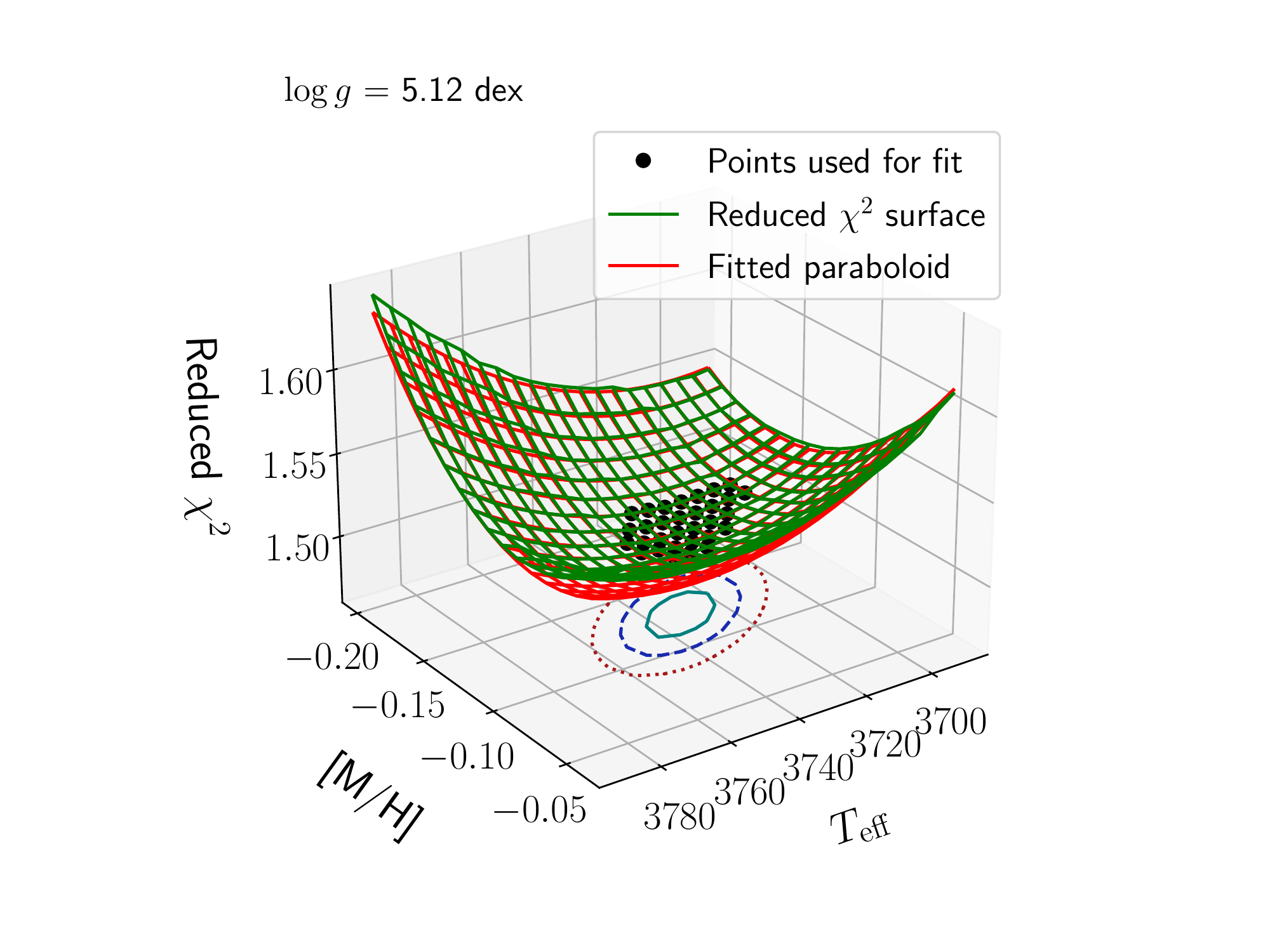}
    \caption{Example of a 2D section of the 3D $\chi^2$ landscape (green) along with the fitted paraboloid (red) 
    derived by comparing
    our Gl~15A template and the grid of PHOENIX spectra. The projected ellipses mark the contours defined by an increase in $\chi^2$ of 1 (solid green), 4 (dashed blue) and 9 (dotted red) from the minimum. The value of $\logg$ is equal to 5.12~dex
    in this particular $\teff$, $\mh$ slice of the 3D $\chi^2$ landscape. }
    \label{fig:paraboloid}
\end{figure}

\begin{figure*}
    \centering
    PHOENIX / PHOENIX\\
    \subfigure{
    \includegraphics[scale=.16]{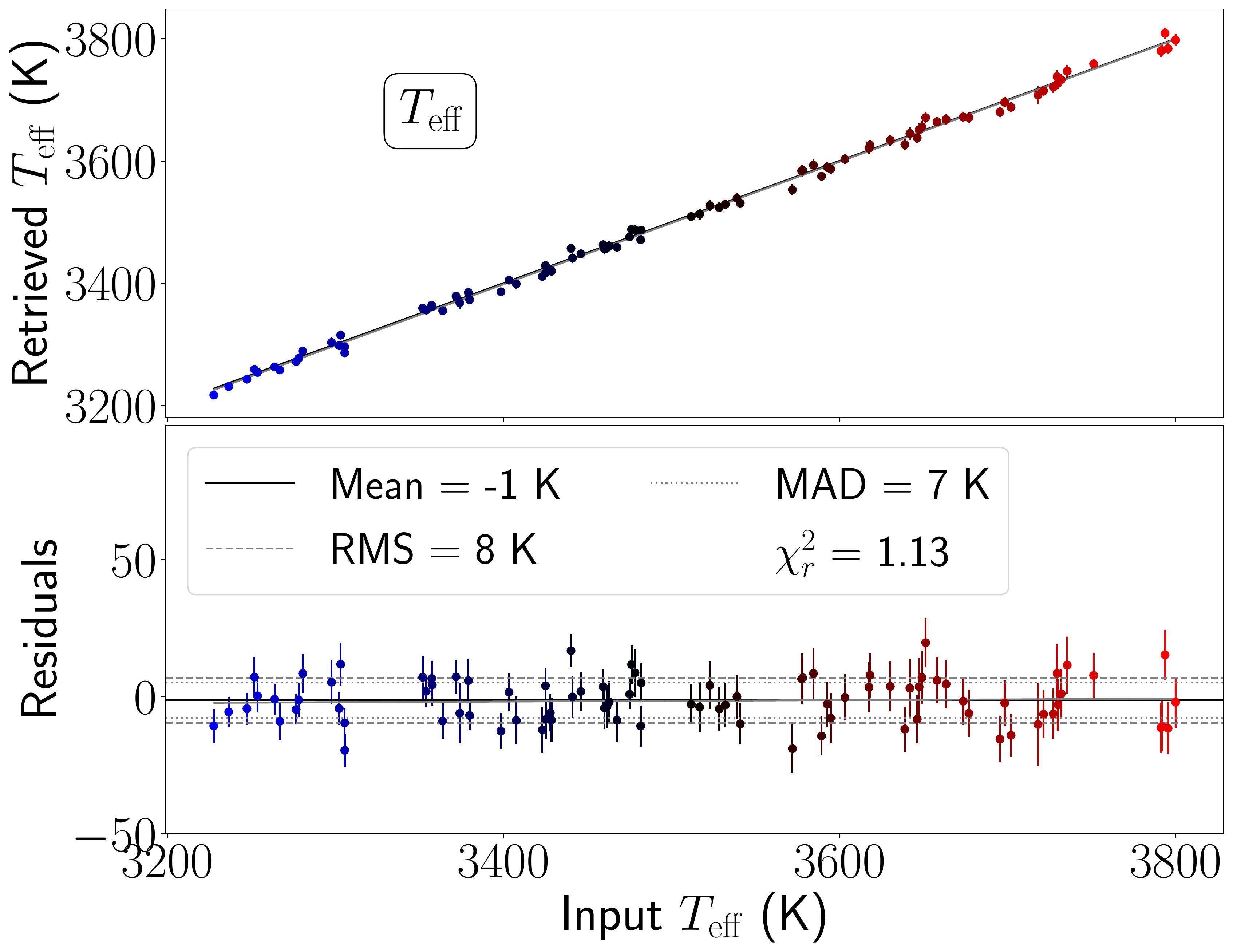}
    }
    \subfigure{
    \includegraphics[scale=.16]{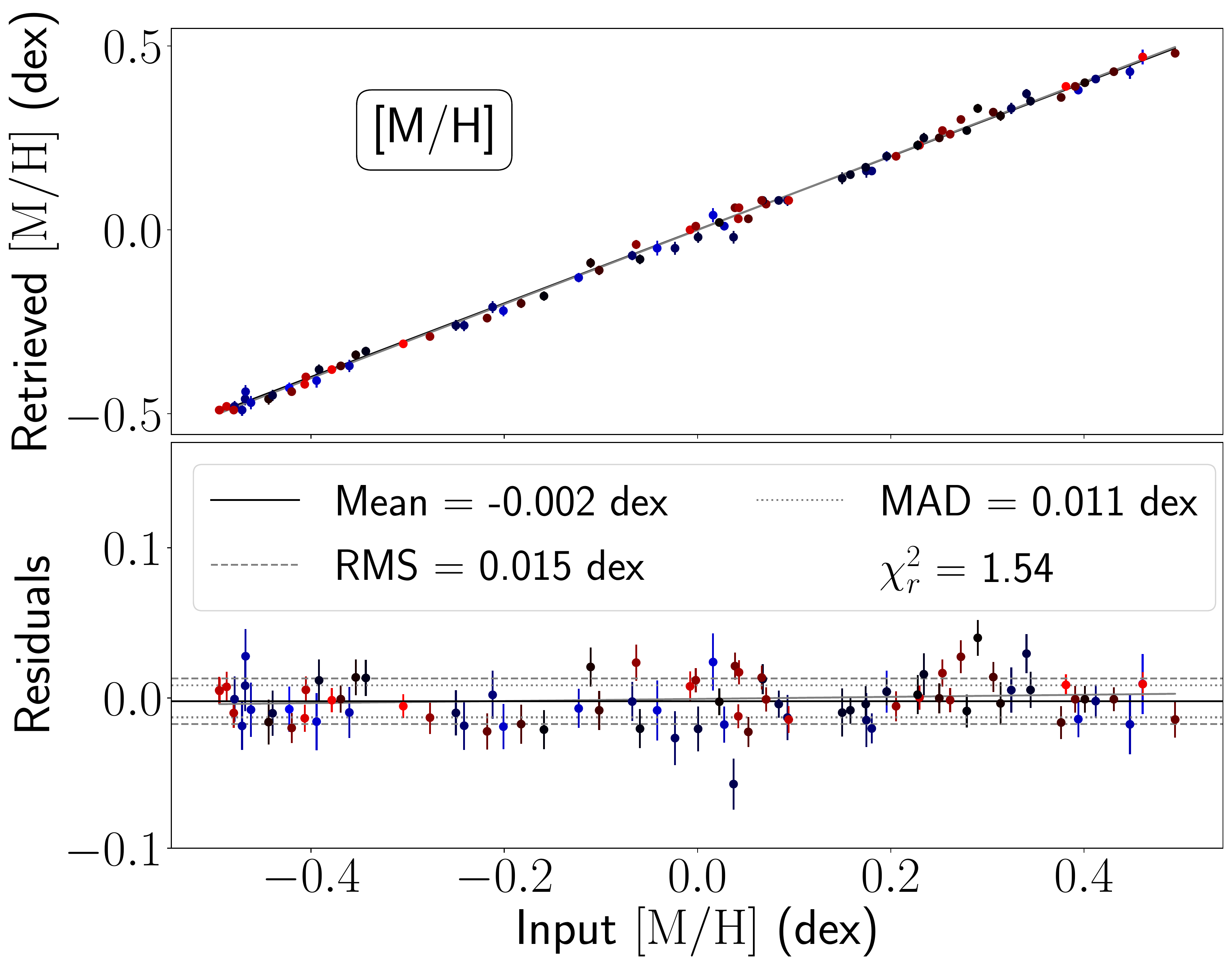}
    }
    \subfigure{
    \includegraphics[scale=.16]{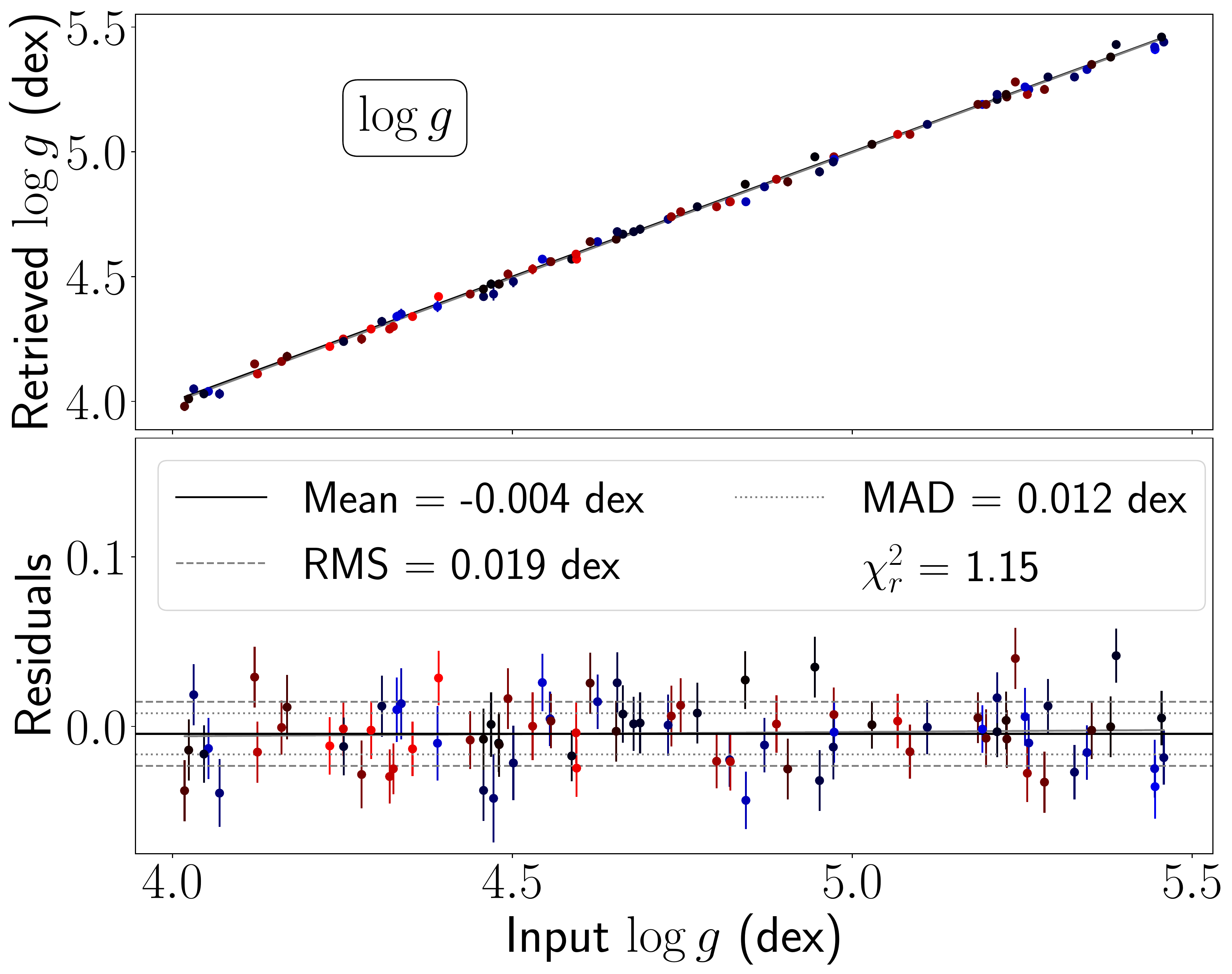}
    }
    MARCS / MARCS\\
    \subfigure{
    \includegraphics[scale=.16]{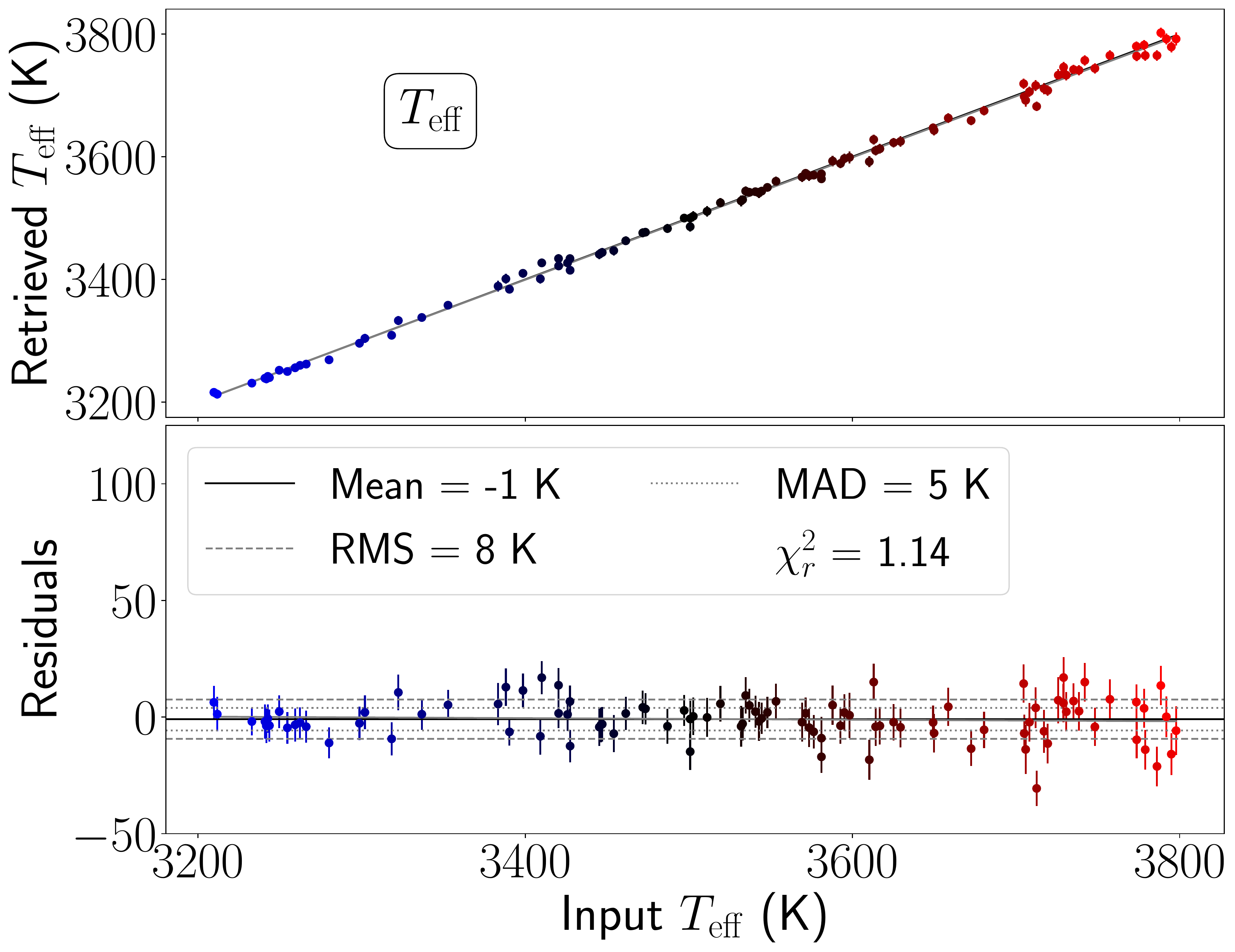}
    }
    \subfigure{
    \includegraphics[scale=.16]{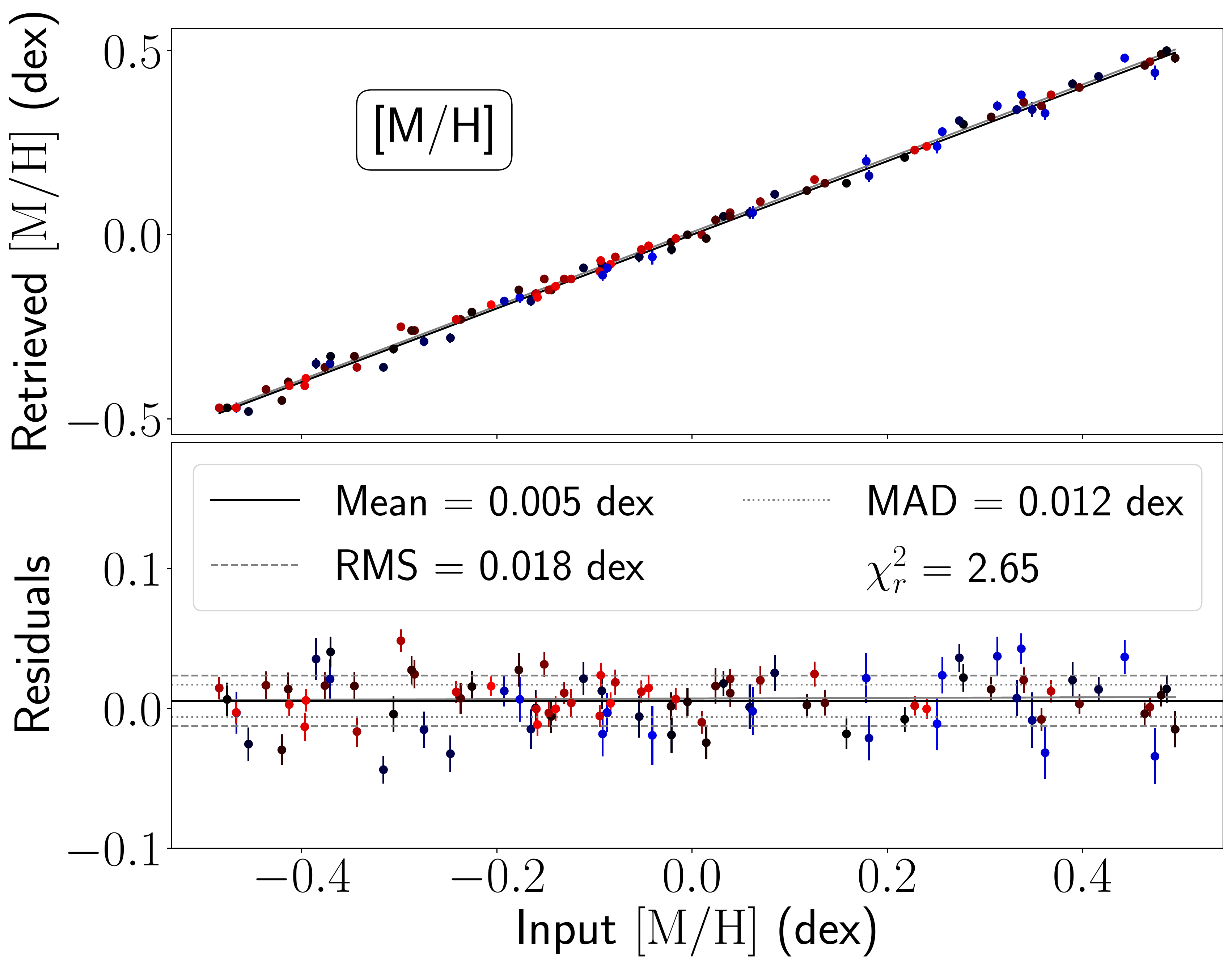}
    }
    \subfigure{
    \includegraphics[scale=.16]{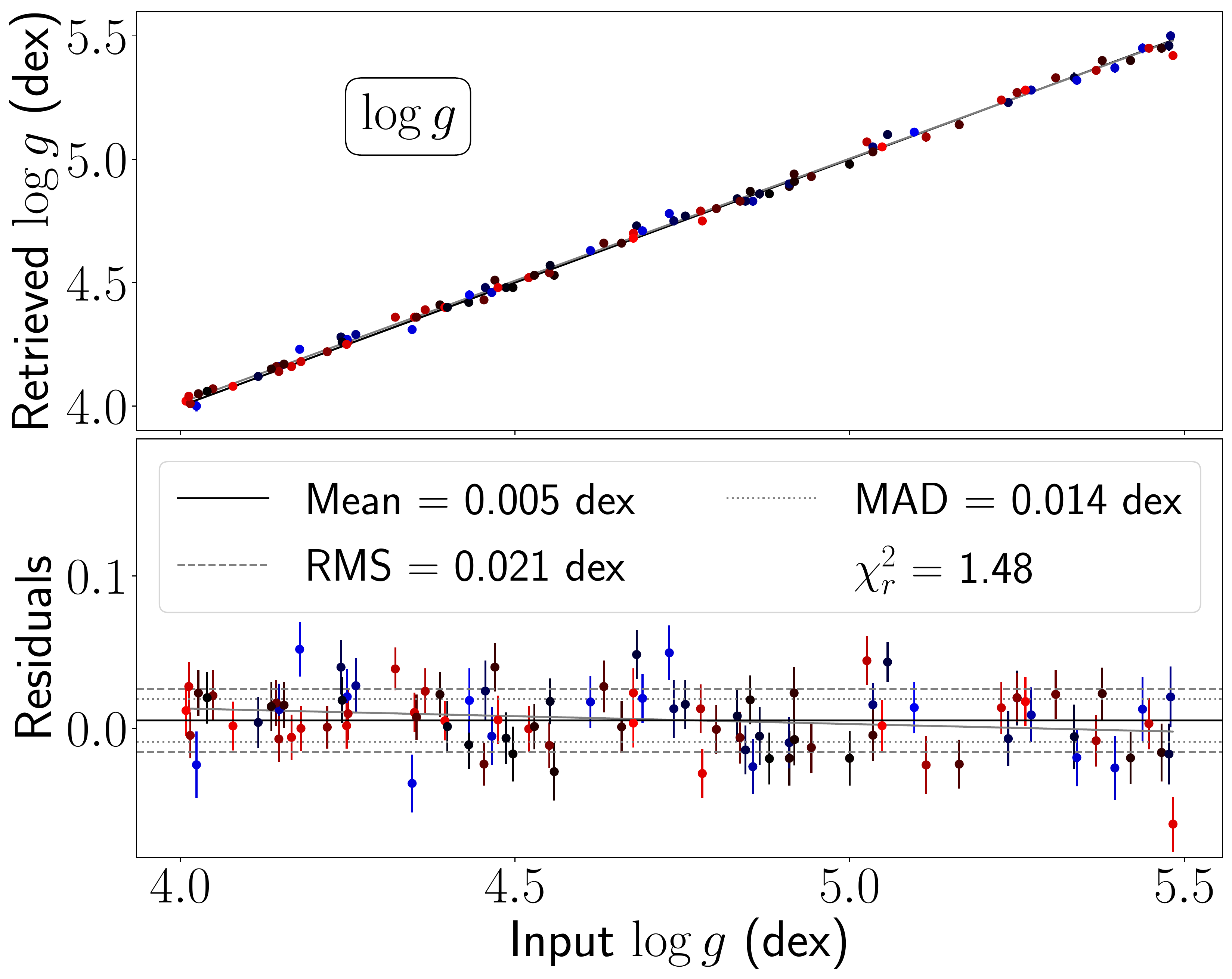}
    }
    PHOENIX / MARCS\\
    \subfigure{
    \includegraphics[scale=.16]{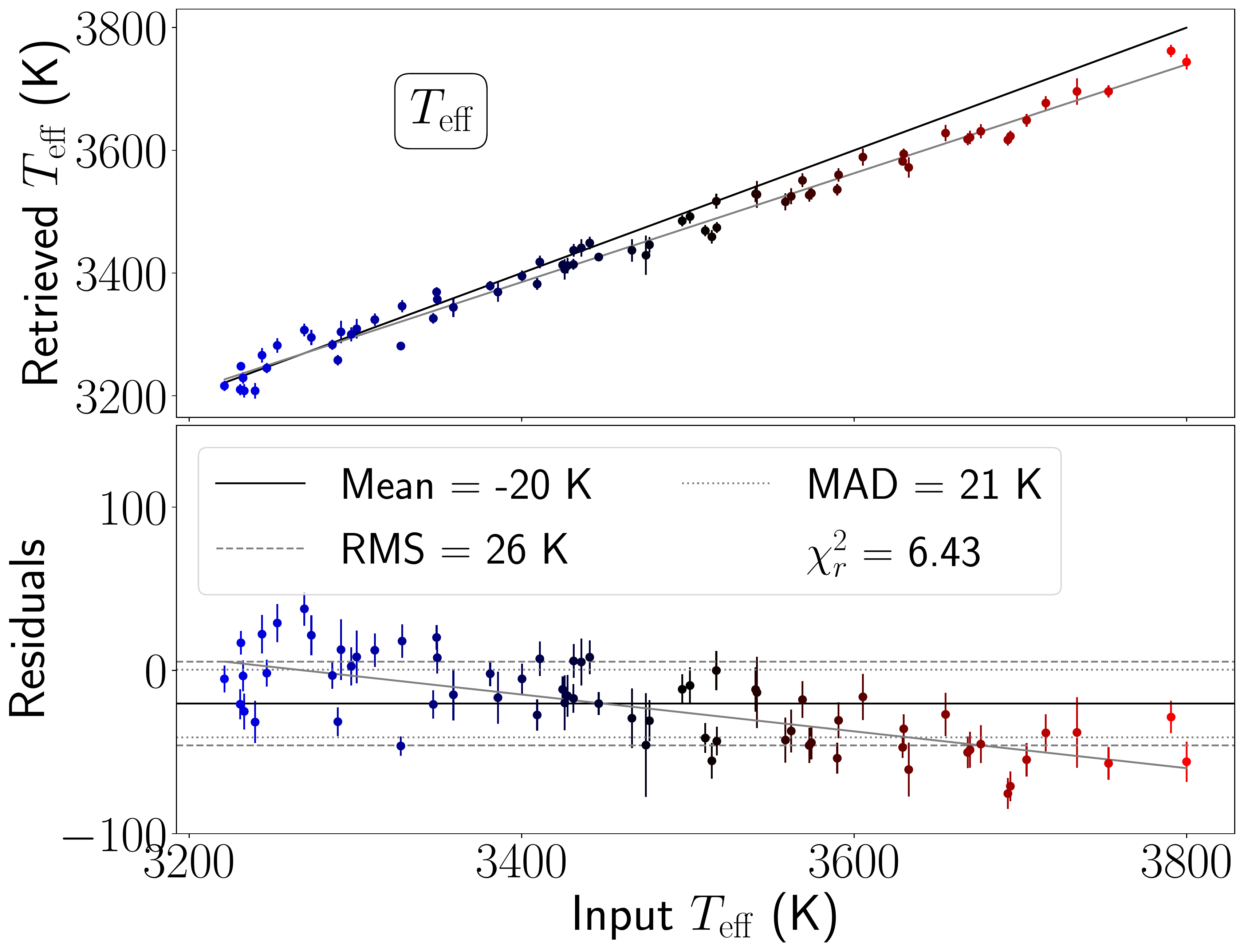}
    }
    \subfigure{
    \includegraphics[scale=.16]{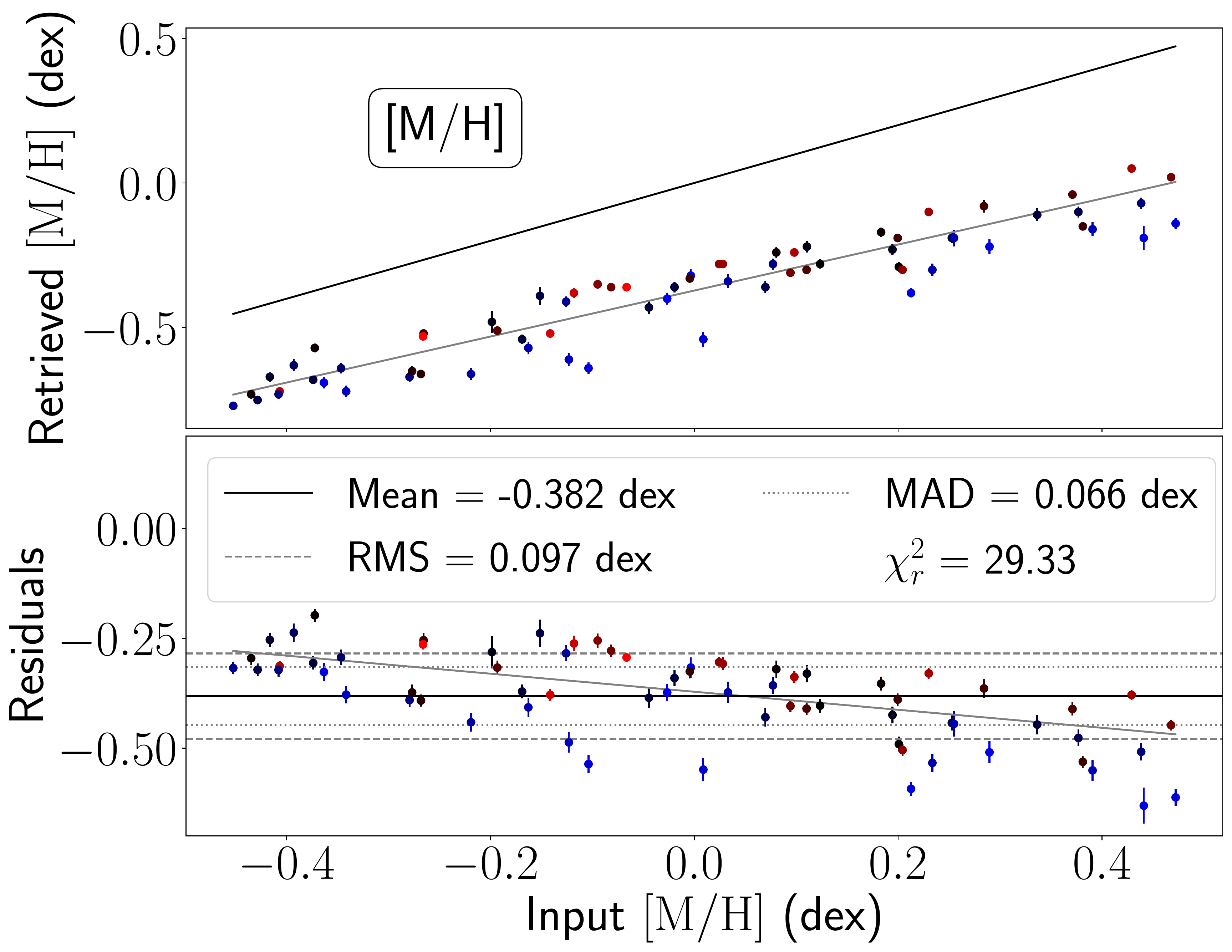}
    }
    \subfigure{
    \includegraphics[scale=.16]{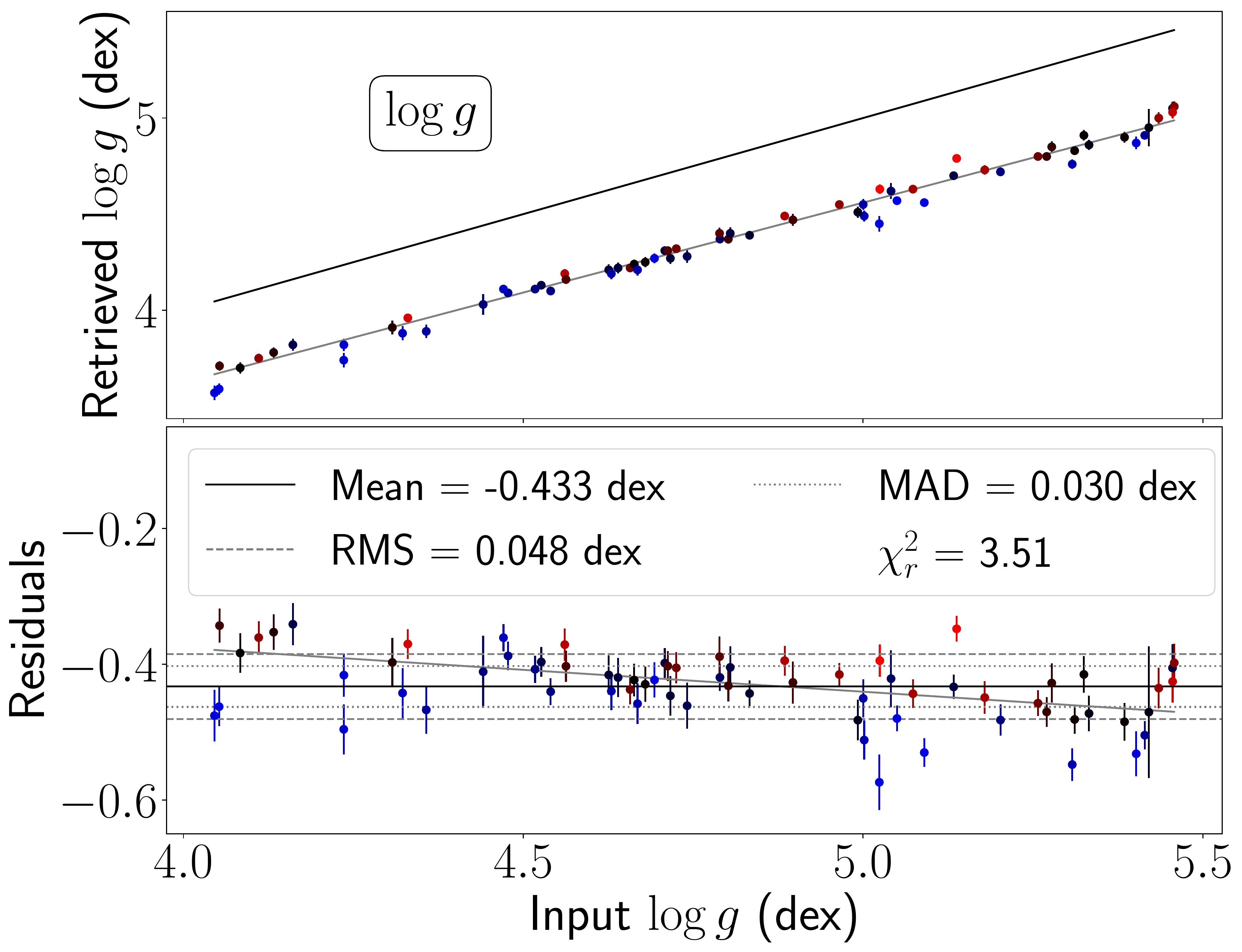}
    }
    MARCS / PHOENIX\\
    \subfigure{
    \includegraphics[scale=.16]{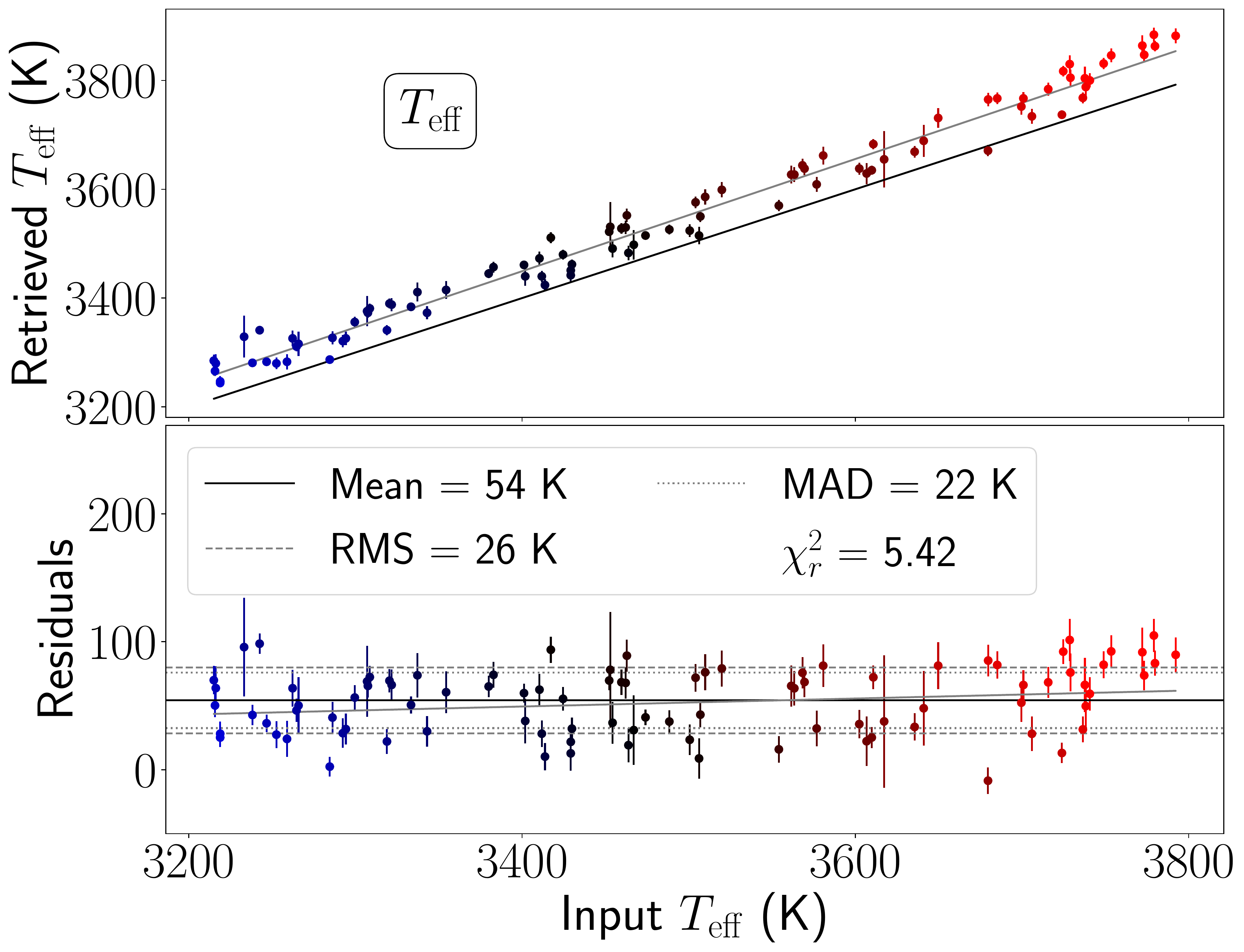}
    }
    \subfigure{
    \includegraphics[scale=.16]{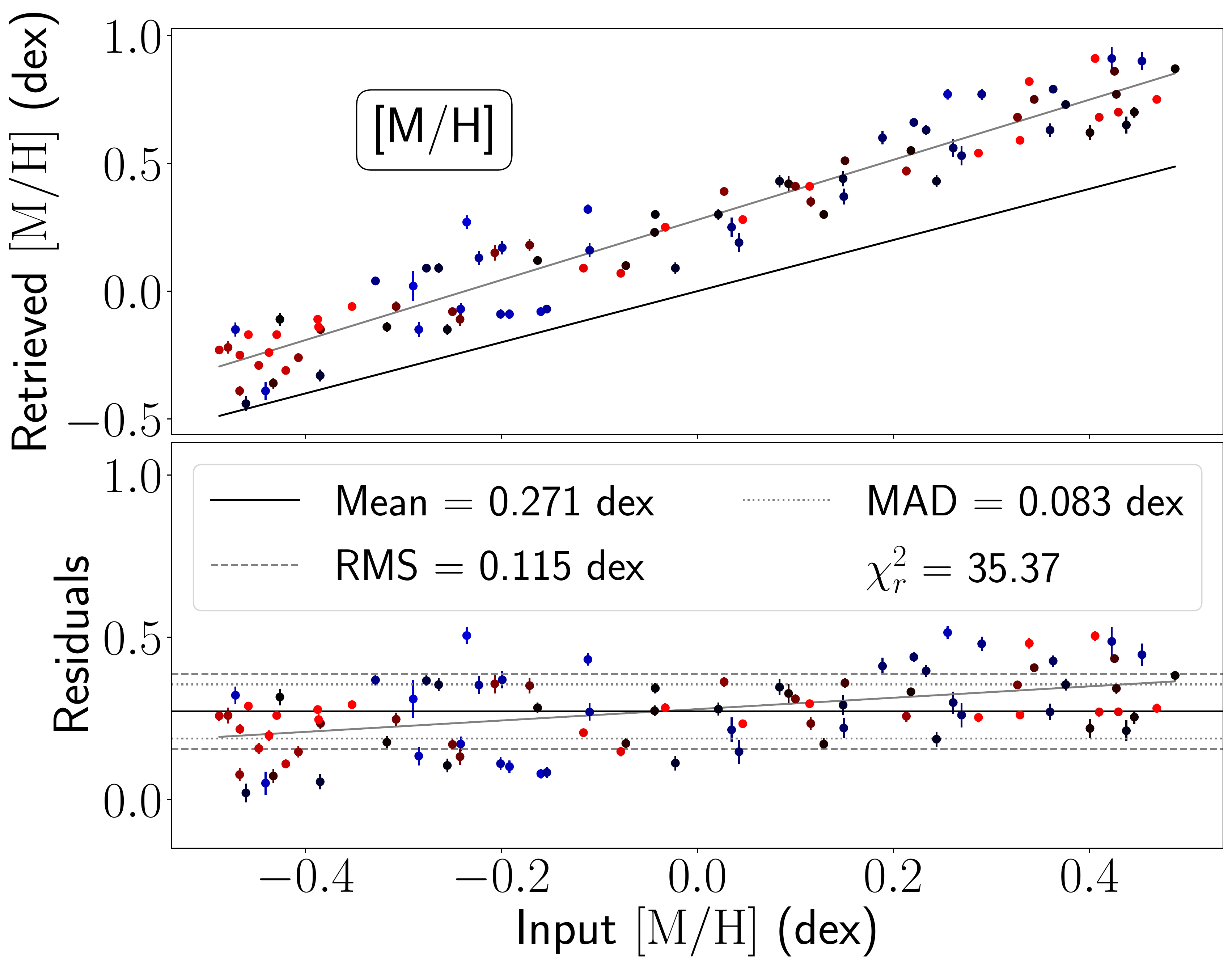}
    }
    \subfigure{
    \includegraphics[scale=.16]{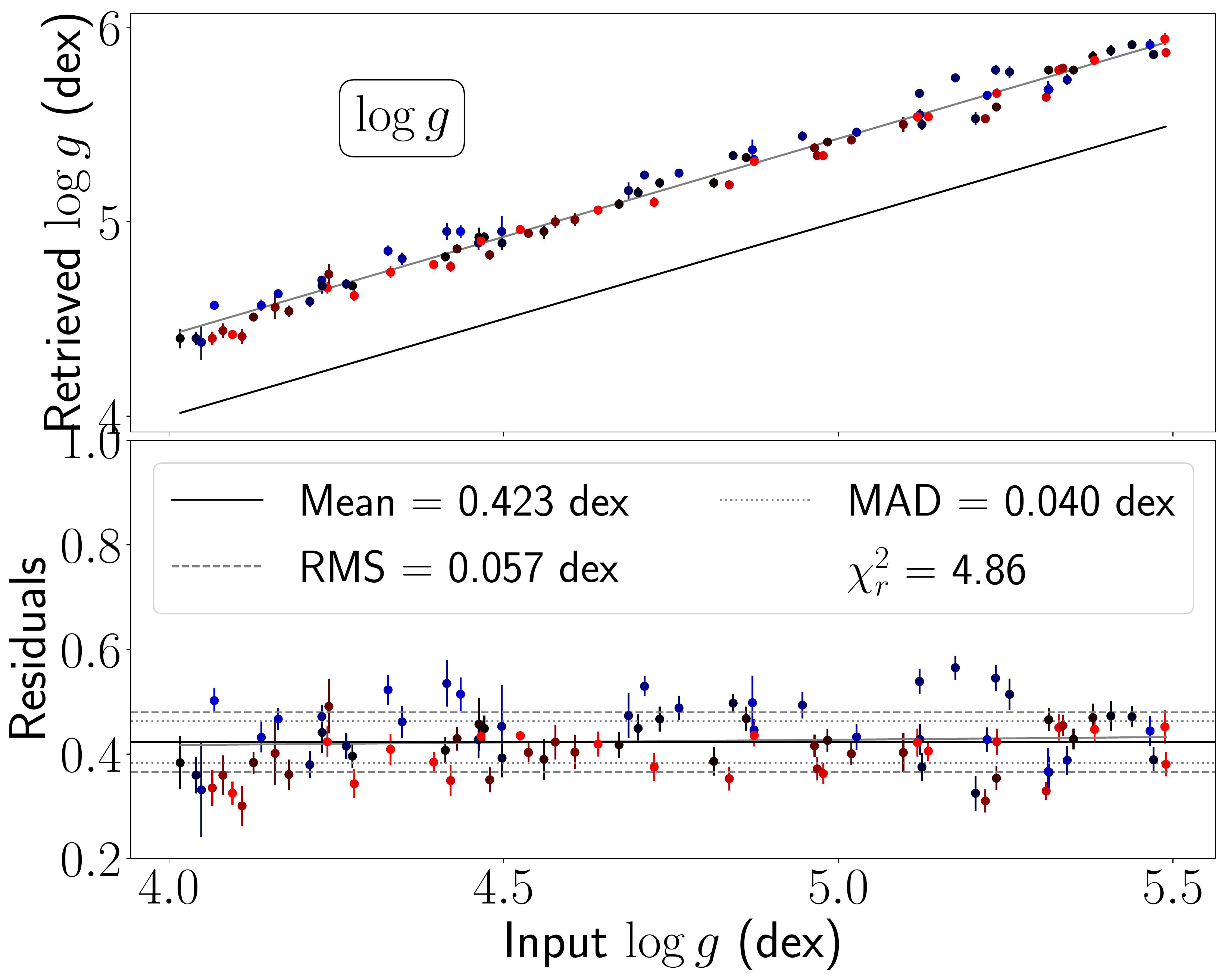}
    }
    \caption{Simulations of parameters determination. The recovered $\teff$, $\logg$ and $\mh$ are plotted against the values used to generate the model spectra. The black solid line marks the equality, and a gray solid line is the result of the linear fit performed on the data points. The coefficients and intercepts of the fits are reported in Table~\ref{tab:coeffs_fit}. All data points are color coded as a function of $\teff$, blue corresponding to the smallest temperature (3200~K), red to the highest temperature (3800~K), and black corresponding to the median $\teff$ of 3500~K. RMS and MAD values are given with respect to the average of the residuals. The models are generated either from PHOENIX (first and third rows) or MARCS (second and fourth rows) synthetic spectra, which parameters chosen randomly, and a Gaussian noise is added to simulate a SNR of $\sim$100 in the H band, accounting for both the blaze function in each order and the SPIRou throughput. For each model the analysis was performed with either the PHOENIX or MARCS grid of synthetic spectra.
    }
    \label{fig:simulations_teff}
\end{figure*}

\begin{table*}
    \caption{Slope and intercepts for the fits obtained on the data presented in Fig.~\ref{fig:simulations_teff}.}
    \begin{tabular}{ccccccc}
        \hline
        & \multicolumn{2}{c}{$\teff$ (K)} & \multicolumn{2}{c}{$\logg$ (dex)} & \multicolumn{2}{c}{$\mh$ (dex)} \\
        & slope & \thead{intercept \\ at 3500~K}  & slope & \thead{intercept \\ at 4.7~dex} & slope & \thead{intercept \\ at 0.0~dex}  \\
        \hline
        PHOENIX / PHOENIX & 1.002 $\pm$ 0.005 & 3498 $\pm$ 19 & 1.002 $\pm$ 0.005 & 4.69 $\pm$ 0.02 & 1.007 $\pm$ 0.005 & -0.0006 $\pm$ 0.0014  \\
        MARCS / MARCS & 0.996 $\pm$ 0.005 & 3499 $\pm$ 17 & 0.989 $\pm$ 0.005 & 4.70 $\pm$ 0.02 & 1.002 $\pm$ 0.005 & 0.0070 $\pm$ 0.0016 \\
        PHOENIX / MARCS & 0.887 $\pm$ 0.014 & 3473 $\pm$ 48 & 0.935 $\pm$ 0.012 & 4.27 $\pm$ 0.06 & 0.794 $\pm$ 0.030 & $-0.3716$ $\pm$ 0.0082 \\
        MARCS / PHOENIX & 1.031 $\pm$ 0.015 & 3552 $\pm$ 52 & 1.00 $\pm$ 0.013 & 5.12 $\pm$ 0.064 & 1.175 $\pm$ 0.030 & 0.2788 $\pm$ 0.0096 \\
        \hline
    \end{tabular}
    \label{tab:coeffs_fit}
\end{table*}

\subsection{Determining stellar parameters}
\label{sec:param_determination}


Each template spectrum is then compared to the whole grid of synthetic spectra following the procedure described in Sec.~\ref{sec:model_description}. We end up with a $\chi^2$ landscape over the full 3D grid of stellar parameters from which we derive the optimal ones and the associated error bars.

More specifically, we begin by comparing the template spectra to the original grid of synthetic spectra sampled in steps of 100~K in $\teff$, 0.5~dex in $\logg$ and 0.5 (resp. 0.25~dex) in $\mh$ with the grid of PHOENIX (resp. MARCS) synthetic spectra, to find a rough minimum $\chi^2$.
We then build a finer grid of synthetic spectra by linear interpolation covering 100~K in $\teff$ and 0.2~dex in $\logg$ and $\mh$ around this minimum, in order to reach steps of $5~K$ in $\teff$ and 0.01~dex in $\logg$ and $\mh$. The interpolation factors and final step sizes are also reported in Table~\ref{tab:marcs_range}.
The optimal parameters and error bars are computed by fitting a 3D paraboloid on the 500 points of smallest $\chi^2$ values. 
Error bars are estimated by measuring the curvature of the  3D paraboloid around its minimum.
We derive the 3D confidence ellipsoid
in which $\chi^2$ increases by no more than 1 with respect to its minimum value, and project it on each parameters axes.
The projected intervals should contain 68.3\%  of the retrieved values for each parameter assuming the noise obeys a Gaussian distribution~\citep{numerical_recipes}.
An example 2D section of a 3D paraboloid fit, along with the 2D confidence ellipsoid is presented in Fig.~\ref{fig:paraboloid}.
These error bars correspond to the minimum uncertainties of our parameter determination process, i.e. the error bars associated to the photon noise. If the minimum reduced $\chi^2$ reached over the map is larger than 1, i.e. if systematic differences exist between the observations and the models, we scale up all the error bars in the spectra to enforce the minimum reduced $\chi^2$ to be 1; this correction should in principle ensure that the derived error bars on the fitted parameters incorporate some of the systematic differences between the observations and the model, assuming that these differences can be treated as uncorrelated noise. The error bars computed in this way will be referred to as formal error bars in the rest of the paper, and are expected to account for the photon noise and some of the systematics.

\begin{table*}
	\centering
	\caption{Retrieved fundamental parameters using the grid of PHOENIX (cols. 2-7) and MARCS (cols 8-13) synthetic spectra with and without fixing $\logg$ to the values presented column 6 of Table~\ref{tab:literature_values}. 
	}
	\resizebox{\textwidth}{!}{
		\begin{tabular}[h]{c|ccc|ccc|ccc|ccc}
			\hline
			& \multicolumn{3}{c|}{PHOENIX} & \multicolumn{3}{c|}{PHOENIX (Fixed $\logg$)} & \multicolumn{3}{c|}{MARCS} & \multicolumn{3}{c}{MARCS (Fixed $\logg$)}\\
			\hline
			Star & $\teff$ (K) & $\logg$ (dex) & $\mh$ (dex) & $\teff$ (K) & $\logg$ (dex) & $\mh$ (dex) & $\teff$ (K) & $\logg$ (dex) & $\mh$ (dex) & $\teff$ (K) & $\logg$ (dex) &  $\mh$ (dex)\\
			\hline
			Gl~846 & 3902 $\pm$ 31 & 5.07 $\pm$ 0.05 & 0.37 $\pm$ 0.10 & 3861 $\pm$ 30 & 4.85 $\pm$ 0.09 & 0.34 $\pm$ 0.10 & 3815 $\pm$ 31 & 4.65 $\pm$ 0.05 & 0.04 $\pm$ 0.10 & 3867 $\pm$ 30 & 4.85 $\pm$ 0.09 & 0.08 $\pm$ 0.10\\
			Gl~880 & 3773 $\pm$ 32 & 5.05 $\pm$ 0.05 & 0.54 $\pm$ 0.10 & 3732 $\pm$ 30 & 4.87 $\pm$ 0.05 & 0.52 $\pm$ 0.10 & 3674 $\pm$ 31 & 4.60 $\pm$ 0.05 & 0.18 $\pm$ 0.10 & 3745 $\pm$ 30 & 4.87 $\pm$ 0.05 & 0.23 $\pm$ 0.10\\
			Gl~15A & 3673 $\pm$ 32 & 5.09 $\pm$ 0.05 & -0.25 $\pm$ 0.10 & 3632 $\pm$ 30 & 4.96 $\pm$ 0.07 & -0.26 $\pm$ 0.10 & 3622 $\pm$ 31 & 4.61 $\pm$ 0.05 & -0.45 $\pm$ 0.10 & 3721 $\pm$ 30 & 4.96 $\pm$ 0.07 & -0.42 $\pm$ 0.10\\
			Gl~411 & 3563 $\pm$ 31 & 4.91 $\pm$ 0.05 & -0.25 $\pm$ 0.10 & 3583 $\pm$ 30 & 4.97 $\pm$ 0.15 & -0.24 $\pm$ 0.10 & 3548 $\pm$ 31 & 4.49 $\pm$ 0.05 & -0.50 $\pm$ 0.10 & 3706 $\pm$ 30 & 4.97 $\pm$ 0.15 & -0.43 $\pm$ 0.10\\
			Gl~752A & 3588 $\pm$ 32 & 5.05 $\pm$ 0.05 & 0.36 $\pm$ 0.10 & 3561 $\pm$ 30 & 4.92 $\pm$ 0.08 & 0.34 $\pm$ 0.10 & 3530 $\pm$ 31 & 4.57 $\pm$ 0.05 & 0.05 $\pm$ 0.10 & 3605 $\pm$ 30 & 4.92 $\pm$ 0.08 & 0.11 $\pm$ 0.10\\
			Gl~849 & 3513 $\pm$ 34 & 5.10 $\pm$ 0.06 & 0.54 $\pm$ 0.10 & 3493 $\pm$ 30 & 4.93 $\pm$ 0.08 & 0.51 $\pm$ 0.10 & 3475 $\pm$ 31 & 4.70 $\pm$ 0.06 & 0.22 $\pm$ 0.10 & 3525 $\pm$ 30 & 4.93 $\pm$ 0.08 & 0.27 $\pm$ 0.10\\
			Gl~436 & 3539 $\pm$ 31 & 5.06 $\pm$ 0.05 & 0.18 $\pm$ 0.10 & 3520 $\pm$ 30 & 4.95 $\pm$ 0.08 & 0.17 $\pm$ 0.10 & 3497 $\pm$ 31 & 4.61 $\pm$ 0.05 & -0.09 $\pm$ 0.10 & 3575 $\pm$ 30 & 4.95 $\pm$ 0.08 & -0.04 $\pm$ 0.10\\
			Gl~725A & 3467 $\pm$ 31 & 4.93 $\pm$ 0.05 & -0.27 $\pm$ 0.10 & 3491 $\pm$ 30 & 5.01 $\pm$ 0.08 & -0.26 $\pm$ 0.10 & 3459 $\pm$ 31 & 4.55 $\pm$ 0.05 & -0.46 $\pm$ 0.10 & 3601 $\pm$ 30 & 5.01 $\pm$ 0.08 & -0.39 $\pm$ 0.10\\
			Gl~725B & 3346 $\pm$ 31 & 4.88 $\pm$ 0.05 & -0.37 $\pm$ 0.10 & 3402 $\pm$ 30 & 5.05 $\pm$ 0.11 & -0.33 $\pm$ 0.10 & 3349 $\pm$ 31 & 4.53 $\pm$ 0.05 & -0.55 $\pm$ 0.10 & 3523 $\pm$ 30 & 5.05 $\pm$ 0.11 & -0.43 $\pm$ 0.10\\
			Gl~15B & 3254 $\pm$ 32 & 5.01 $\pm$ 0.06 & -0.58 $\pm$ 0.10 & 3295 $\pm$ 30 & 5.13 $\pm$ 0.09 & -0.52 $\pm$ 0.10 & 3257 $\pm$ 31 & 4.66 $\pm$ 0.05 & -0.67 $\pm$ 0.10 & 3404 $\pm$ 30 & 5.13 $\pm$ 0.09 & -0.54 $\pm$ 0.10\\
			Gl~699 & 3190 $\pm$ 32 & 4.71 $\pm$ 0.06 & -0.70 $\pm$ 0.10 & 3329 $\pm$ 30 & 5.13 $\pm$ 0.14 & -0.61 $\pm$ 0.10 & 3259 $\pm$ 47 & 4.58 $\pm$ 0.12 & -0.80 $\pm$ 0.11 & 3440 $\pm$ 30 & 5.13 $\pm$ 0.14 & -0.62 $\pm$ 0.10\\
			Gl~905 & 2994 $\pm$ 32 & 4.99 $\pm$ 0.06 & -0.07 $\pm$ 0.11 & 3028 $\pm$ 30 & 5.14 $\pm$ 0.11 & 0.04 $\pm$ 0.10 & 3023 $\pm$ 35 & 4.67 $\pm$ 0.08 & -0.09 $\pm$ 0.11 & 3140 $\pm$ 30 & 5.14 $\pm$ 0.11 & 0.22 $\pm$ 0.10\\
			\hline
		\end{tabular}
	}
	\label{tab:retrieved_parameters_phoenix}
\end{table*}

\subsection{Benchmarking the precision of our parameter determination}
\label{sec:simulations}
To better assess the precision of the derived parameters, and the reliability of the derived error bars, we carried out a benchmark using synthetic spectra to simulate SPIRou templates, that we analyzed in a second step with the procedure outlined in Sec~\ref{sec:model_description} to Sec~\ref{sec:param_determination}.

To achieve this, we randomly generated 100 spectra with parameters ranging from 3000~K to 4000~K in $\teff$, from 3.5~dex to 5.5~dex in $\logg$ and from -0.5~dex to 0.5~dex in $\mh$. We added Gaussian noise to these spectra to simulate a signal-to-noise ratio (SNR) of~$\sim$100 in the H band, accounting for both the blaze in each order and the throughput of SPIRou~\citep{donati_2020}. We then ran the procedure described in Sec.~\ref{sec:param_determination} on the simulated spectra to recover optimal values and corresponding error bars for $\teff$, $\logg$ and $\mh$ for each of these spectra. 
The test was performed with either PHOENIX or MARCS models to simulate SPIRou templates and carry out the analysis, leading to 4 cases to study.
Fig.~\ref{fig:simulations_teff} presents the results of the different cases along with the corresponding residuals. Linear trends are fitted on the retrieved parameters, with the slopes and intercepts listed in Table~\ref{tab:coeffs_fit}.

Performing the simulations with the same model (PHOENIX or MARCS) used to  
produce the input spectra and to run the analysis, we compute a minimum reduced $\chi^2$ close to 1, and we are able to assess the precision of the formal error bars computed as described in Sec.~\ref{sec:param_determination}. With the PHOENIX (respectively MARCS) synthetic spectra, we compute a RMS on the residuals of 8.2~K in $\teff$, 0.019~dex in $\logg$ and 0.015~dex in $\mh$ (respectively 8.4~K in $\teff$, 0.020~dex in $\logg$ and 0.018~dex in $\mh$), slightly larger than the formal error bars of the order of 7.9~K in $\teff$, 0.017~dex in $\logg$ and 0.012~dex in $\mh$ (respectively, 7.7~K in $\teff$, 0.017~dex in $\logg$ and 0.010~dex in $\mh$). These results tend to indicate that the formal error bars are overestimated by about 10-20\%, maybe up to 60\% on the metallicity with the MARCS models. These error bars are those one could expect if the only source of uncertainty on the spectrum was the photon noise.

When using the PHOENIX models to simulate the template-like spectra and running the analysis with the grid of MARCS spectra, or vice-versa, we compute a typical minimum reduced $\chi^2$ of 1.8. Ensuring a reduced $\chi^2$ of 1 as described in Sec~\ref{sec:param_determination}, we compute formal error bars of the order of about 10~K in $\teff$, 0.025~dex in $\logg$ and 0.015~dex in $\mh$.
The RMS of the residuals is of the order of 30~K in $\teff$, 0.05~dex in $\logg$ and 0.1~dex in $\mh$, significantly larger than the computed formal error bars, which demonstrates that rescaling the $\chi^2$ to 1 is not a sufficient correction
to fully account for the uncertainty added by the systematic differences between the models.
We therefore define updated error bars, which we will refer to as empirical error bars,
as the quadratic sum of the formal error bars and estimates of the RMS computed when comparing the models, i.e. 30~K in $\teff$, 0.05~dex in $\logg$ and 0.1~dex in $\mh$.

We additionally observe systematic shifts and trends when comparing the retrieved parameters to the expected values. In particular, the grid of MARCS spectra leads to systematic underestimates of $\logg$ (by about 0.4~dex) and $\mh$ (by about 0.3~dex) when compared to the values adopted for the PHOENIX models, and vice-versa.

\begin{figure*}
    \subfigure{
    \includegraphics[width=\columnwidth]{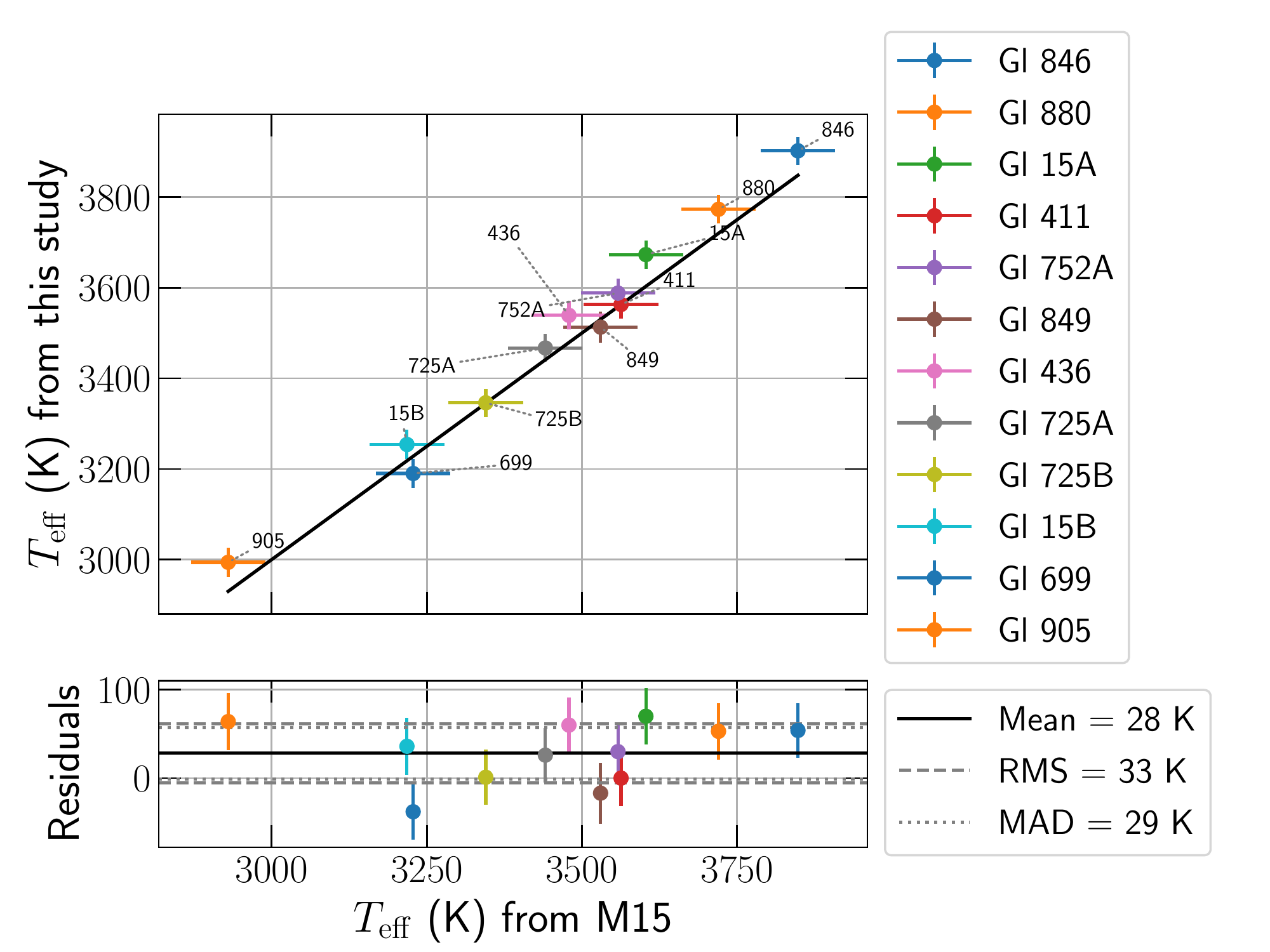}
    }
    \subfigure{
    \includegraphics[width=\columnwidth]{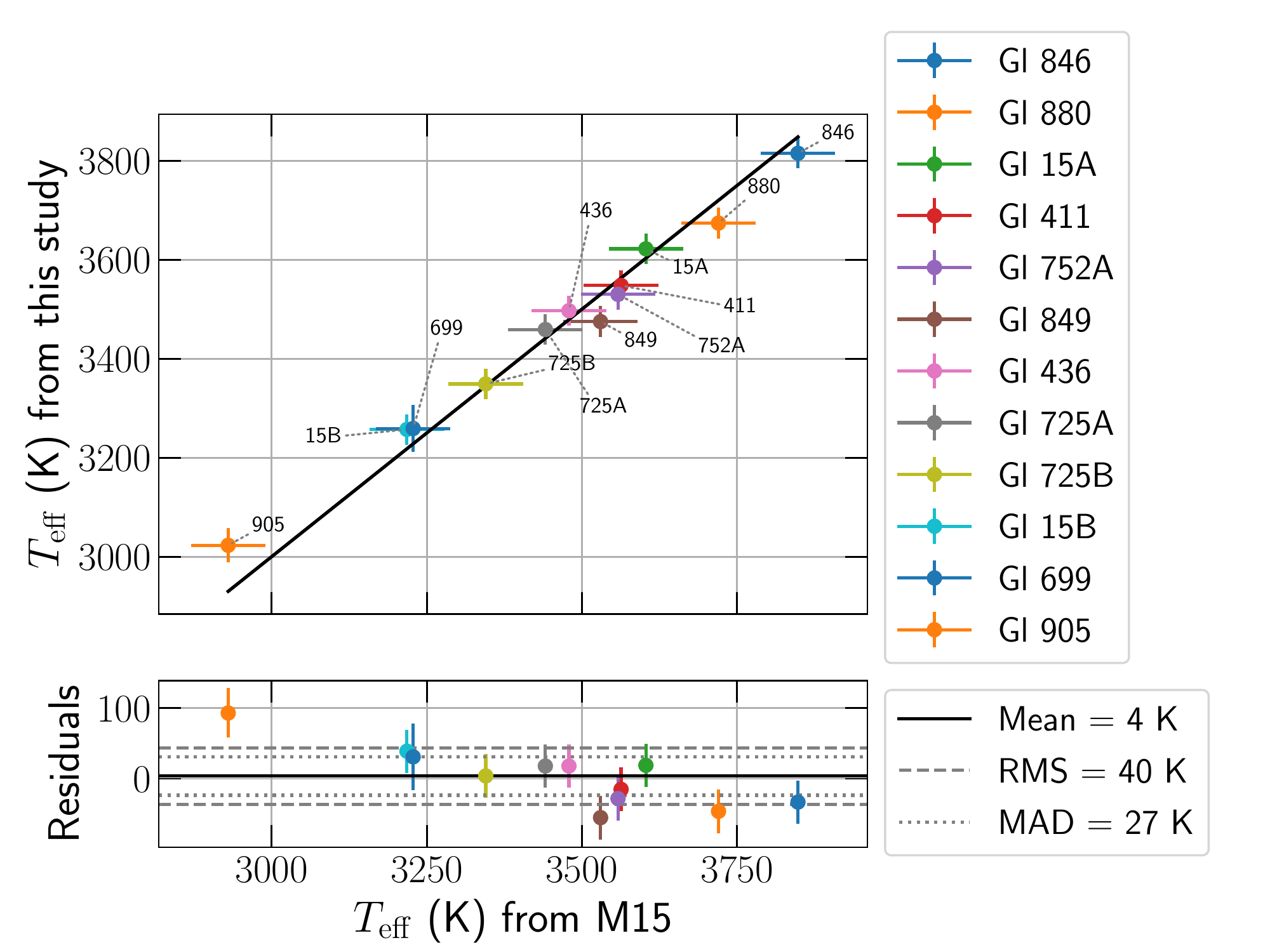}
    }
    \caption{Retrieved $\teff$ using the grid of PHOENIX (left) and MARCS (right) spectra plotted against values published by~\mann{}. The bottom plot presents the residuals, i.e. the retrieved values minus literature values. RMS and MAD values are computed after application of a sigma clipping function on the residuals with a threshold at 5~$\sigma$.}
    \label{fig:results_teff}
\end{figure*}

\begin{figure*}
    \subfigure{
    \includegraphics[width=\columnwidth]{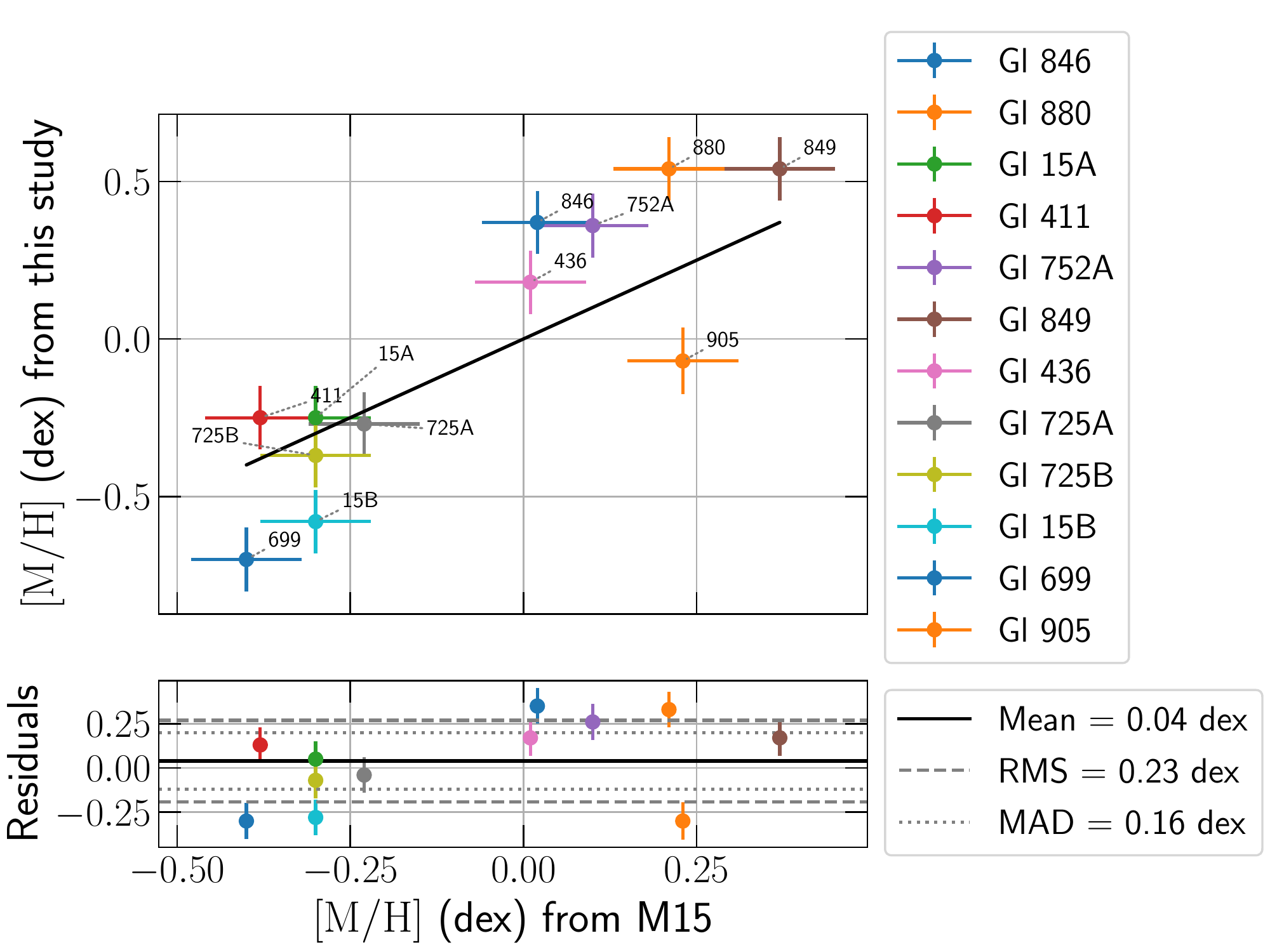}
    }
    \subfigure{
    \includegraphics[width=\columnwidth]{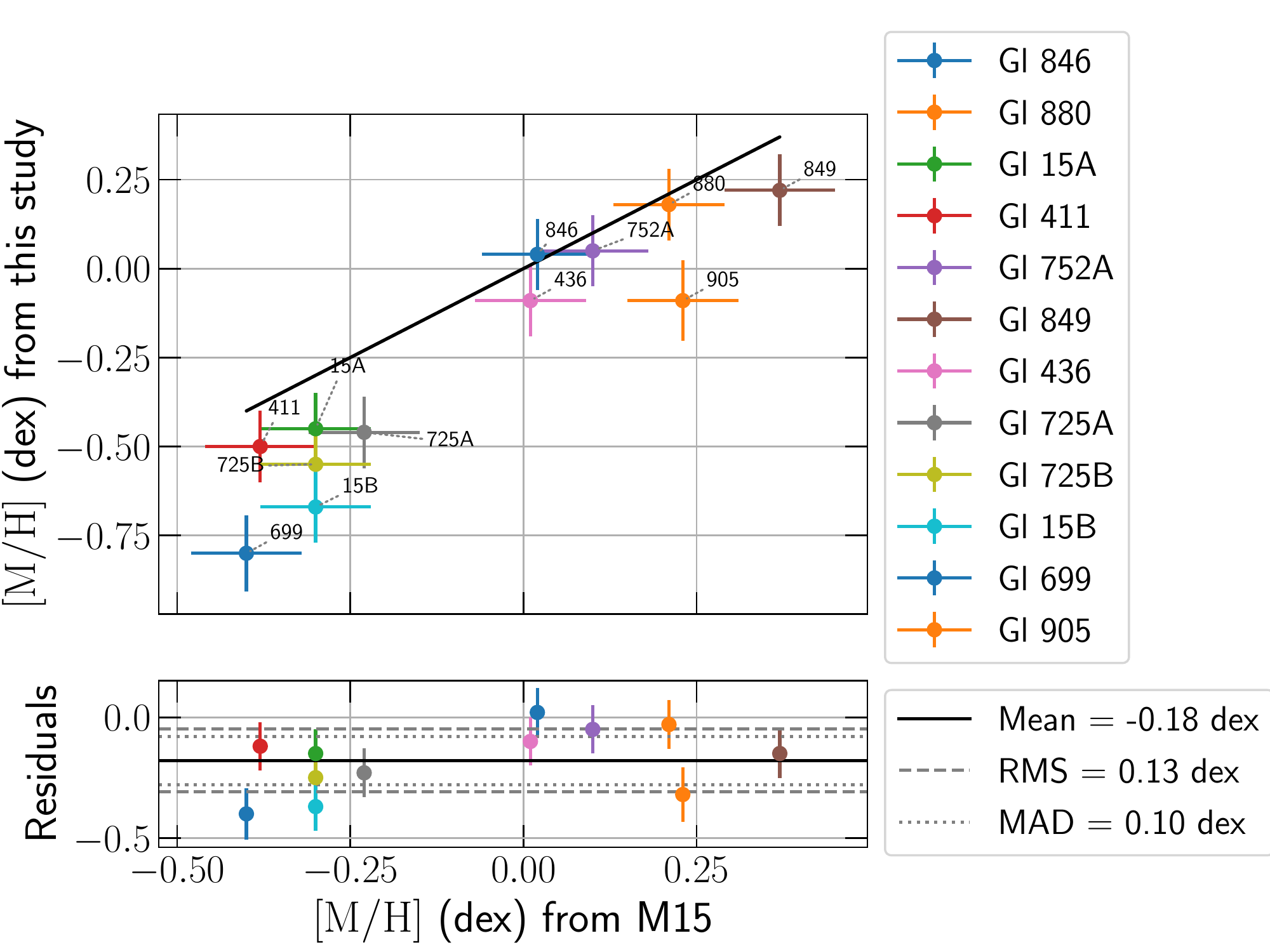}
    }
    \caption{Same as Fig.~\ref{fig:results_teff} but for $\mh$.}
    \label{fig:results_mh}
\end{figure*}
\begin{figure*}
    \subfigure{
    \includegraphics[scale=0.4]{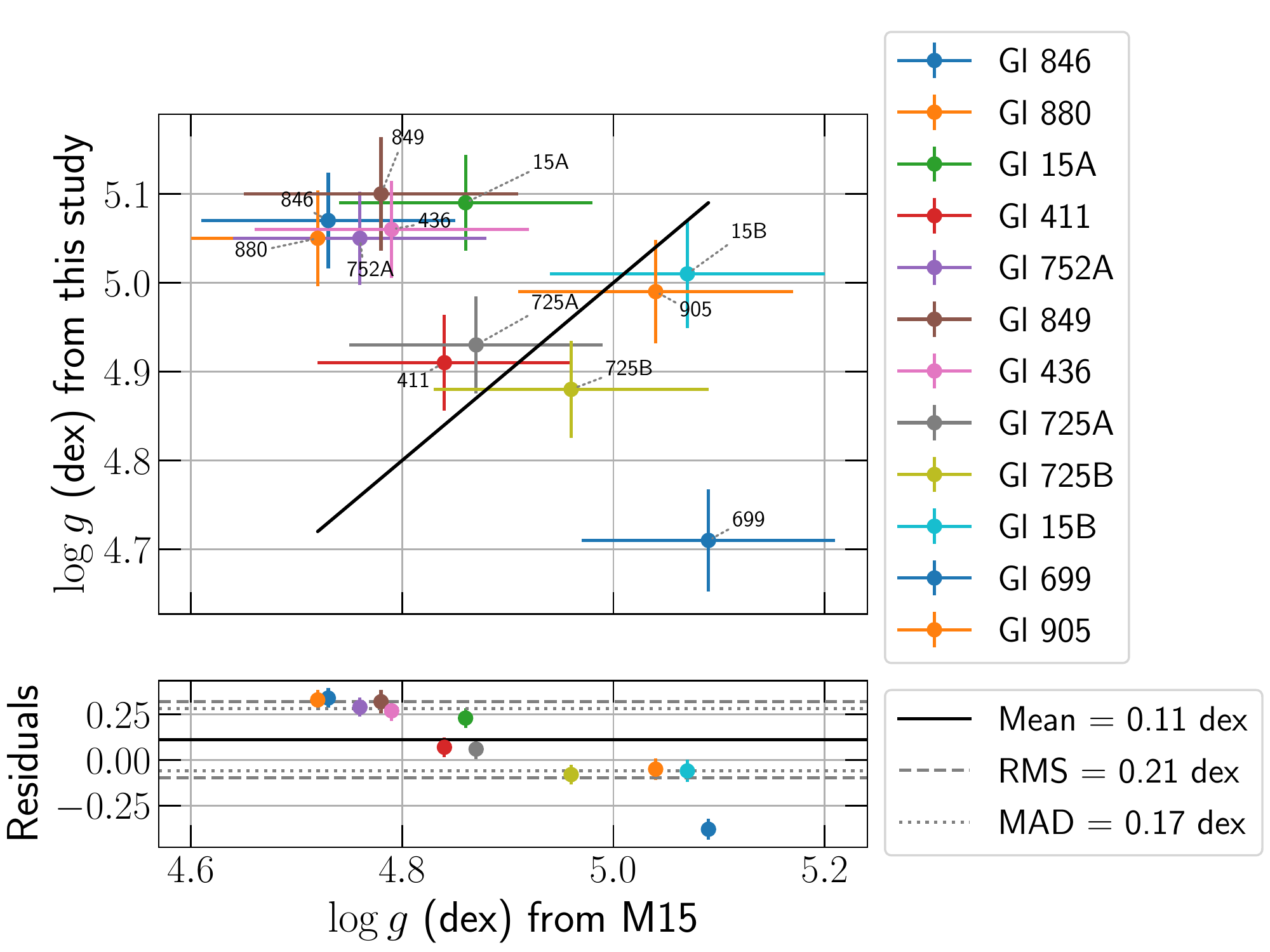}
    }
    \subfigure{
    \includegraphics[scale=0.4]{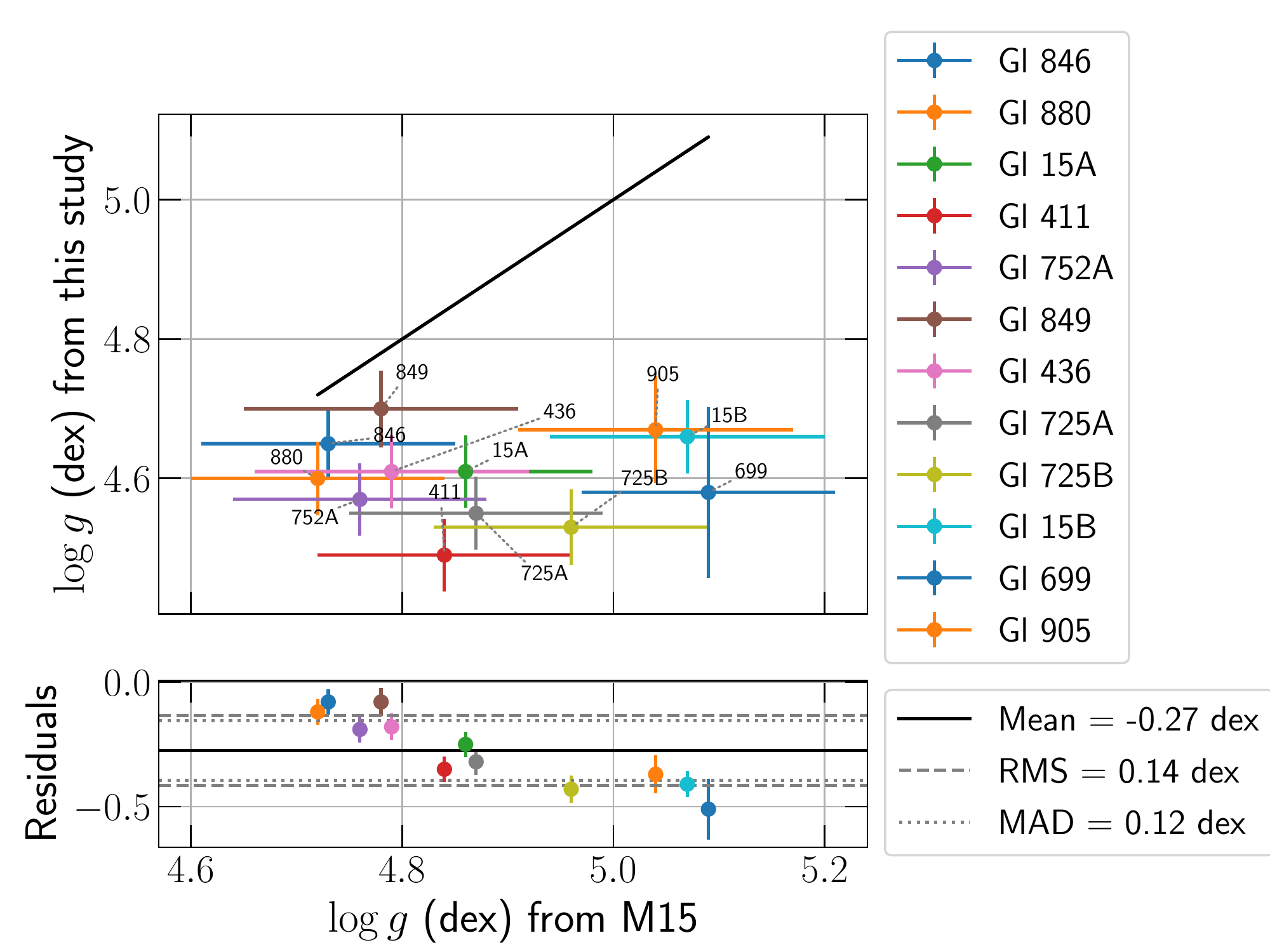}
    }
    \caption{Same as Fig.~\ref{fig:results_teff} but for $\logg$.}
    \label{fig:results_logg}
\end{figure*}

\section{Results}
\label{sec:results}

We performed the analysis described in Sec.~\ref{sec:spectral_anaysis} for the twelve stars in our sample assuming a broadening kernel of FWHM $\vbroad=3$~$\kms$ (corresponding to a velocity of $\xi=1.8$~$\kms$ if the broadening is fully attributed to macroturbulence). The retrieved parameters are reported in Table~\ref{tab:retrieved_parameters_phoenix}, and presented among literature values in Fig.~\ref{fig:teff_literature}.

We find that, for each SPIRou template, 
the minimum $\chi^2$ value ($\chi^2_{\rm min}$) retrieved for the best fit is systematically larger than the number of used data points (N, typically 1200), reflecting systematic differences between observations and synthetic spectra that are not accounted for.
More specifically, the reduced $\chi^2$ computed when comparing SPIRou templates to observation is on average of 250, much larger than the 1.8 found when comparing synthetic models (see Sec~\ref{sec:spectral_anaysis}). Here again, we ensure that our formal error bars account for some of these differences by forcing the $\chi^2$ to 1, as described in Sec.~\ref{sec:param_determination}.

The typical level to which our template spectra are fitted is equal to 2 to 3~\% of the continuum.

\subsection{Effective temperature}

Fig.~\ref{fig:results_teff} presents a comparison between 
the $\teff$ values derived
using the grid of PHOENIX and MARCS synthetic spectra and the values published in \mann{}. Fig.~\ref{fig:results_teff_pass} presents the same results compared to the values published by~\pass{}.
With both models, we found $\teff$ values in good agreement with the values published by~\mann{}, with empirical error bars of the order of 30~K. We also compute a RMS value of the order of 40~K, smaller that the typical uncertainties reported by~\mann{}. Additionally, we find that the values recovered with the grid of PHOENIX models are on average about 30~K higher than with the MARCS models, comparable to the difference observed when running the simulations (see Sec.~\ref{sec:simulations}).


Looking more specifically at how our $\teff$ values derived with the grids of PHOENIX and MARCS spectra vary with those of~\mann{}, we find trends whose slopes are not exactly one, but rather 1.02~$\pm$~0.04 and 0.85~$\pm$~0.03 respectively, and with RMS dispersion about this trend equal to 33~K and 21~K respectively, close to the computed empirical error bars.
These trends are in fair agreement with those computed when comparing the two models with simulated data (see Sec.~\ref{sec:simulations}).

\subsection{Metallicity}

The values of $\mh$ estimated from both the PHOENIX and MARCS spectral grids are compared to the values published by~\mann{} in Fig.~\ref{fig:results_mh}. Fig.~\ref{fig:results_mh_pass} presents a similar comparison of our results to the values published by~\pass{}. The typical empirical error bars obtained for $\mh$ are about 0.10~dex with the two grids, i.e. about 1.5 to 2.5 times smaller than the RMS between our values and those of \mann{} (equal to 0.14~dex with the grid of MARCS spectra and 0.23~dex with the grid of PHOENIX spectra), and of the order of the $\mh$ uncertainties derived by \mann{} (equal to 0.08~dex).
We also find that the estimated $\mh$ derived with the MARCS spectra are on average 0.18~dex smaller than the values published by~\mann{}. The large offset in the average values retrieved with the PHOENIX and MARCS models, of about 0.4~dex, is in good agreement with the offsets computed in our simulations introduced in Sec.~\ref{sec:simulations}

For the two binary stars in our sample, we compare the metallicities of both components. With the grid of MARCS spectra, for the Gl~15AB and the Gl~725AB binaries, we find differences in the metallicities of 0.21~dex and 0.09~dex, respectively. The values derived with this model agree at a 2$\sigma$ level with the computed empirical error bars.
With the grid of PHOENIX spectra, the retrieved $\mh$ values differ by 0.10~dex for Gl~725A and Gl~725B, again in good agreement with our empirical error bars; but the difference in $\mh$ values reaches 0.33~dex for  Gl~15A and Gl~15B, i.e. 3.3 times our empirical error bars.



\subsection{Surface gravity}

Fig.~\ref{fig:results_logg} presents a comparison between the $\logg$ estimates derived with the grid of PHOENIX and MARCS spectra and the values published by \mann{}. 
The $\logg$ values recovered with the grid of PHOENIX spectra are largely scattered around the equality line, with a computed RMS of the residuals of about 0.2~dex, 3 to 4 times the typical empirical error bars. With the grid of MARCS spectra, the values of $\logg$ appear to be systematically underestimated by about 0.30~dex with respect to~\mann{}, and the RMS of residuals is of 0.16~dex, close to the uncertainties published by~\mann{} for these parameters (of 0.12~dex). 
{\paul We also notice that the retrieved $\logg$ values do not fully agree with those expected from the mass luminosity relations and interferometric data (see Sec~\ref{sec:empirical_logg}), although we remind that they span only a small range of values (smaller than the step size in $\logg$ within the grid of synthetic spectra, equal to 0.5~dex).}

\begin{table*}
    \centering
    \caption{Mean, RMS and MAD values of the residuals -- i.e. parameter values of this study minus values published by~\mann{} -- derived with the PHOENIX and MARCS spectral grids. The label `Fixed $\logg$' specifies that we adopted the values presented column 6 of Table~\ref{tab:literature_values} for this parameter, and the values in parentheses therefore do not originate from fits.}
    \begin{tabular}{c ccc ccc ccc}
    \hline
         Model used & \multicolumn{3}{c}{$\teff$ (K)} & \multicolumn{3}{c}{$\logg$} & \multicolumn{3}{c}{[M/H]} \\
        \hline
        & MEAN & RMS & MAD & MEAN & RMS & MAD & MEAN & RMS & MAD  \\
		PHOENIX & 28 & 33 & 29 & 0.11 & 0.21 & 0.17 & 0.04 & 0.23 & 0.16 \\ 
		MARCS & 4 & 40 & 27 & -0.27 & 0.14 & 0.12 & -0.18 & 0.13 & 0.10 \\ 
		PHOENIX (Fixed $\logg$) & 21 & 48 & 36 & 0.04 & 0.03 & 0.02 & 0.04 & 0.19 & 0.13 \\ 
		MARCS (Fixed $\logg$) & 111 & 82 & 72 & 0.04 & 0.03 & 0.02 & -0.11 & 0.09 & 0.06 \\ 
        \hline
    \end{tabular}
    \label{tab:rms_mad}
\end{table*}


Given that $\logg$ is apparently difficult to constrain reliably, at least from the list of stellar lines used, we attempted to improve the precision on the other parameters by fixing the value of $\logg$ to the values derived from mass-radius relations and evolutionary models (see Sec.~\ref{sec:empirical_logg}). Our approach is similar to that used by~\citet{mann_2015}, who derived masses from mass-luminosity relations and radii from bolometric flux and parallaxes.  The estimated $\teff$ and $\mh$ with both the PHOENIX and MARCS spectral grids are listed in Table~\ref{tab:retrieved_parameters_phoenix}.

This additional constraint has little impact on the $\teff$ and $\mh$ derived with the grid of PHOENIX spectra. With this grid, the most notable change is a trend in the recovered $\teff$ of slope 0.83 ± 0.03 with respect to the values of~\mann{}, which causes an increase in the computed RMS for this parameter.
With the grid of MARCS spectra, we observe a significant offset of about 100~K in the retrieved $\teff$ values, along with a RMS of about 85~K, about twice the RMS computed when fitting all three parameters. 
Moreover, we observe that fixing $\logg$ does not reduce significantly the gap between the recovered $\mh$ for Gl~15A and Gl~15B with the grid of PHOENIX spectra.

All RMS and MAD values are listed in Table~\ref{tab:rms_mad}.

\begin{figure*}
    \centering
    \subfigure{
     \includegraphics[width=.9\columnwidth, trim={1 1 1 1}, clip]{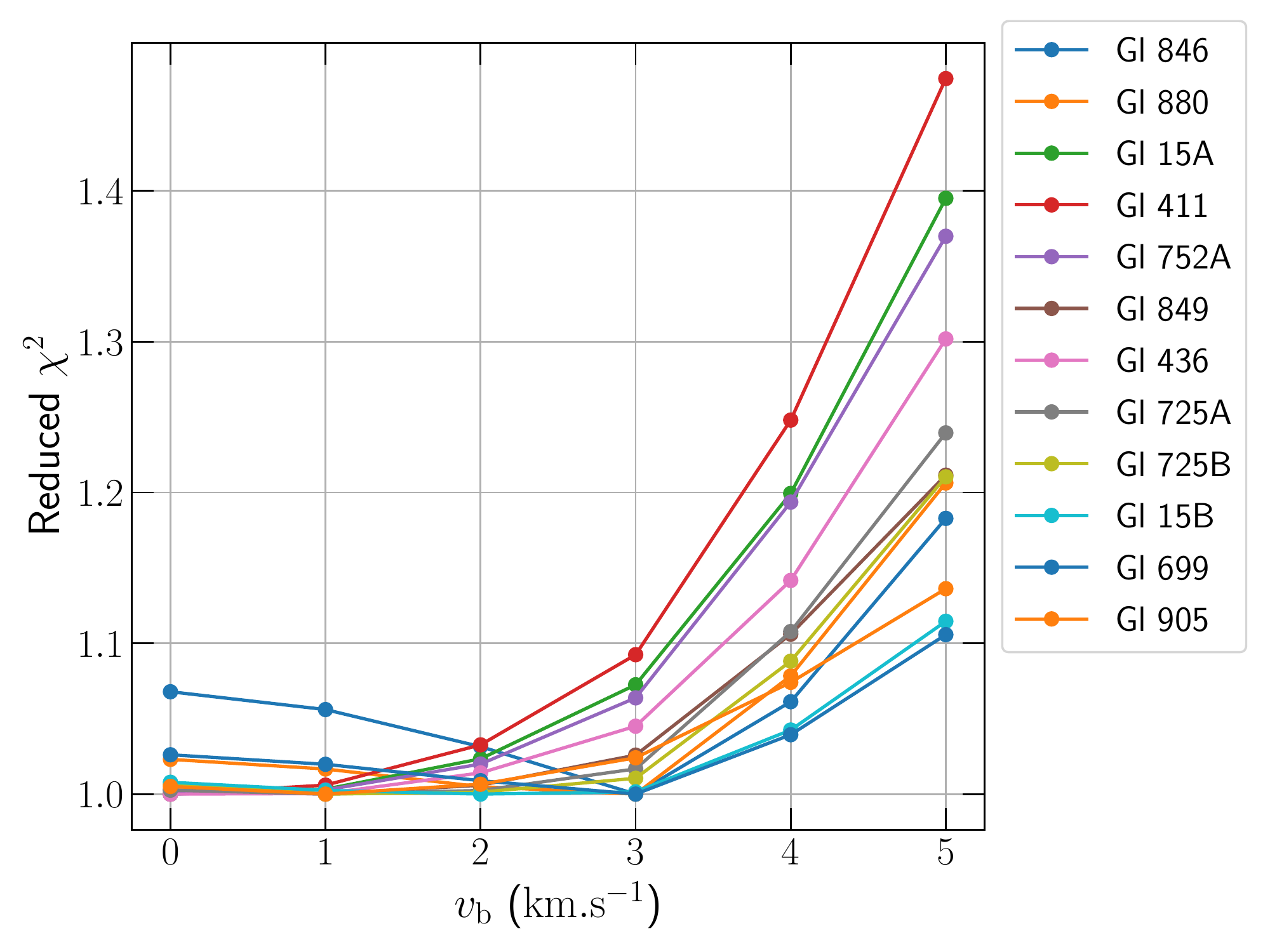}
     }
    \subfigure{
     \includegraphics[width=.9\columnwidth, trim={1 1 1 1}, clip]{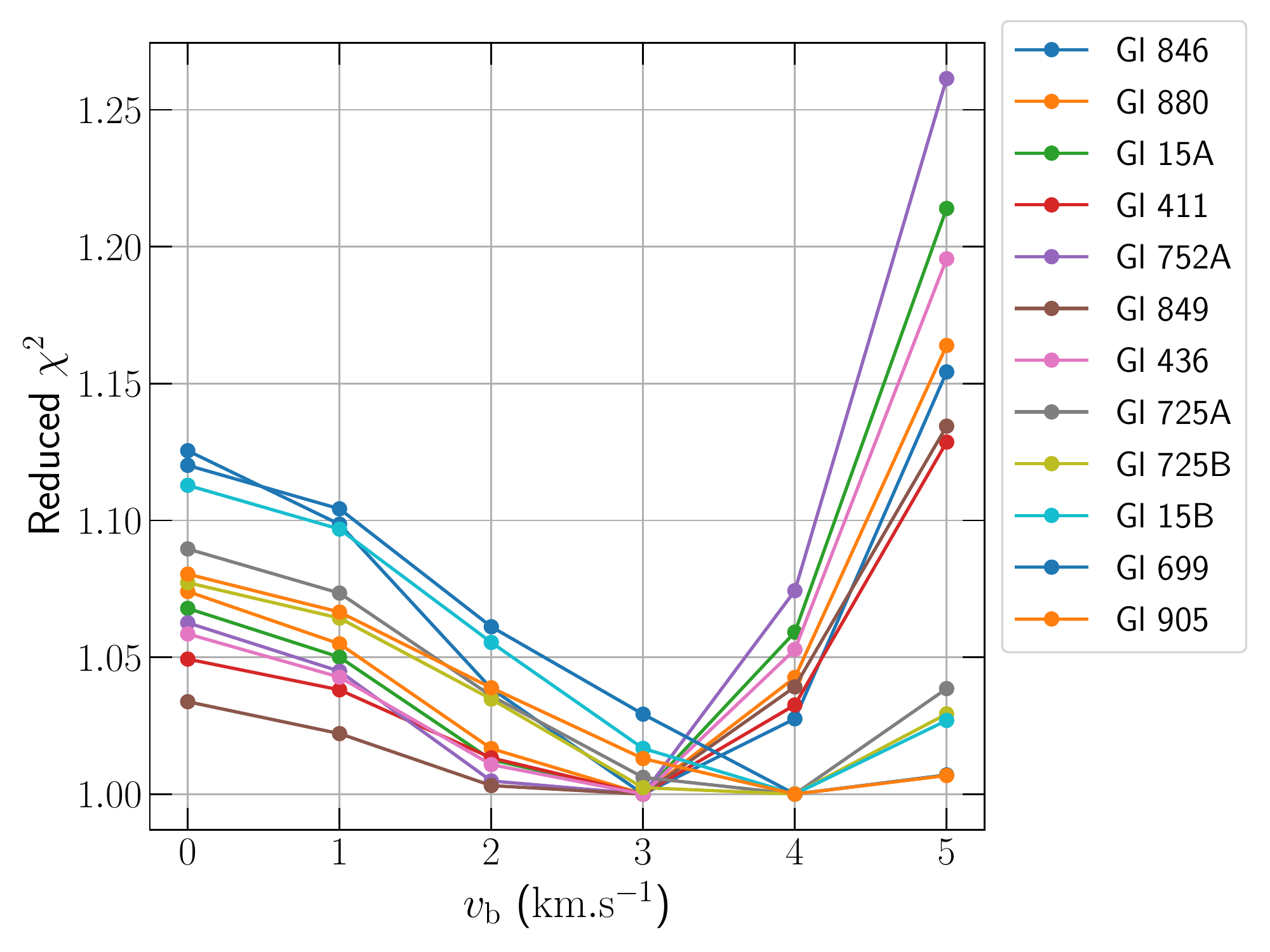}
     }
    \caption{Reduced $\chi^2$ as a function of the FWHM of the considered Gaussian profile, $\vbroad$, obtained with the grid of PHOENIX (left) and MARCS (right) synthetic spectra, for each target of our sample. The reduced $\chi^2$ for each star are rescaled to the minimum value reached over the range of $\vbroad$, to ease comparison.}
    \label{fig:chi2_fun_vmac}
\end{figure*}

\subsection{Assessing the influence of \texorpdfstring{$\vbroad$}{TEXT} on the recovered parameters}
We repeated our analysis for several values of the FWHM ($\vbroad$) considered for the Gaussian profile used to broaden the synthetic spectra,
 which accounts for the joint contributions of  $\vsini$,  $\vmac$, and any other underestimated broadening effect.
As shown in Fig.~\ref{fig:chi2_fun_vmac}, we find that the value of $\vbroad$ providing an optimal fit to the observed spectra falls in the range 1-3~$\kms$
for most of the stars in our sample,
and is lower with the grid of PHOENIX spectra than with the grid of MARCS spectra. As already stressed, being FWHM, these values should be compared with care to $\vsini$ or $\vmac$ estimates reported in the literature.
We also report that the assumed value of $\vbroad$ has no more than a weak impact on the retrieved parameters.
The mean, RMS and MAD computed for different values of $\vbroad$ are presented in Table~\ref{tab:rms_fun_vb}.


\subsection{Estimating the precision of formal error bars}
\label{sec:uncertainties}
\begin{figure}
    \centering
    \includegraphics[width=\columnwidth]{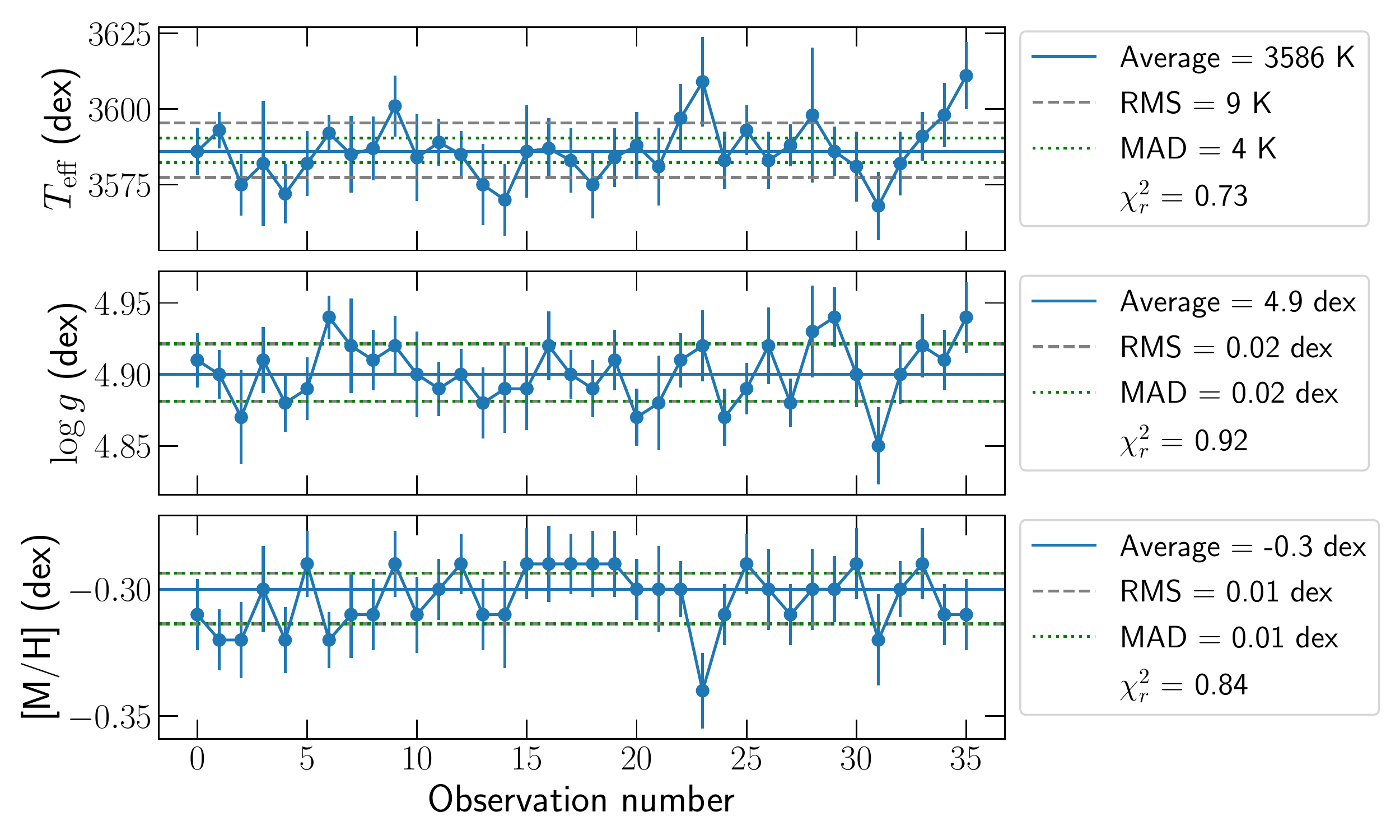}
    \caption{Optimal estimates and corresponding error bars retrieved for a series of SPIRou spectra of Gl~411. The mean value, standard deviation and median absolute deviation are indicated for each series. We also compute the reduced $\chi^2$ to the mean value of each series.}
    \label{fig:series_gl15a}
\end{figure}







To further assess the precision of the method, we performed the analysis on numerous SPIRou spectra acquired for a single target.
Fig.~\ref{fig:series_gl15a} presents the parameters retrieved for our series of Gl~411 spectra along the computed formal error bars, which only account for photon noise and part of the systematics.
We observe fluctuations in the retrieved parameters, and estimate their deviation to the mean with respect to our formal error bars by computing the reduced $\chi^2$ on the series of retrieved values.
The computed reduced $\chi^2$ show that the formal error bars on the retrieved parameters seem to provide a reliable value on the internal measurement precision.
We repeated this test on a series of high-SNR spectra of Gl~699 and recovered a reduced $\chi^2$ of 0.66 for $\teff$, 0.85 on $\logg$ and 0.89 on $\mh$, again suggesting that our formal error bars properly account for the deviation found in the parameters for a given star, i.e. at a given point of the $\teff$-$\logg$-$\mh$ parameter space.
As a sanity check, we performed MCMC computations to explore the $\chi^2$ grid, performing linear interpolation within the grid to retrieve the $\chi^2$ values at each MCMC step. We find that the derived parameters and error bars are in good agreement with those obtained with our main method (described in Sec~\ref{sec:param_determination}).

\section{Discussion and conclusions}
\label{sec:conclusions}

In this paper, we presented the results of a method aimed at determining the fundamental parameters of M dwarfs (i.e. $\teff$, $\logg$, $\mh$) from nIR high-resolution spectra acquired with SPIRou. 

We built high-SNR template spectra of 12 inactive M dwarfs from 40 to 80
observed SPIRou spectra collected for each star at a wide range of BERV values, allowing us to reliably correct these spectra
for telluric features and sky lines.
The correction is performed by iteratively fitting a synthetic model of the Earth atmosphere's transmission (TAPAS) on each observed spectrum, and taking advantage of the numerous observations acquired at various epochs for each target. PCA is also used to further improve telluric correction and remove emission lines from the sky at the same time.  
We then selected spectral regions that are sensitive to the stellar parameters to be retrieved and best reproduced by two
sets of synthetic spectra derived from PHOENIX and MARCS model atmospheres.

The analysis of the template spectra relies on their direct comparison to the synthetic spectra in the selected regions. 
Only small regions of the synthetic spectra reproduce observed features well enough to constrain parameters because of the lack of precision of the models and line lists currently used, especially in the nIR. We were therefore led to restrict our analysis to about 30 atomic lines, 2 OH lines
and 40 CO lines from the bands redward of 2293~nm,
in spite of the thousands of spectral lines present in the SPIRou spectra.
Moreover, remaining discrepancies are observed between the models and template spectra, even for the selected lines, leading to differences between the parameters recovered with both models.
The MARCS models rely on the most recent VALD line lists, updated since the publication of the PHOENIX models, which may partially explain the observed differences. 

To assess the reliability and precision of our method, 
we carried out a benchmark, substituting the template spectra with synthetic spectra generated for random parameters, and adding Gaussian noise to simulate a SNR per pixel of~$\sim$~100 in the H band.
These simulations allowed to confirm that the formal error bars computed with our procedure provide a fair estimate of the uncertainties associated with photon noise. 
By confronting the PHOENIX synthetic spectra to the MARCS synthetic spectra through our simulations, 
we observed a larger dispersion on the retrieved parameters than our formal error bars can account for, which can be attributed to the systematic differences between the models. We therefore chose to provide a more realistic estimation of the  error bars by taking the quadratic sum of these systematic error bars and our computed formal error bars.
Performing the analysis on our SPIRou templates, we derive empirical error bars of the order of 30~K in $\teff$, 0.05~dex in $\logg$ and 0.10~dex in $\mh$, smaller than the typically published uncertainties on these parameters.


In order to estimate the accuracy of our method with respect to values published in the literature, we compared our
results to the pseudo-empirical parameters estimated by~\citet{mann_2015}. In particular, we compute a standard deviation of about 30~K to 50~K in $\teff$, and 0.15~dex to 0.20~dex in $\logg$ and $\mh$ with the two grids of synthetic spectra, about twice larger than our empirical error bars, and comparable to the typical uncertainties published by~\pass{}.



Additionally, we find significant differences in the results obtained with the two grids of synthetic spectra, of about 30~K in $\teff$, 0.2~dex in $\mh$ and 0.4~dex in $\logg$. These observed offsets are in good agreement with these observed when comparing the PHOENIX and MARCS synthetic spectra through our simulations, and can therefore be attributed to the systematic differences in the line profiles predicted by the two models.
We also find trends between our retrieved $\teff$ and those of~\mann{}, with slopes that are not exactly equal to one (1.04 $\pm$ 0.04 and 0.86 $\pm$ 0.04 with the grids of PHOENIX and MARCS spectra respectively) and with RMS about these trends very close to the empirical error bars computed with both models ($\sim$30~K). These trends are also in good agreement with those retrieved when comparing the two models through our simulations.

Because constraining the surface gravity appears to be difficult,  we investigated the effect of fixing the values of $\logg$ to derive $\teff$ and $\mh$.
 This constraint caused a significant increase in the average and scattering of $\teff$ values derived with the grid of MARCS spectra, and did not bring significant improvement to the analysis relying on the grid of PHOENIX spectra.
 


Binary stars provide an additional way of testing the precision of our method, as we expect to retrieve similar metallicities for both components. 
For the 2 binaries included in our study and with both synthetic grids, these discrepancies tend to be of the order of 0.2~dex or lower, in rough agreement with our empirical error bars, except for the Gl~15AB binary when modeled with PHOENIX spectra, for which we find a difference of about 0.33~dex.
We also report that fixing $\logg$ to derive $\mh$ does not significantly reduce this gap.

The results presented in this paper demonstrate the viability of the approach, i.e. of extracting stellar parameters from nIR SPIRou spectra, and show that the necessary assumptions on which this study relies (such as the choice of broadening kernel and normalization strategies) have a much smaller impact on the results than the discrepancies found between synthetic models and observations.
A close comparison of the line parameters used by PHOENIX and MARCS (when available) shows significant differences for some lines. Our line selection procedure is however based on a comparison of both models, 
which likely led us to select lines for which parameters best agree between the two lists. Large differences however remain between the PHOENIX and MARCS synthetic spectra, even for the selected lines, which may indicate that the choice of model atmospheres, and modeling assumptions, may be responsible for most of the observed discrepancies.
A subsequent work will attempt to better understand these differences, in order to improve our modeling strategies and the accuracy of our analysis.


In a future study, 
we will attempt to produce PHOENIX spectra using the latest line lists available. This will allow us to carry out a more precise comparison of the PHOENIX and MARCS models and to assess the impact of the line lists on the produced spectra.  In parallel, we plan to improve the analysis by 
identifying more lines capable of constraining the stellar parameters, in particular $\logg$ and $\mh$. A second step will include the modification of the line list in the regions selected for the analysis, guided by the SPIRou high resolution spectra of reference stars, allowing to further calibrate the analysis method. 

Following works will then aim at
applying the procedure discussed in this paper to all M dwarfs observed with SPIRou as part of the SLS, in order to build a self-consistent database of stellar properties. We will also focus on
other classes of stars of interest for the SLS, in particular active pre-main-sequence (PMS) low-mass stars. These stars are known to be difficult to model because of the presence of large star spots and strong magnetic fields at their surface, for which a 2-temperature model seems to be required to obtain a proper fit to the spectra~\citep{santiago_2017}. 
By improving the spectral modeling of low-mass stars, we should be able to pinpoint their stellar properties with a higher precision than what is currently achieved, directly from nIR SPIRou spectra. In turn, such constraints will help to better characterize planets orbiting these stars, and to guide us towards more reliable atmospheric models of M dwarfs and PMS stars.




\section*{Acknowledgements}

This project received funding from the European Research Council under the H2020 research \& innovation program (grant \#740651 NewWorlds).

This work is based on observations obtained at the Canada-France-Hawaii Telescope (CFHT) which is operated by the National Research Council (NRC) of Canada, the Institut National des Sciences de l'Univers of the Centre National de la Recherche Scientifique (CNRS) of France, and the University of Hawaii. The observations at the CFHT were performed with care and respect from the summit of Maunakea which is a significant cultural and historic site. 

This work made use of TAPAS models acquired through the ETHER center (\url{http://ether.ipsl.jussieu.fr/tapas/ }).

This research has made use of the SIMBAD database~\citep{wenger_2000}, operated at CDS, Strasbourg, France

This work has made use of the VALD database, operated at Uppsala University, the Institute of Astronomy RAS in Moscow, and the University of Vienna.
We also acknowledge funding from the French National Research Agency (ANR) under contract number ANR-18-CE31-0019 (SPlaSH)

{\paul TM acknowledges financial support from the Spanish Ministry of Science and Innovation (MICINN)
through the Spanish State Research Agency, under the Severo Ochoa Program
2020-2023 (CEX2019-000920-S).}

{\paul We thank an anonymous referee for suggesting modifications that improved the manuscript.}

\section*{Data Availability}
The data used in the present work was acquired in the context of the SLS, and will be publicly available from the Canadian Astronomy Data Center one year following the completion of the SLS.




\bibliographystyle{mnras}
\bibliography{mnras_template} 



\appendix


\section{Selected lines compared synthetic models}
{\p Fig.~\ref{fig:lines_1} presents the templates spectra and best fitted PHOENIX and MARCS models for 8 of the selected regions used in the analysis. All the regions used for the analysis are presented in Fig.~A2 available as supplementary material.} The atomic line parameters considered by the models are presented in Table~\ref{tab:line_list}.


\begin{figure*}
    \centering
    \includegraphics[width=.5\columnwidth]{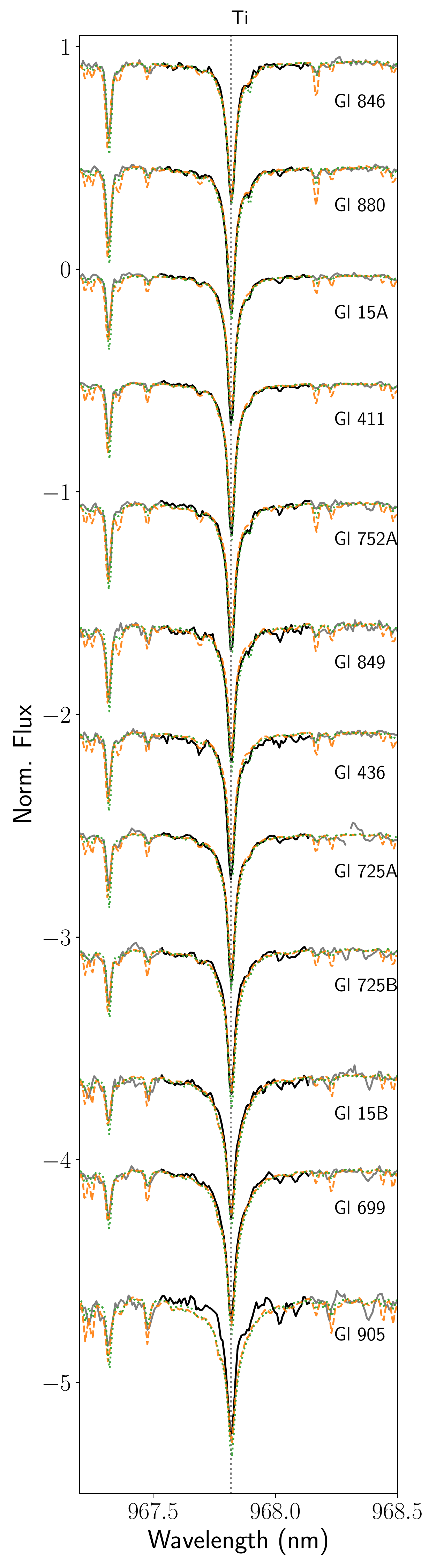}
    \includegraphics[width=.5\columnwidth]{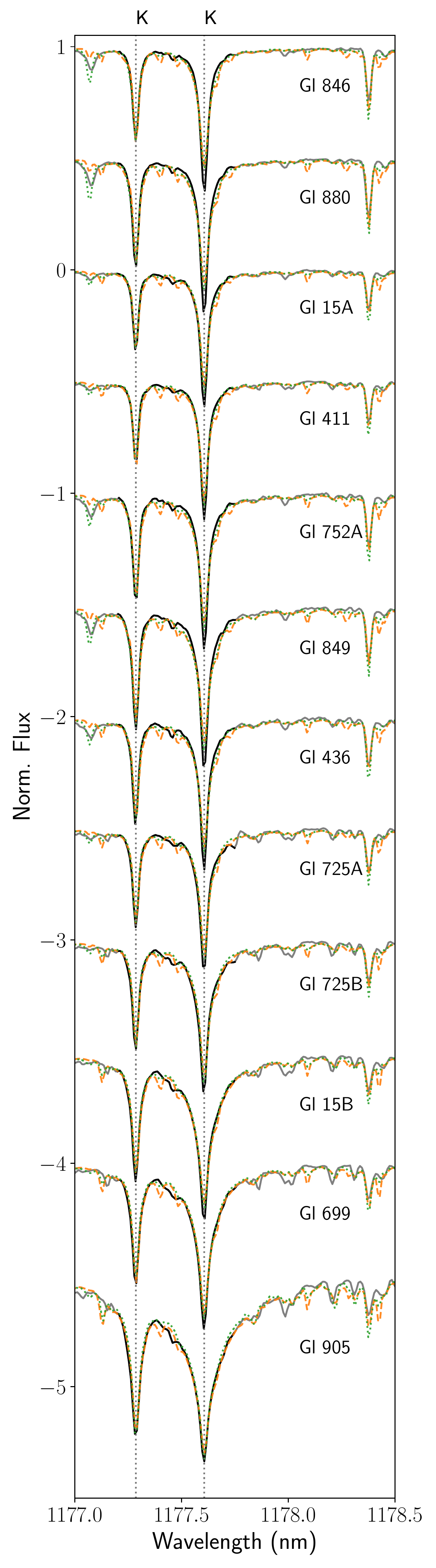}
    \includegraphics[width=.5\columnwidth]{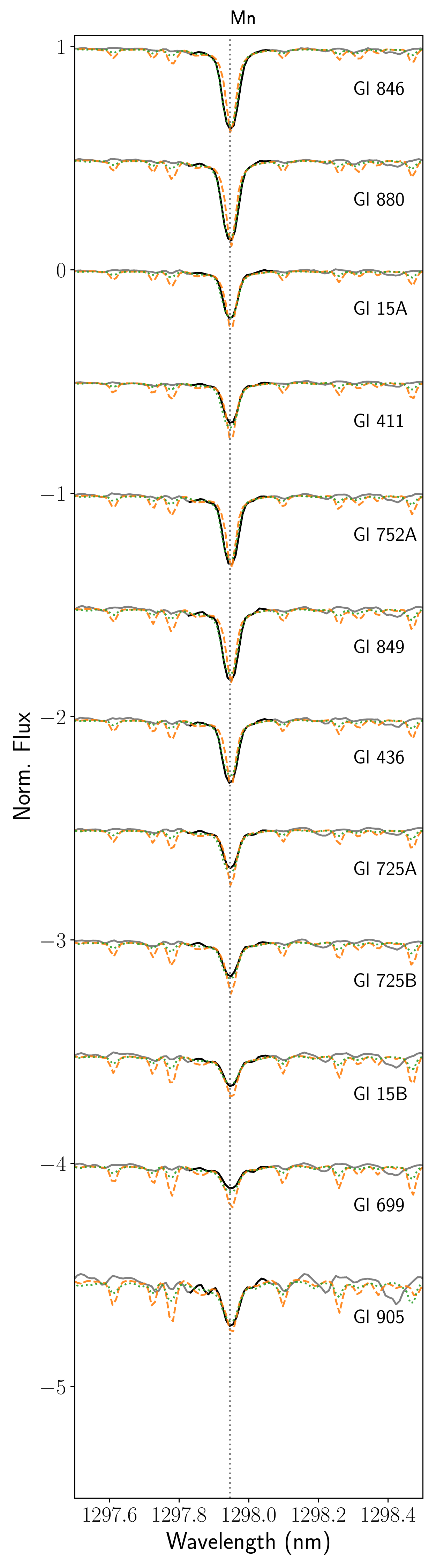}
    \includegraphics[width=.5\columnwidth]{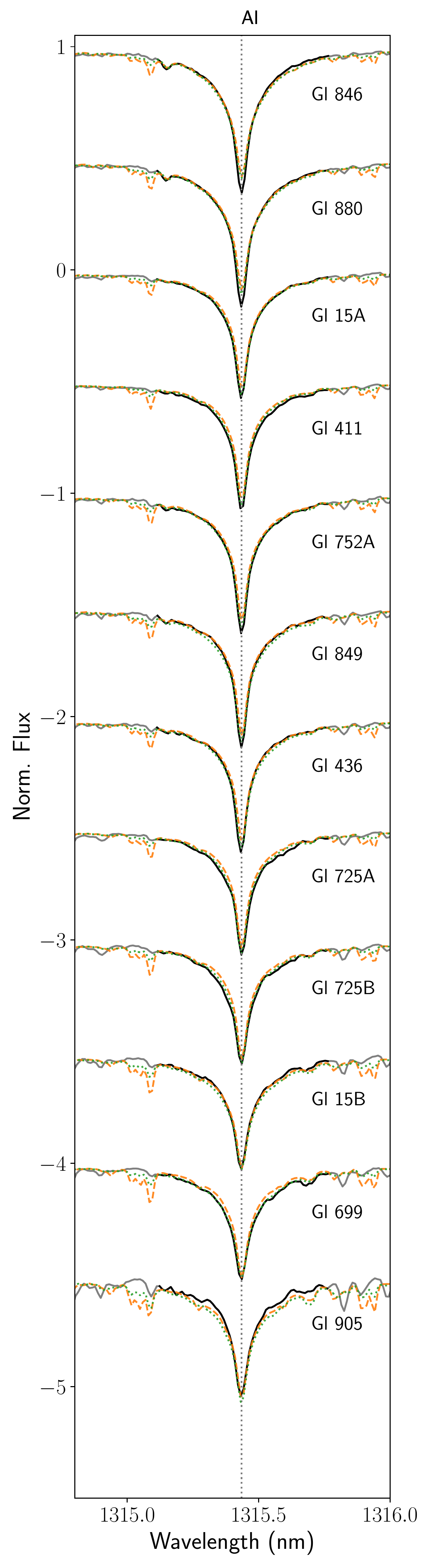}
    \caption{Template spectra (grey) along with the fitted PHOENIX model (dashed orange) and MARCS model (dotted green) for {\paul eight} spectral windows. Selected regions of the template spectra over which the comparison was carried out are shown in black. {\paul From top to bottom: Gl~846, Gl~880, Gl~15A, Gl~411, Gl~752A, Gl~849, Gl~436, Gl~725A, Gl~725B, Gl~15B, Gl~699 and Gl~905. Every spectrum but the first one is shifted by a multiple of 0.5 for better readability. {\p Figure A2 (available as supplementary material) shows all the spectral windows used for the analysis.}}}
    \label{fig:lines_1}
\end{figure*}

\begin{figure*}
    \centering
    \includegraphics[width=.5\columnwidth]{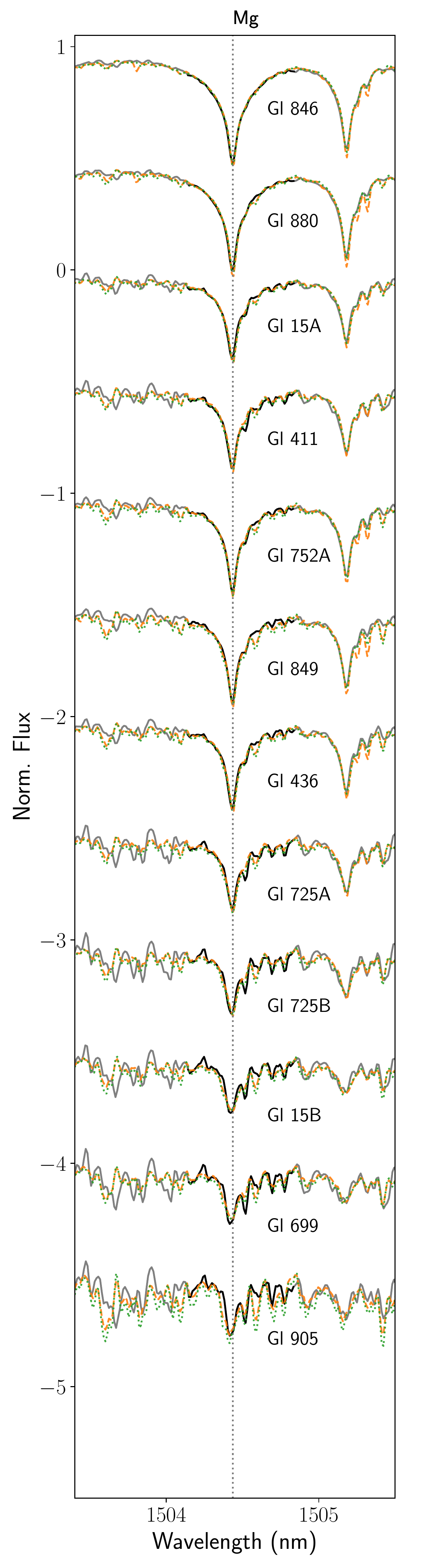}
    \includegraphics[width=.5\columnwidth]{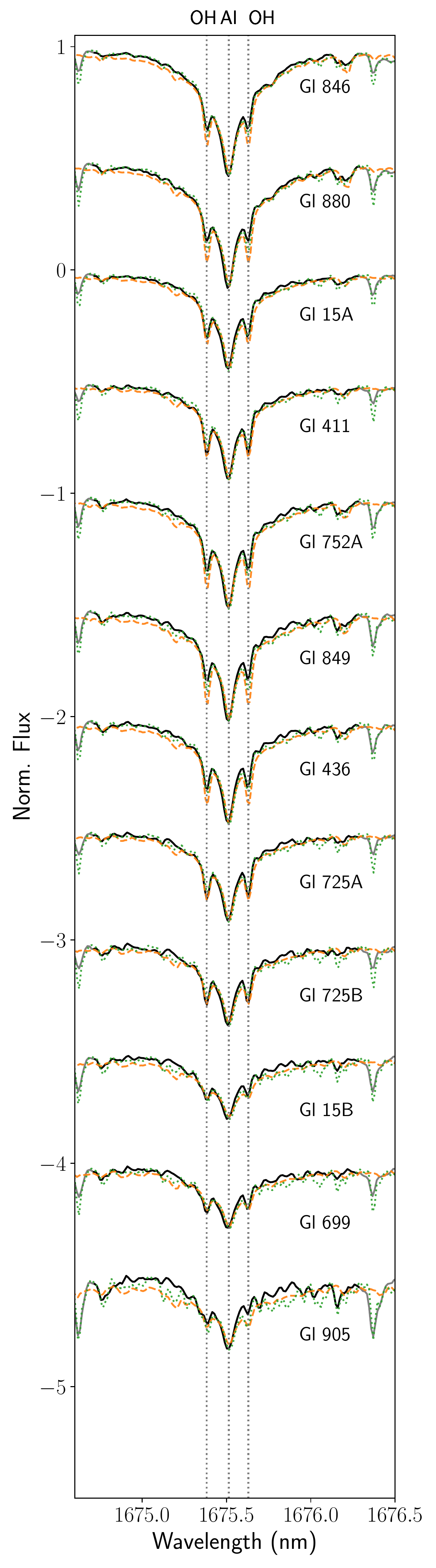}
    \includegraphics[width=.5\columnwidth]{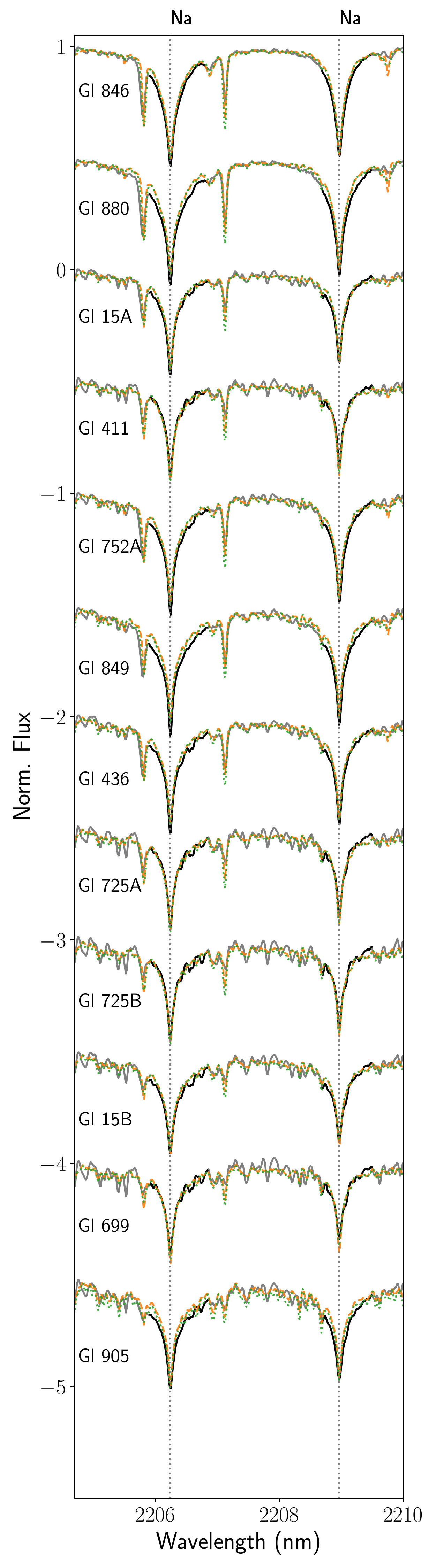}
    \includegraphics[width=.5\columnwidth]{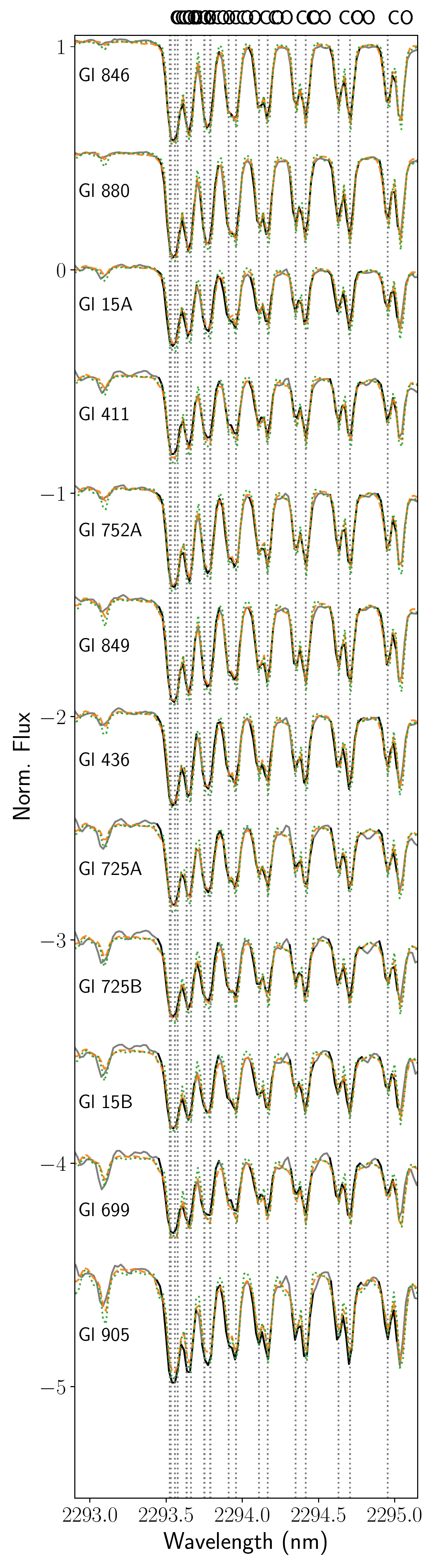}
    \contcaption{}
\end{figure*}

\begin{table*}
    \centering
       \caption{Fundamental parameters of the atomic lines included in the analysis as found in the lists used by the PHOENIX and MARCS models. For the lines with hyperfine structure, we display the parameters of all components. For the van der Waals parameter, values below 0 correspond to commonly reported $\log{\gamma_6}$, values between 0 and 5 correspond to the Unsöld factor, and values above 0 encode the two parameters defined in~\citet{barklem_2000}, with the integer part being the cross section for collisions by neutral hydrogen, and the fractional part being the velocity parameter $\alpha$.}
   \label{tab:line_list}
     \resizebox*{!}{0.90\textheight}{
    \begin{tabular}{c|ccc|cccccc}
    \hline
& \multicolumn{3}{c|}{PHOENIX} & \multicolumn{6}{c}{MARCS} \\
\hline
 & & & & & & & \multicolumn{3}{c}{damping parameters} \\
Species & Vacuum wavelength (nm) & $\chi_{\rm low}$ (eV) & $\log{gf}$ & Vacuum wavelength (nm) & $\chi_{\rm low}$ (eV) & $\log{gf}$ & Van der Waals & {\rm Rad.} & {\rm Stark}   \\
\hline
Na I & 2206.245 & 3.187 & 0.289 & 2206.324 & 3.191 & -0.519, & 2.000 & 5.000 & -- \\
Na I & 2208.969 & 3.187 & -0.019 & 2209.057 & 3.191 & -0.518, & 2.000 & 5.000 & -- \\
&  & & & 2209.052 & 3.191 & -1.217, & 2.000 & 5.000 & -- \\
&  & & & 2209.051 & 3.191 & -0.518, & 2.000 & 5.000 & -- \\
&  & & & 2206.331 & 3.191 & -1.218, & 2.000 & 5.000 & -- \\
&  & & & 2206.331 & 3.191 & -0.519, & 2.000 & 5.000 & -- \\
&  & & & 2206.330 & 3.191 & -0.072, & 2.000 & 5.000 & -- \\
&  & & & 2206.324 & 3.191 & -0.917, & 2.000 & 5.000 & -- \\
&  & & & 2206.324 & 3.191 & -0.519, & 2.000 & 5.000 & -- \\
&  & & & 2209.058 & 3.191 & -0.518, & 2.000 & 5.000 & -- \\
Mg I & 1504.436 & 5.098 & 0.119 & 1504.527 & 5.108 & 0.115, & -7.200 & 8.170 & -- \\
Al I & 1675.514 & 4.087 & 0.407 & 1675.709 & 4.087 & -0.506, & -7.220 & 7.560 & -- \\
Al & 1672.353 & 4.077 & 0.152 & 1672.547 & 4.085 & -0.55, & -7.150 & 7.560 & -- \\
Al I & 1315.435 & 3.136 & -0.030 & 1315.608 & 3.143 & -0.519, & 2.500 & 5.000 & -- \\
&  & & & 1315.609 & 3.143 & -1.063, & 2.500 & 5.000 & -- \\
&  & & & 1315.616 & 3.143 & -0.519, & 2.500 & 5.000 & -- \\
&  & & & 1315.615 & 3.143 & -0.616, & 2.500 & 5.000 & -- \\
K I & 1177.606 & 1.616 & 0.509 & 1177.866 & 1.617 & -1.87, & 649.270 & 7.810 & -5.170 \\
&  & & & 1177.866 & 1.617 & 0.084, & 649.270 & 7.810 & -5.170 \\
&  & & & 1177.866 & 1.617 & -0.724, & 649.270 & 7.810 & -5.170 \\
K I & 1243.568 & 1.608 & -0.438 & 1243.781 & 1.610 & -0.944, & 1258.183 & 7.790 & -4.880 \\
&  & & & 1243.781 & 1.610 & -1.643, & 1258.183 & 7.790 & -4.880 \\
&  & & & 1243.782 & 1.610 & -0.944, & 1258.183 & 7.790 & -4.880 \\
&  & & & 1177.866 & 1.617 & -1.694, & 649.270 & 7.810 & -5.170 \\
&  & & & 1243.781 & 1.610 & -0.944, & 1258.183 & 7.790 & -4.880 \\
&  & & & 1177.866 & 1.617 & -0.627, & 649.270 & 7.810 & -5.170 \\
&  & & & 1177.866 & 1.617 & -0.694, & 649.270 & 7.810 & -5.170 \\
&  & & & 1177.866 & 1.617 & -0.74, & 649.270 & 7.810 & -5.170 \\
K I & 1169.342 & 1.608 & 0.249 & 1169.609 & 1.610 & -0.556, & 648.269 & 7.810 & -5.170 \\
&  & & & 1169.609 & 1.610 & -0.556, & 648.269 & 7.810 & -5.170 \\
&  & & & 1169.609 & 1.610 & -0.954, & 648.269 & 7.810 & -5.170 \\
&  & & & 1169.609 & 1.610 & -0.109, & 648.269 & 7.810 & -5.170 \\
&  & & & 1169.609 & 1.610 & -0.556, & 648.269 & 7.810 & -5.170 \\
&  & & & 1169.609 & 1.610 & -1.255, & 648.269 & 7.810 & -5.170 \\
K I & 1177.286 & 1.616 & -0.449 & 1177.546 & 1.617 & -1.654, & 649.270 & 7.810 & -5.170 \\
&  & & & 1177.866 & 1.617 & -0.122, & 649.270 & 7.810 & -5.170 \\
&  & & & 1177.546 & 1.617 & -1.45, & 649.270 & 7.810 & -5.170 \\
&  & & & 1177.546 & 1.617 & -1.654, & 649.270 & 7.810 & -5.170 \\
&  & & & 1177.546 & 1.617 & -1.508, & 649.270 & 7.810 & -5.170 \\
&  & & & 1177.546 & 1.617 & -1.353, & 649.270 & 7.810 & -5.170 \\
&  & & & 1177.546 & 1.617 & -1.45, & 649.270 & 7.810 & -5.170 \\
&  & & & 1177.546 & 1.617 & -0.906, & 649.270 & 7.810 & -5.170 \\
&  & & & 1177.546 & 1.617 & -1.508, & 649.270 & 7.810 & -5.170 \\
K I & 1516.721 & 2.669 & -0.660 & 1516.802 & 2.670 & 0.632, & -6.820 & 7.640 & -- \\
&  & & & 1177.866 & 1.617 & -0.372, & 649.270 & 7.810 & -5.170 \\
&  & & & 1177.546 & 1.617 & -2.052, & 649.270 & 7.810 & -5.170 \\
&  & & & 1516.802 & 2.670 & -1.04, & -6.980 & 7.640 & -- \\
Ca I & 1034.665 & 2.927 & -0.407 & 1035.145 & 2.932 & -0.3, & -7.480 & 8.500 & -5.060 \\
Ti I & 967.820 & 0.834 & -0.898 & 968.633 & 0.836 & -0.804, & -7.800 & 6.250 & -6.090 \\
Ti I & 1281.498 & 2.160 & -1.364 & 1281.692 & 2.160 & -1.39, & -7.750 & 7.990 & -6.010 \\
Ti I & 1197.712 & 1.460 & -1.443 & 1197.956 & 1.460 & -1.39, & -7.790 & 6.870 & -6.100 \\
Ti & 1189.613 & 1.427 & -1.739 & 1189.863 & 1.430 & -1.73, & -7.790 & 6.930 & -6.100 \\
Ti I & 1066.454 & 0.817 & -1.996 & 1066.857 & 0.818 & -1.915, & -7.810 & 5.130 & -6.090 \\
Ti I & 1058.753 & 0.825 & -1.858 & 1059.172 & 0.826 & -1.775, & -7.810 & 5.130 & -6.090 \\
Ti & 972.162 & 1.501 & -1.257 & 972.941 & 1.503 & -1.181, & -7.780 & 6.161 & -6.110 \\
Ti I & 970.833 & 0.825 & -1.100 & 971.622 & 0.826 & -1.009, & -7.800 & 6.241 & -6.090 \\
Ti I & 969.153 & 0.812 & -1.707 & 969.955 & 0.813 & -1.61, & -7.800 & 6.241 & -6.090 \\
Ti I & 1571.987 & 1.872 & -1.287 & 1571.950 & 1.873 & -1.28, & -7.440 & 6.380 & -- \\
Ti I & 2296.961 & 1.885 & -1.616 & 2297.041 & 1.887 & -1.53, & -7.790 & 6.810 & -6.060 \\
Mn I & 1297.948 & 2.886 & -0.940 & 1298.133 & 2.888 & -1.797, & 2.500 & 5.000 & -- \\
Fe I & 1197.632 & 2.175 & -1.499 & 1197.877 & 2.176 & -1.483, & -7.820 & 7.190 & -6.220 \\
Fe I & 1169.317 & 2.220 & -2.076 & 1169.584 & 2.223 & -2.068, & -7.820 & 7.149 & -6.220 \\
\hline
    \end{tabular}
      }
\end{table*}

\newpage
{\textcolor{white}{.}}
\newpage
{\textcolor{white}{.}}
\newpage
{\textcolor{white}{.}}
\newpage

\section{{\p Results compared to other references}}

\begin{figure*}
    \centering
    \includegraphics[width=0.8\linewidth, trim={0 2.5cm 0 0cm}, clip]{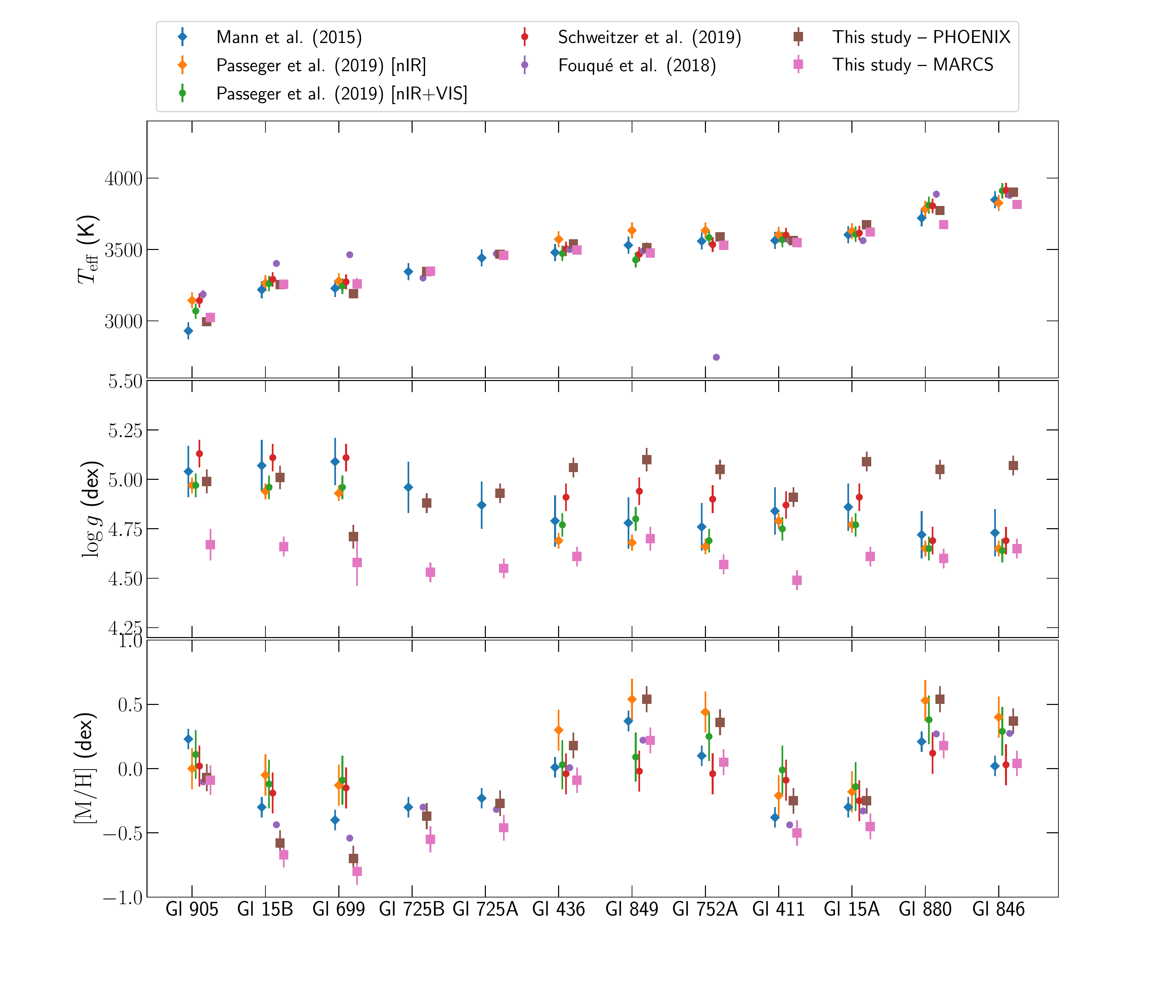}
    \caption{$\teff$, $\logg$ and $\mh$ values derived in this work, along with the values published by various studies for the stars in our sample.}
    \label{fig:teff_literature}
\end{figure*}
\begin{figure*}
    \subfigure{
    \includegraphics[width=\columnwidth]{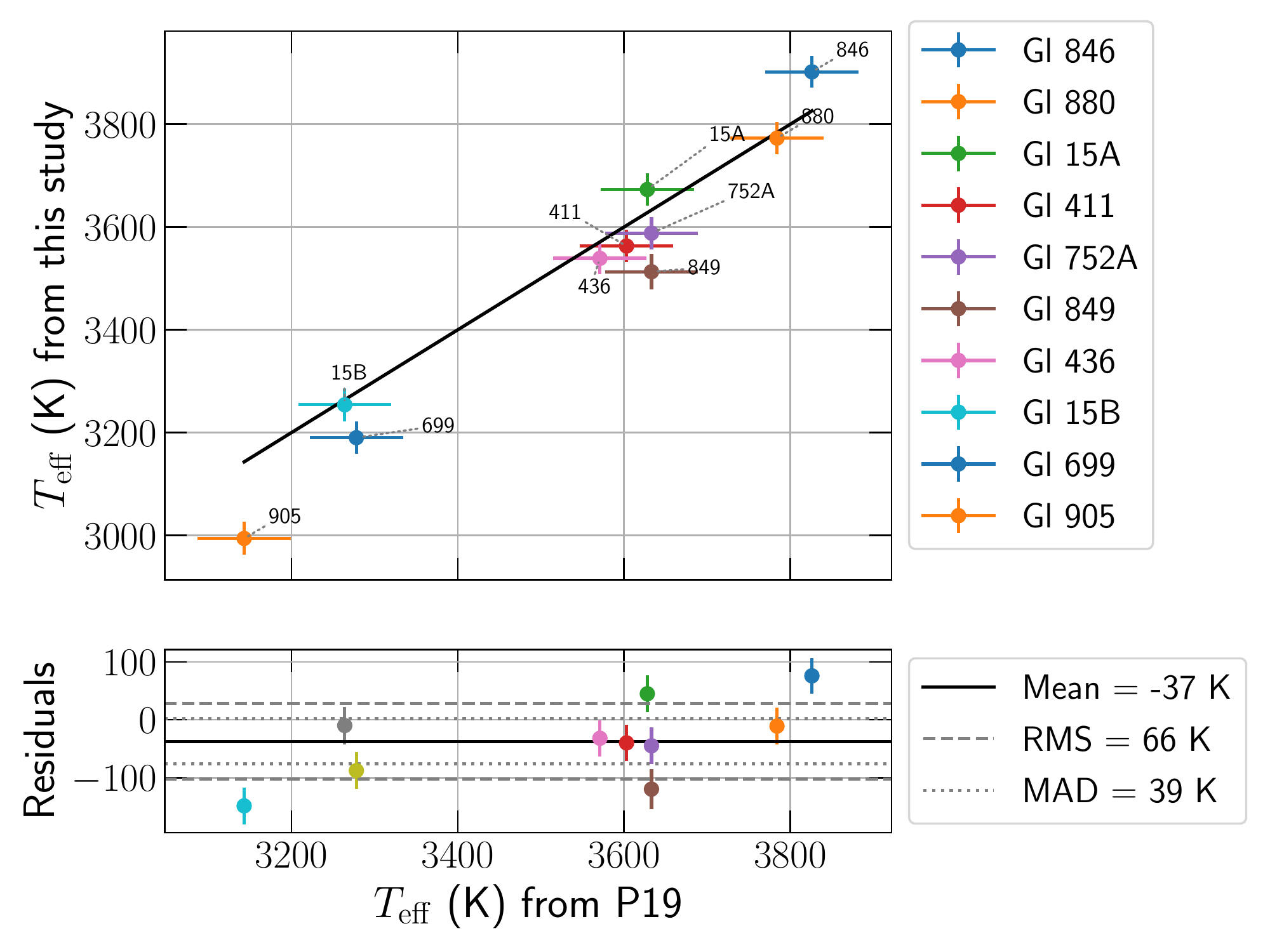}
    }
    \subfigure{
    \includegraphics[width=\columnwidth]{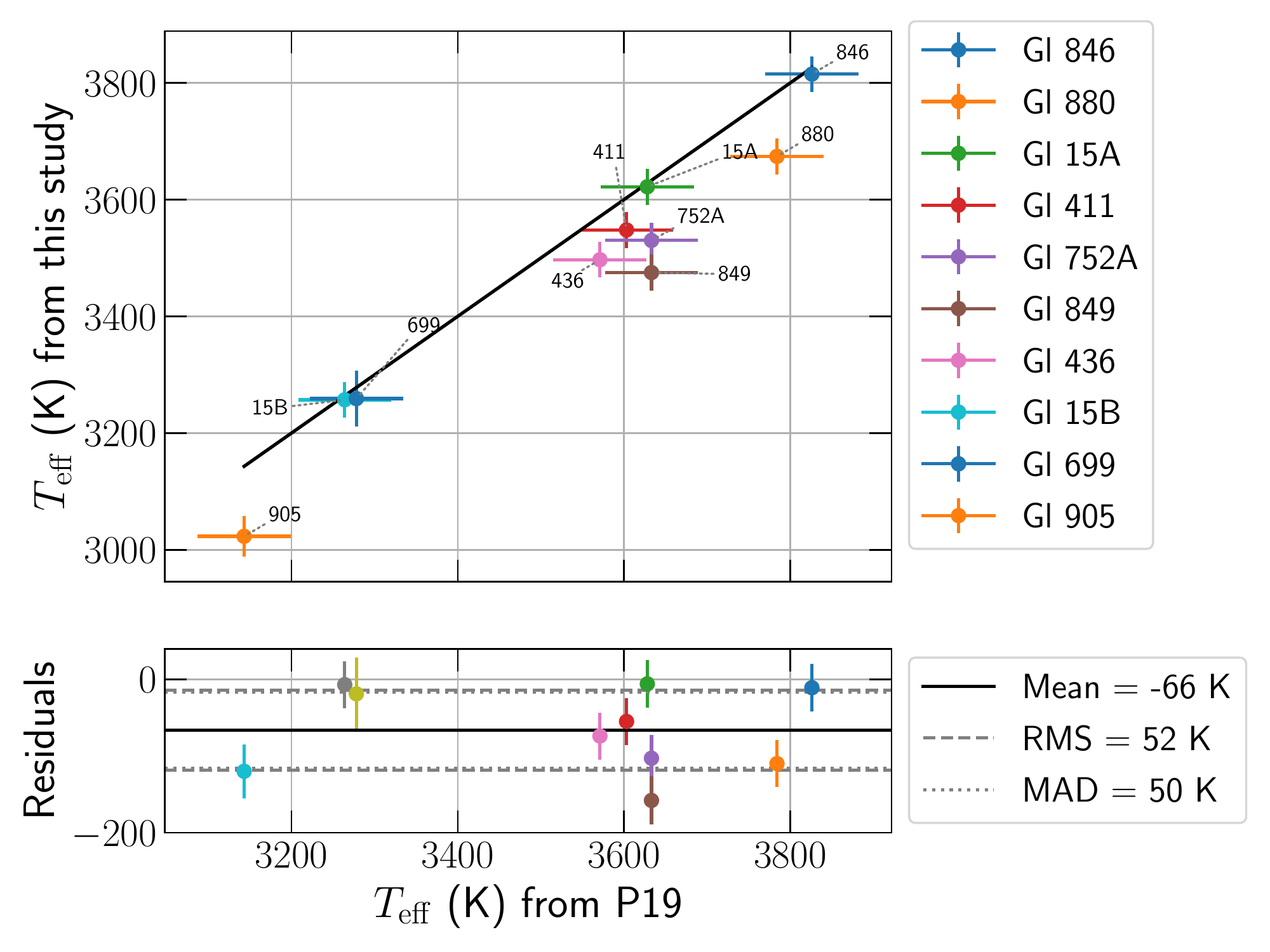}
    }
    \caption{Retrieved $\teff$ using the grid of PHOENIX (top) and MARCS (bottom) spectra plotted against values published by~\pass{}. The bottom plot presents the residuals, i.e. the retrieved values minus literature values. RMS and MAD values are computed after application of a sigma clipping function on the residuals with a threshold at 5~$\sigma$.}
    \label{fig:results_teff_pass}
\end{figure*}
\begin{figure*}
    \subfigure{
    \includegraphics[width=\columnwidth]{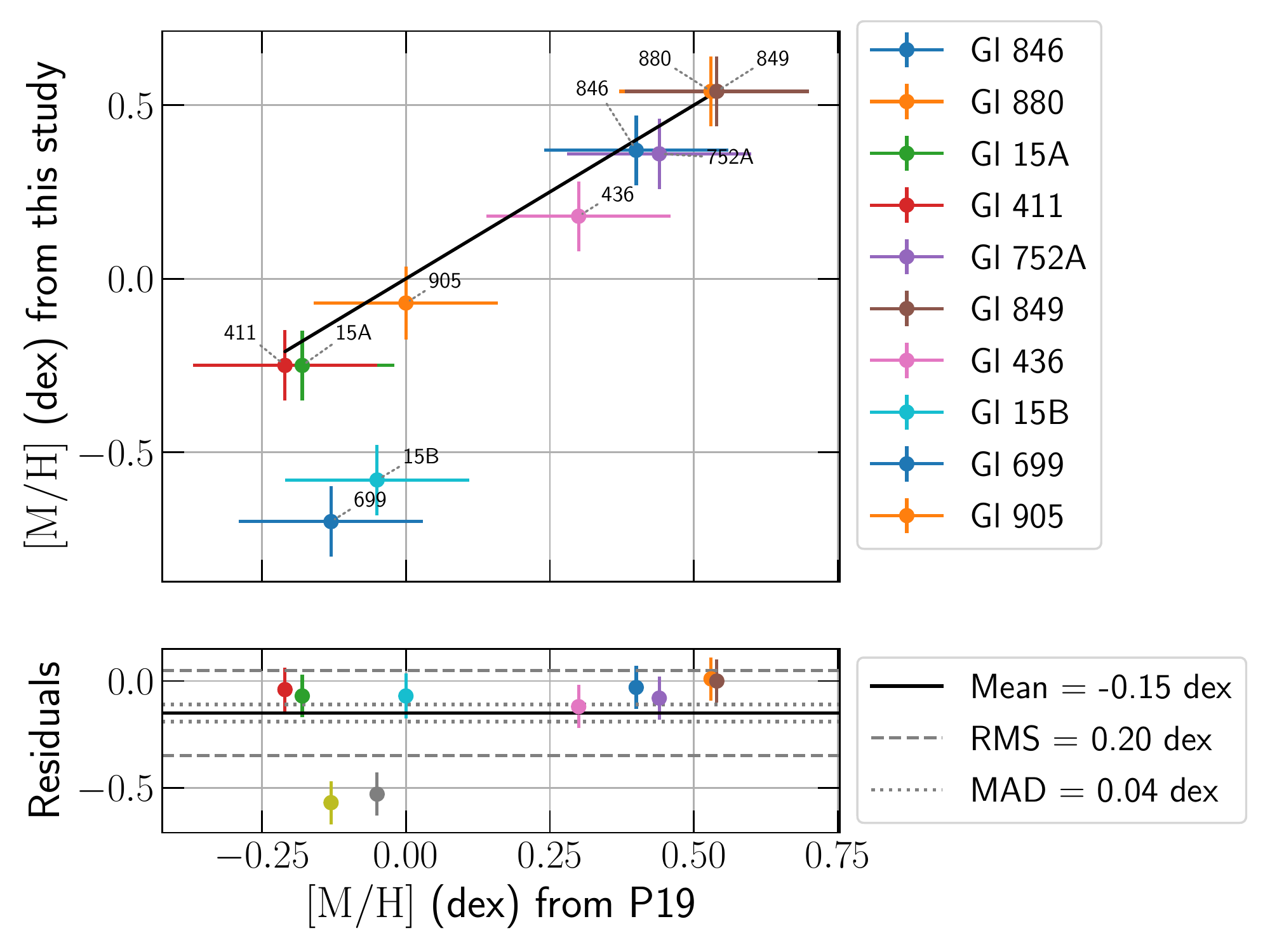}
    }
    \subfigure{
    \includegraphics[width=\columnwidth]{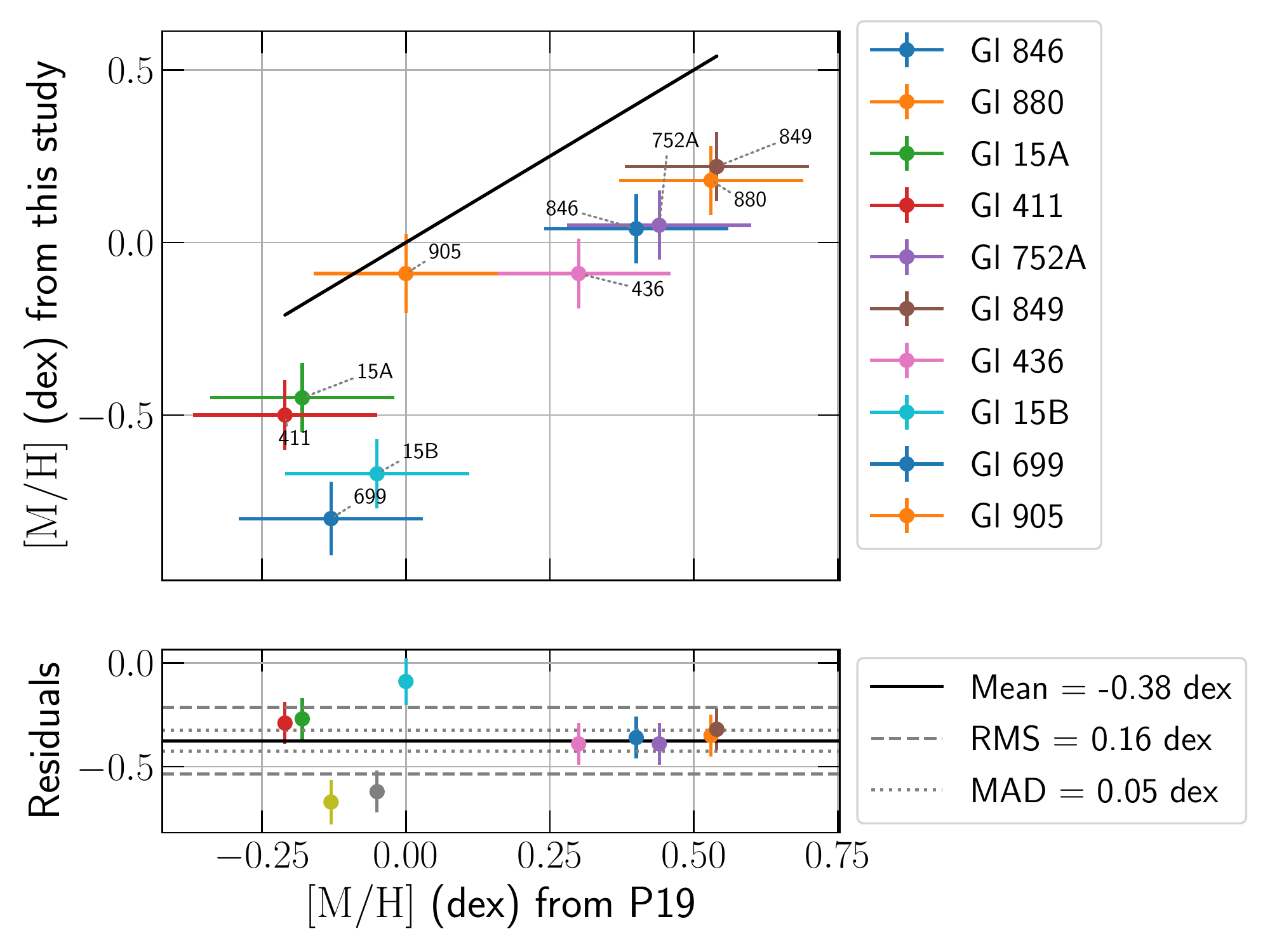}
    }
    \caption{Same as Fig.~\ref{fig:results_teff} but for $\mh$.}
    \label{fig:results_mh_pass}
\end{figure*}

{\p Fig.~\ref{fig:teff_literature} presents the $\teff$, $\logg$ and $\mh$ values published by several authors~\citep{mann_2015, passegger_2019, schweitzer_2019, fouque_2018} along with the parameters derived in this study. Fig.~\ref{fig:results_teff_pass} and Fig.~\ref{fig:results_mh_pass} present a comparison of the retrieved $\teff$ and $\mh$ values and those of~\pass{}.}

\newpage
{\textcolor{white}{.}}
\newpage

\section{Recovered parameters as a function of \texorpdfstring{$\vbroad$}{TEXT}}

{\p Table~\ref{tab:rms_fun_vb} presents the mean, RMS and MAD values of the residuals obtained with various values of $v_{\rm b}$.}

\begin{table*}
\caption{Mean, standard deviation and median absolute deviation values of the residuals for various values of $\vbroad$.}
\centering
\begin{tabular}[h]{c ccc ccc ccc c}
\hline
Model used & \multicolumn{3}{c}{$\teff$ (K)} & \multicolumn{3}{c}{$\logg$} & \multicolumn{3}{c}{[M/H]} & $\vbroad$ ($\kms$) \\
\hline
& MEAN & RMS & MAD & MEAN & RMS & MAD & MEAN & RMS & MAD &  \\
 \hline
 \multirow{7}{*}{PHOENIX} 
  & 56 & 33 & 34 & 0.16 & 0.23 & 0.21 & -0.03 & 0.21 & 0.15 & 0 \\ 
  & 52 & 29 & 24 & 0.17 & 0.24 & 0.21 & 0.01 & 0.23 & 0.16 & 1 \\ 
  & 43 & 31 & 28 & 0.15 & 0.23 & 0.19 & 0.02 & 0.22 & 0.16 & 2 \\ 
  & 28 & 33 & 29 & 0.11 & 0.21 & 0.17 & 0.04 & 0.23 & 0.16 & 3 \\ 
  & 19 & 35 & 28 & 0.08 & 0.22 & 0.18 & 0.06 & 0.23 & 0.16 & 4 \\ 
  & 10 & 42 & 27 & 0.08 & 0.21 & 0.11 & 0.11 & 0.21 & 0.17 & 5 \\ 
  & -11 & 43 & 36 & 0.0 & 0.21 & 0.13 & 0.1 & 0.23 & 0.14 & 6 \\ 
  \hline
 \multirow{7}{*}{MARCS} 
  & 32 & 40 & 24 & -0.2 & 0.15 & 0.13 & -0.21 & 0.14 & 0.11 & 0\\ 
  & 28 & 41 & 24 & -0.21 & 0.15 & 0.13 & -0.2 & 0.13 & 0.1 & 1\\ 
  & 17 & 40 & 26 & -0.24 & 0.15 & 0.12 & -0.2 & 0.13 & 0.1 & 2\\ 
  & 4 & 40 & 27 & -0.27 & 0.14 & 0.12 & -0.18 & 0.13 & 0.1 & 3\\ 
  & -17 & 40 & 25 & -0.33 & 0.15 & 0.12 & -0.16 & 0.13 & 0.1 & 4\\ 
  & -30 & 47 & 30 & -0.36 & 0.16 & 0.12 & -0.14 & 0.12 & 0.07 & 5\\ 
  & -47 & 45 & 36 & -0.4 & 0.16 & 0.12 & -0.11 & 0.12 & 0.06 & 6\\
  \hline
 \end{tabular}
\label{tab:rms_fun_vb}
\end{table*}

\bsp	
\label{lastpage}
\end{document}



\begin{figure*}[ht]
    \centering
    \hspace{-1.2cm}
    \includegraphics[width=.26\columnwidth]{Figures/region_0.pdf}
    \includegraphics[width=.26\columnwidth]{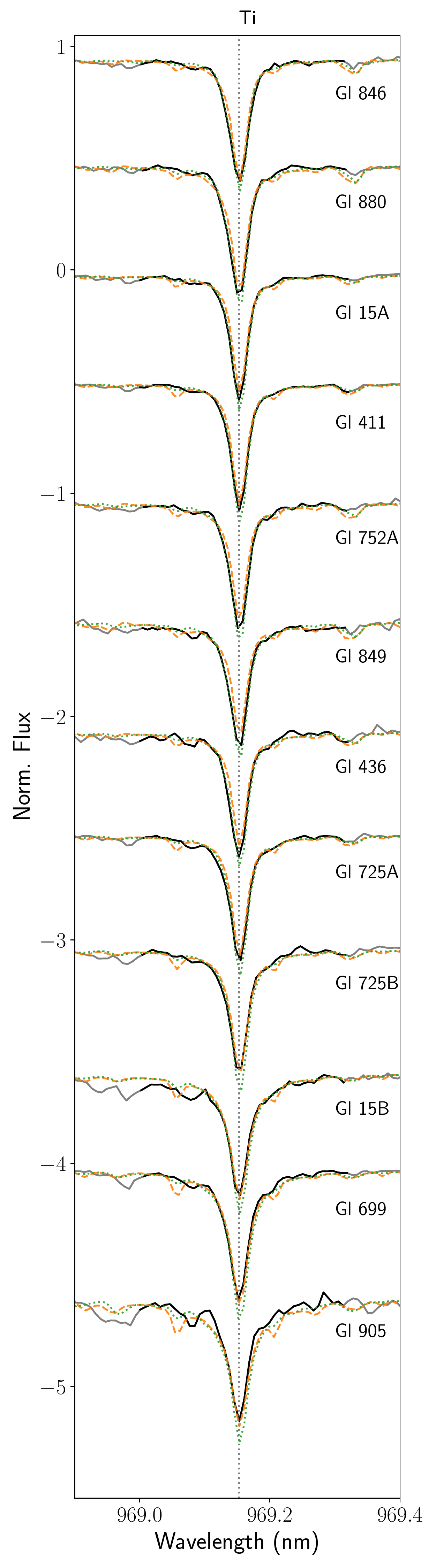}
    \includegraphics[width=.26\columnwidth]{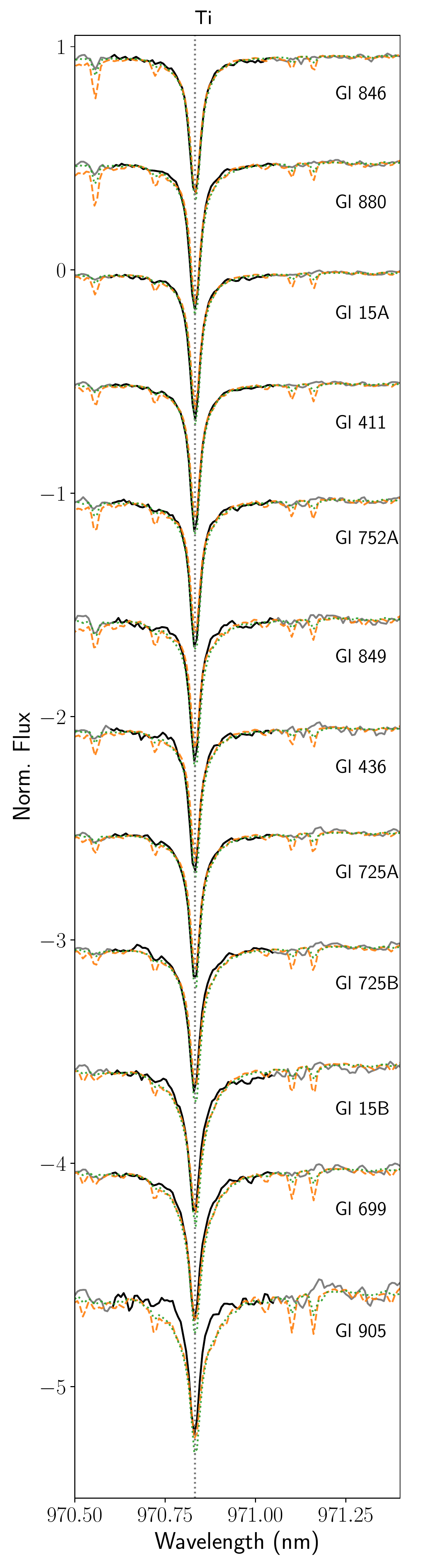}
    \includegraphics[width=.26\columnwidth]{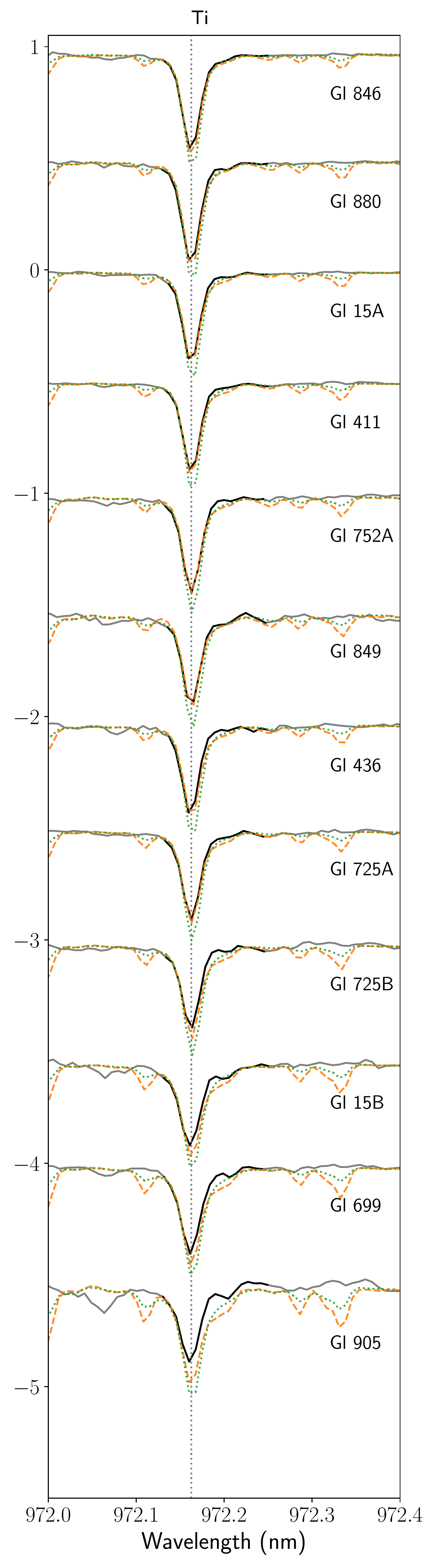}
    \caption*{\textbf{Figure A2.} Template spectra (grey) along with the fitted PHOENIX model (dashed orange) and MARCS model (dotted green) for the different spectral windows. Selected regions of the template spectra over which the comparison was carried out are shown in black. From top to bottom: Gl~846, Gl~880, Gl~15A, Gl~411, Gl~752A, Gl~849, Gl~436, Gl~725A, Gl~725B, Gl~15B, Gl~699 and Gl~905. Every spectrum but the first one is shifted by a multiple of 0.5 for better readability.}
    \label{fig:lines_1}
\end{figure*}

\begin{figure*}[ht]
    \centering
        \hspace{-1.2cm}
    \includegraphics[width=.26\columnwidth]{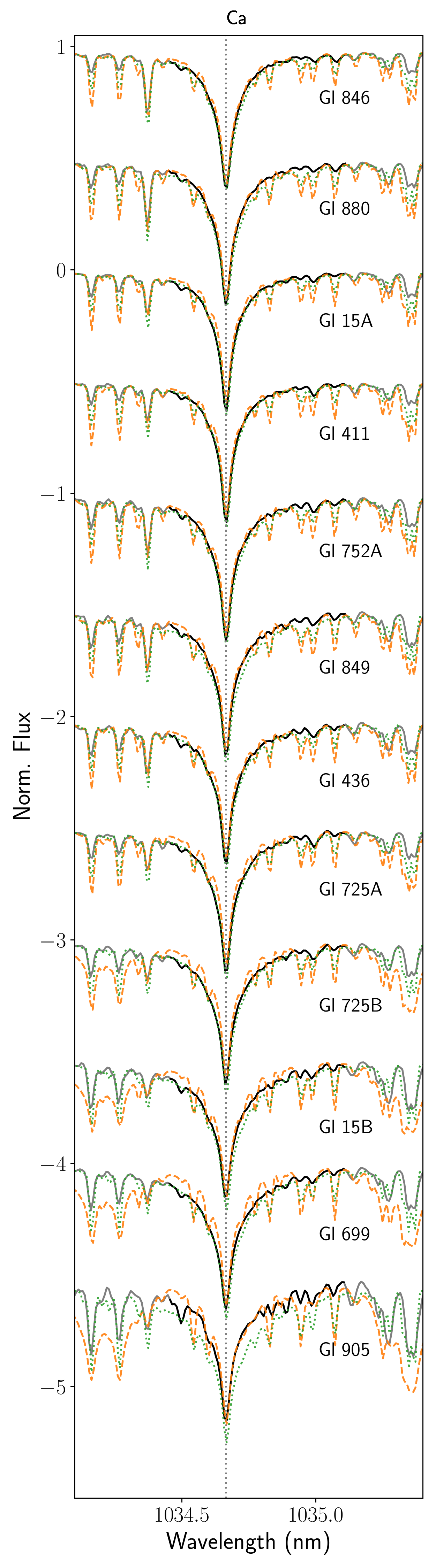}
    \includegraphics[width=.26\columnwidth]{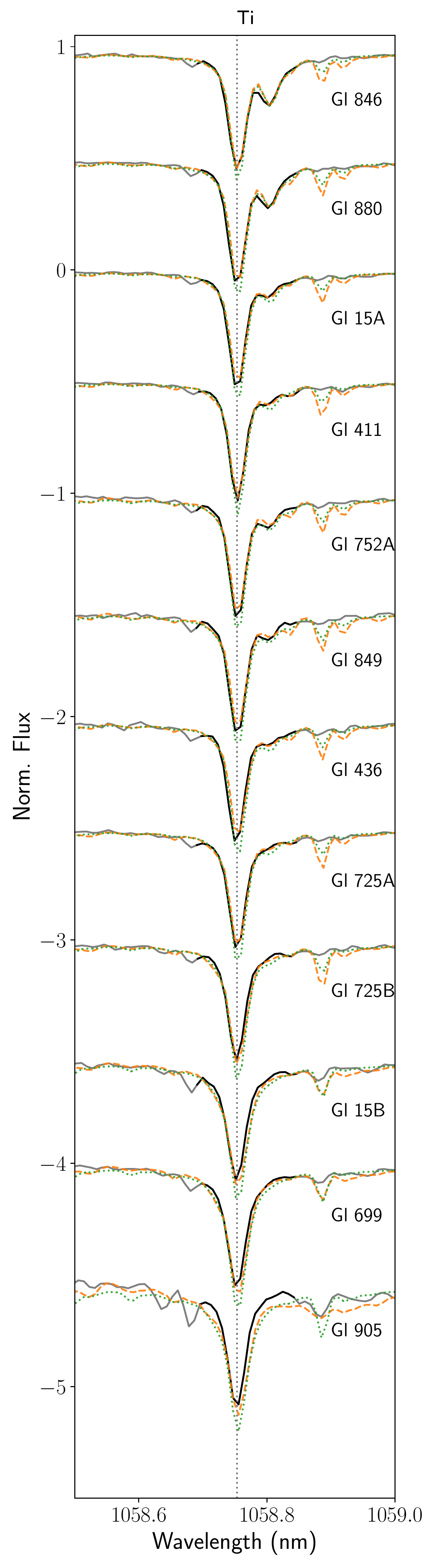}
    \includegraphics[width=.26\columnwidth]{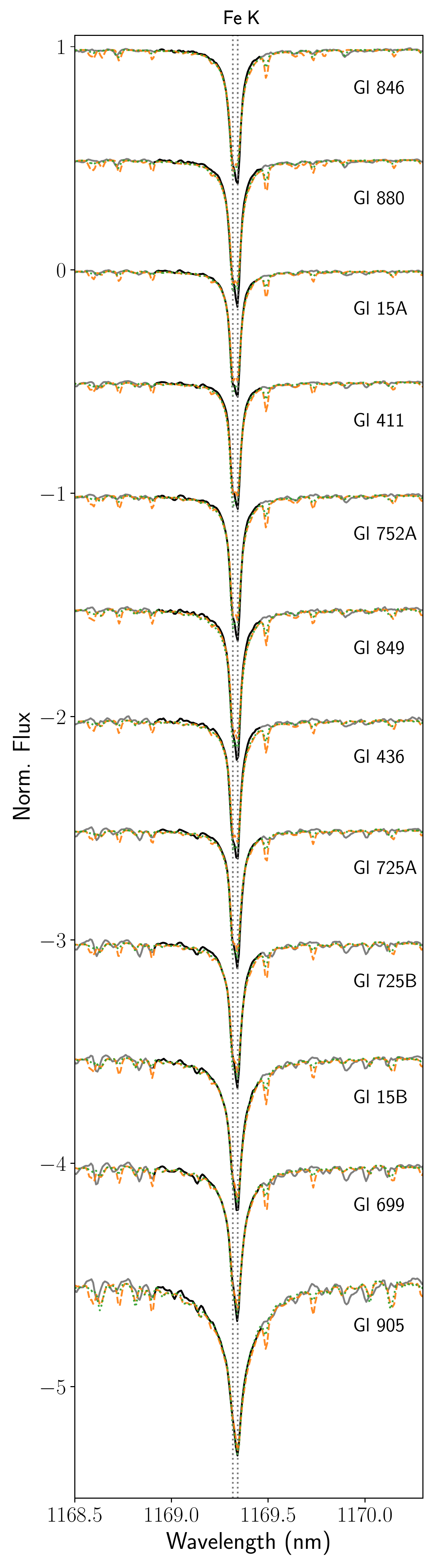}
    \includegraphics[width=.26\columnwidth]{Figures/region_7.pdf}
    \caption*{\textbf{Figure A2} -- \textit{continued}}
\end{figure*}

\begin{figure*}[ht]
    \centering
        \hspace{-1.2cm}
    \includegraphics[width=.26\columnwidth]{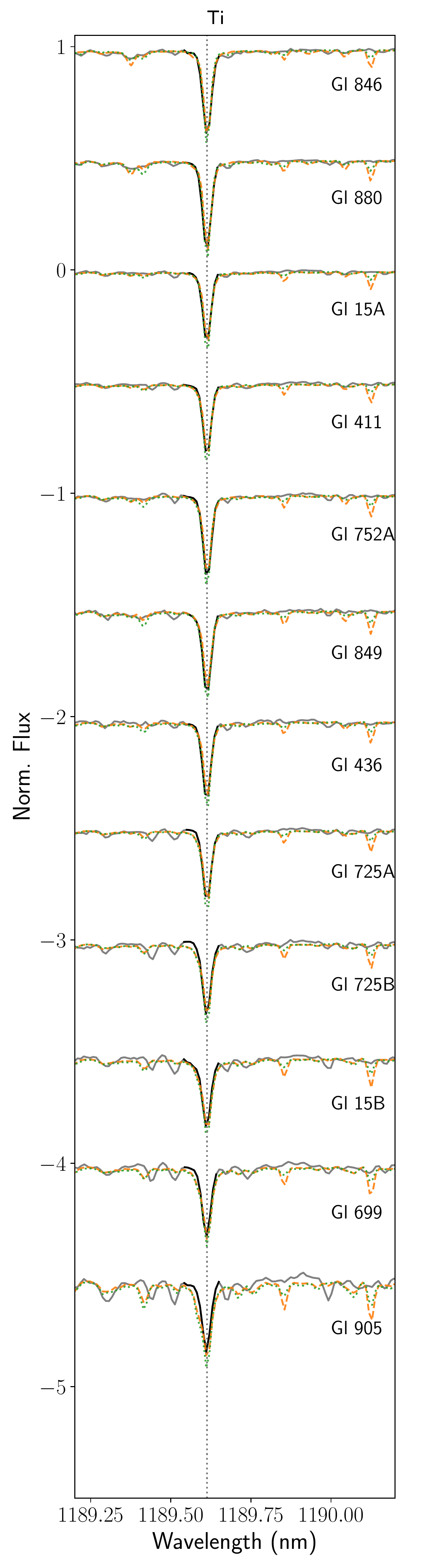}
    \includegraphics[width=.26\columnwidth]{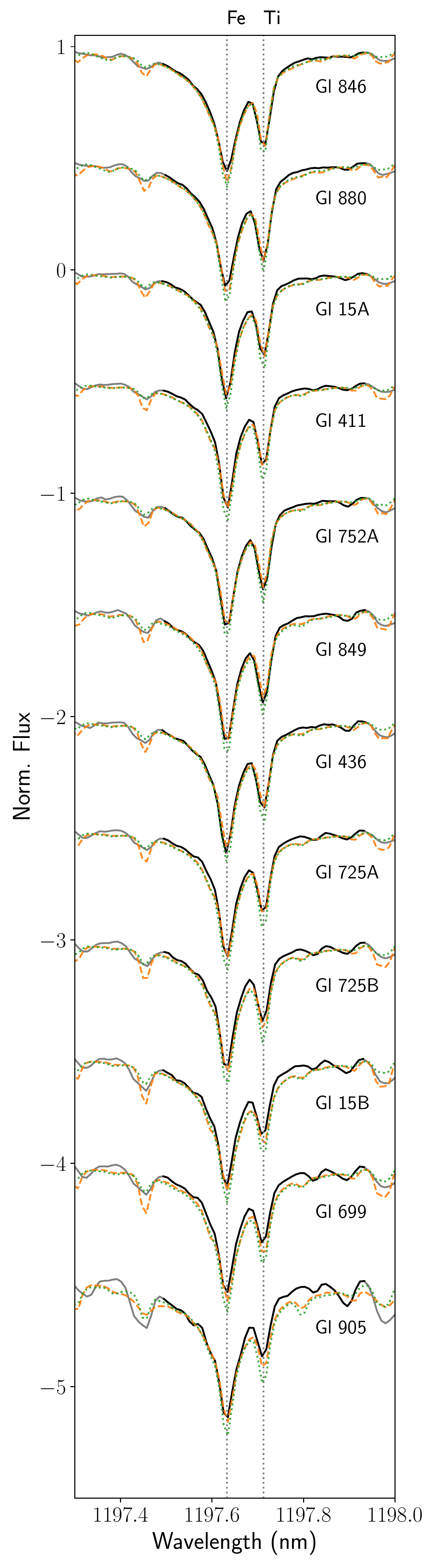}
    \includegraphics[width=.26\columnwidth]{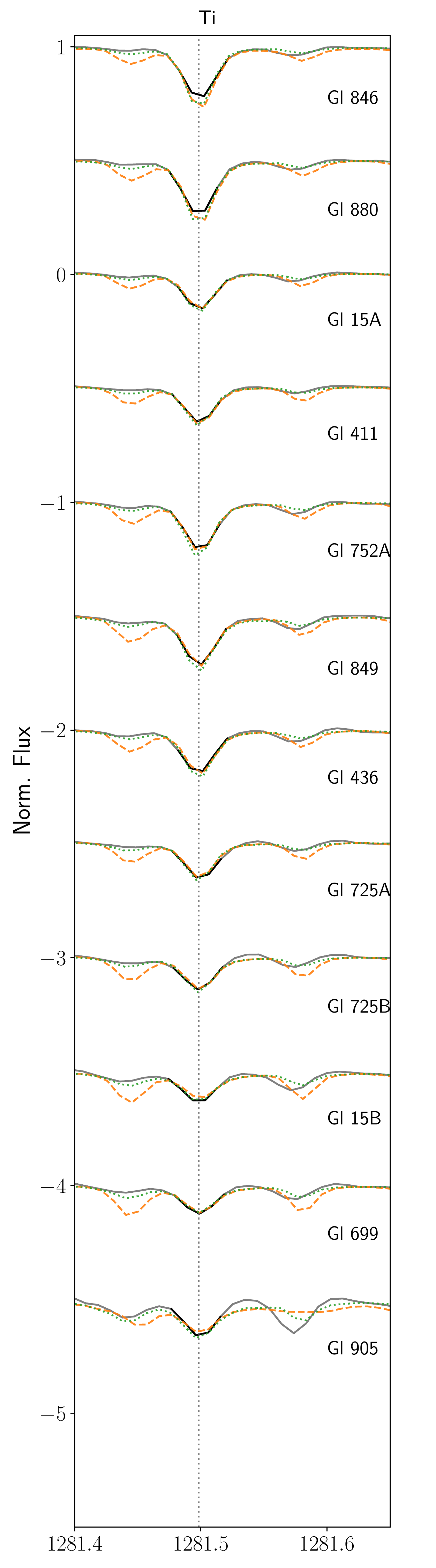}
    \includegraphics[width=.26\columnwidth]{Figures/region_11.pdf}
    \caption*{\textbf{Figure A2} -- \textit{continued}}
\end{figure*}

\begin{figure*}[ht]
    \centering
        \hspace{-1.2cm}
    \includegraphics[width=.26\columnwidth]{Figures/region_12.pdf}
    \includegraphics[width=.26\columnwidth]{Figures/region_13.pdf}
    \includegraphics[width=.26\columnwidth]{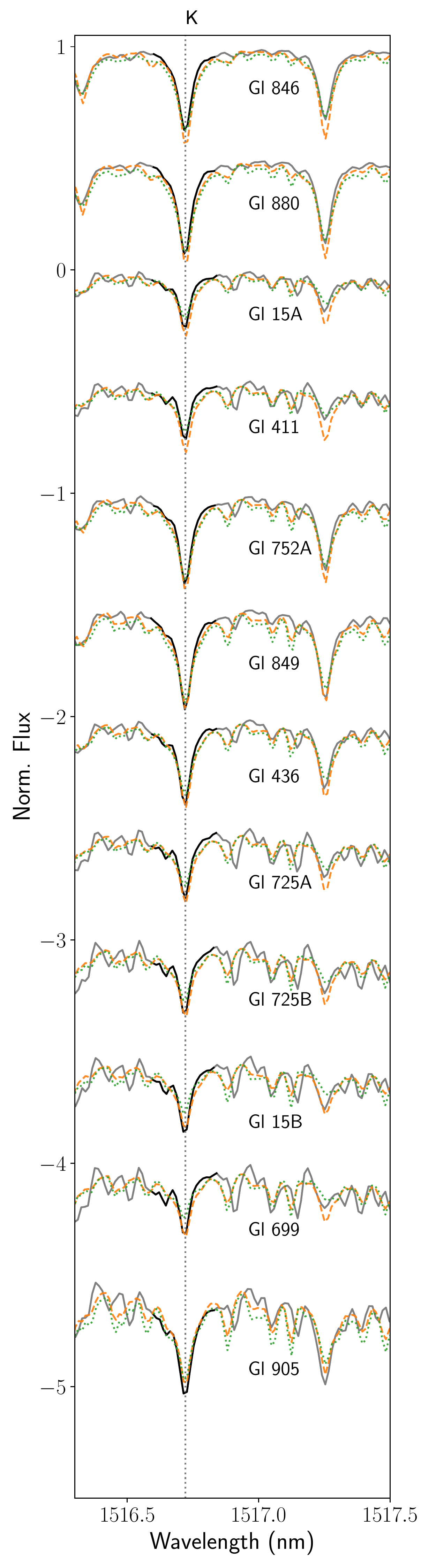}
    \includegraphics[width=.26\columnwidth]{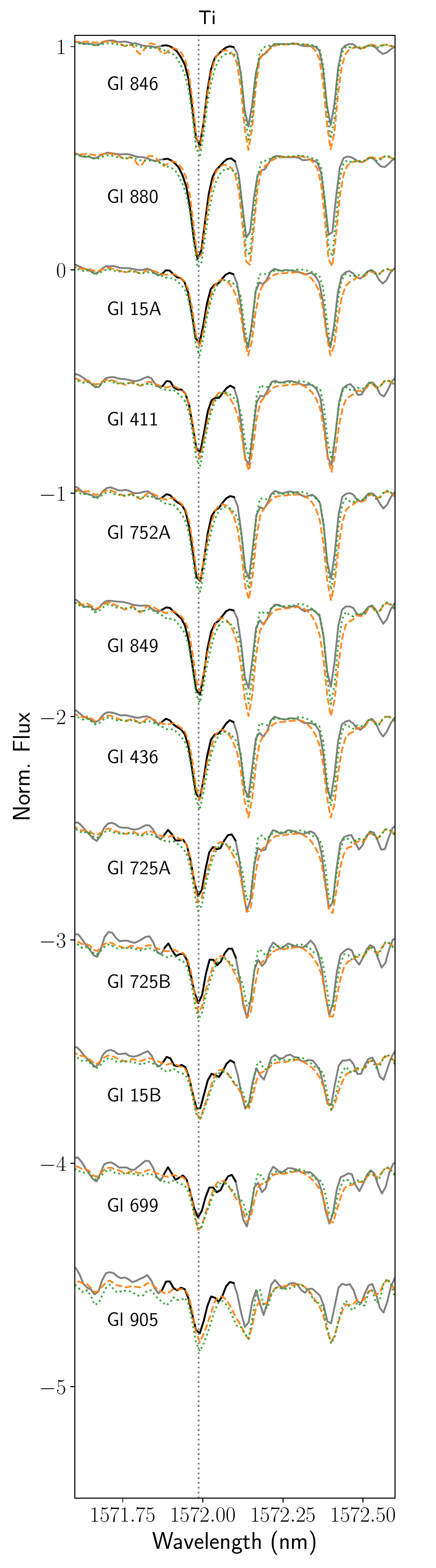}
    \caption*{\textbf{Figure A2} -- \textit{continued}}
\end{figure*}

\begin{figure*}[ht]
    \centering
        \hspace{-1.2cm}
    \includegraphics[width=.26\columnwidth]{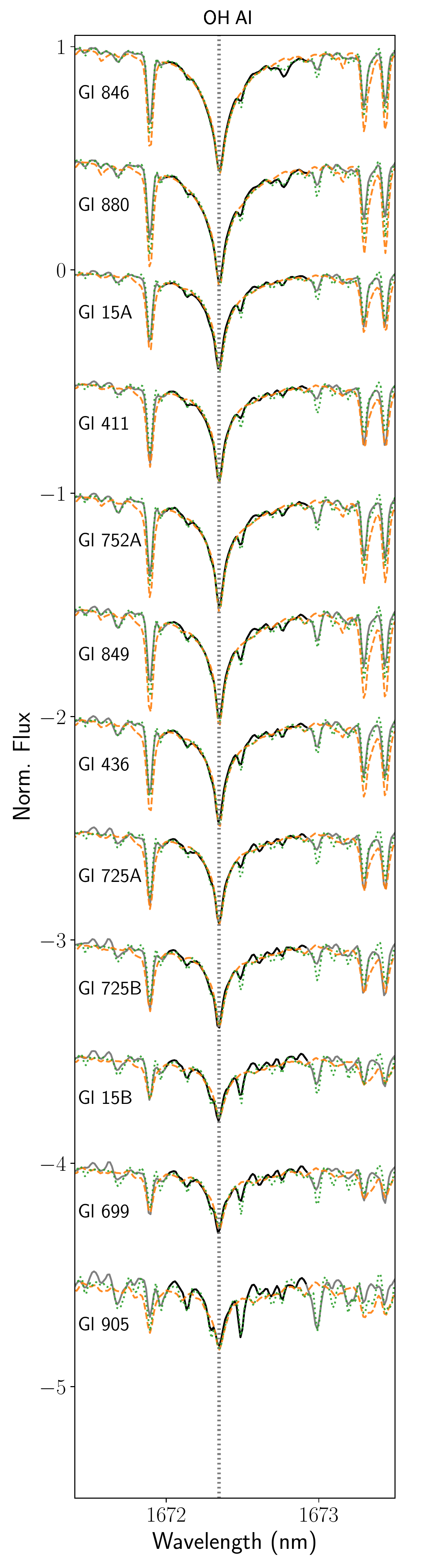}
    \includegraphics[width=.26\columnwidth]{Figures/region_17.pdf}
    \includegraphics[width=.26\columnwidth]{Figures/region_18.pdf}
    \includegraphics[width=.26\columnwidth]{Figures/region_19.pdf}
    \caption*{\textbf{Figure A2} -- \textit{continued}}
\end{figure*}

\begin{figure*}[ht]
    \centering
        \hspace{-1.2cm}
    \includegraphics[width=.26\columnwidth]{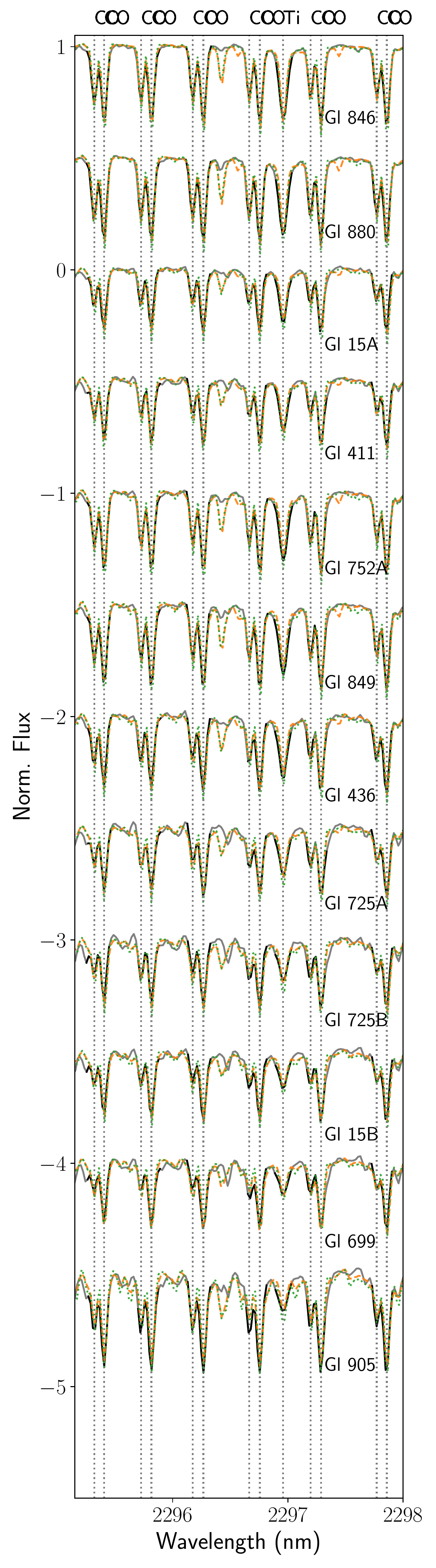}
    \includegraphics[width=.26\columnwidth]{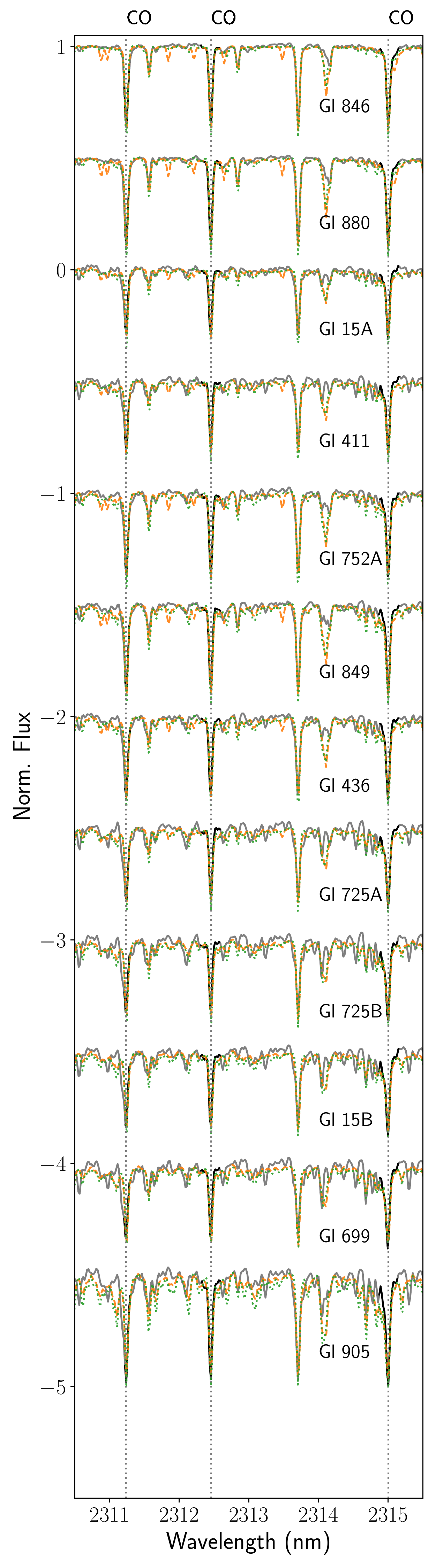}
    \caption*{\textbf{Figure A2} -- \textit{continued}}
\end{figure*}